\documentclass[11pt,letterpaper]{article}
\usepackage[margin=1in]{geometry}
\usepackage{amsmath,amsthm,amssymb,amsfonts,amscd,amstext}
\usepackage{graphicx} 
\usepackage[table]{xcolor} 
\usepackage{mathtools}
\usepackage{url} 
\usepackage{authblk} 
\usepackage{algorithm} 
\usepackage{algpseudocode}
\usepackage{natbib} 
\usepackage{multirow}
\usepackage[font=scriptsize]{caption}
\usepackage{enumitem}
\usepackage[htt]{hyphenat} 
\usepackage{longtable}
\usepackage{thmtools,thm-restate}
\usepackage{setspace}

\providecommand{\keywords}[1]{\par\medskip\noindent\textbf{Keywords:} #1\par\medskip}

\newcounter{savealgorithm}

\makeatletter
\newcommand{\algmargin}{\the\ALG@thistlm}
\makeatother

\algrenewcommand\algorithmicrequire{\textbf{Input:}}
\algrenewcommand\algorithmicensure{\textbf{Output:}}

\algnewcommand{\parState}[1]{\State%
    \parbox[t]{\dimexpr\linewidth-\algmargin}{\strut\hangindent=\algorithmicindent \hangafter=1 #1\strut}}

\newcommand\independent{\protect\mathpalette{\protect\independenT}{\perp}}
\def\independenT#1#2{\mathrel{\rlap{$#1#2$}\mkern2mu{#1#2}}}

\theoremstyle{plain}
\newtheorem{theorem}{Theorem}
\newtheorem{lemma}{Lemma}

\newtheorem{Assumption}{Assumption}

\newtheorem{Definition}{Definition}

\newtheorem{Lemma}{Lemma}

\newtheorem{Corollary}[Lemma]{Corollary}

\theoremstyle{remark}
\newtheorem{Remark}{Remark}

\newcommand{\IP}{{\rm I}\kern-0.18em{\rm P}}
\newcommand{\1}{{\rm 1}\kern-0.24em{\rm I}}
\newcommand{\E}{{\rm I}\kern-0.18em{\rm E}}
\newcommand{\R}{{\rm I}\kern-0.18em{\rm R}}

\allowdisplaybreaks

\begin{document}

\title{A semi-supervised framework for diverse multiple hypothesis testing scenarios}

\author[1*]{Jack Freestone}
\author[2,3]{William Stafford Noble}
\author[4*]{Uri Keich}
\affil[1]{School of Mathematical and Physical Sciences, Macquarie University, New South Wales 2109, Australia}
\affil[2]{Department of Genome Sciences, University of Washington}
\affil[3]{Paul G.\ Allen School of Computer Science and Engineering, University of Washington}
\affil[4]{School of Mathematics and Statistics F07, University of Sydney, New South Wales 2006, Australia}

\date{}

\maketitle

$^*$Corresponding authors: jack.freestone@mq.edu.au, uri.keich@sydney.edu.au

\begin{abstract}
Standard multiple testing procedures are designed to report a list of discoveries, or suspected false null hypotheses,
given the hypotheses' p-values or test scores. Recently there has been a growing interest in enhancing such procedures
by combining additional information with the primary p-value or score.
In line with this idea, we develop RESET (REScoring via Estimating and Training), which splits the data into a training part and an estimating part so that any semi-supervised learning approach can factor in the available side information while maintaining finite-sample error-rate control. Our practical implementation, RESET Ensemble, selects from an ensemble of classification algorithms so that it is compatible with a range of multiple testing scenarios without the need for the user to select the appropriate one. We apply RESET to both p-value and competition based multiple testing problems and show that RESET is (1) power-wise competitive, (2) fast compared to most tools and (3) able to uniquely achieve finite sample false discovery rate or false discovery exceedance control, depending on the user's preference.
\end{abstract}
\keywords{multiple hypothesis testing, semi-supervised learning, FDR control, FDX control, competition, proteomics}


\section{Introduction}

Scientists are often interested in testing many null hypotheses, $\{H_i : i = 1, \dots, m \}$. 
Large-scale hypothesis testing is usually achieved by assigning a p-value, $p_i$, or a score, $Z_i$, to each null hypothesis. 
The collection of p-values, or scores, then undergoes a filtering process that rejects a subset of these null hypotheses subject to type-1 error rate control to minimize the number of misreported true null hypotheses. 

The most common choice of error rate is the false discovery rate (FDR), which is defined as the expectation of the false discovery proportion (FDP)
--- the fraction of true null hypotheses in the reported list of rejections, or \textit{discoveries}~(\cite{benjamini:controlling}). 
In this case, the filtering procedure ensures that the FDR is $\leq \alpha$ where $\alpha \in \left(0, 1 \right)$ is a prespecified threshold. 
Alternatively, we might be interested in controlling false discovery exceedance (FDX), which is the upper tail probability of the FDP. 
The FDX is said to be controlled with threshold $\alpha$ at confidence $1 - \gamma$ if $\mathbb{P}(FDP > \alpha ) \leq \gamma$. 
Evidently, such FDX control reduces the chance that the realized FDP exceeds $\alpha$, which is not guaranteed by FDR control~(\cite{guo:generalized,katsevich:simultaneous,goeman:only,li2024simultaneous,luo:competition,ebadi:bounding}).

Recently, researchers have focused on developing methods that leverage side information $\mathbf{x}_i \in \mathcal{X}$, which complements each p-value, or score, to better discriminate between true and false null hypotheses.
These methods utilize a variety of strategies but ultimately operate with the same goal in mind: the side information is used to weight or rescore the hypotheses according to how confidently we believe that each hypothesis is a false null. 

In this paper, we offer a novel approach for multiple testing with side information. 
The idea is to randomly split a set of suspected true nulls into two sets, where one set is used for training a semi-supervised learning model to discriminate between the true and false null hypotheses, and the other for estimating the number of false discoveries. 
Accordingly, we refer to our new approach as \textit{RESET}---REScoring via Estimating and Training. 
Our method flexibly incorporates any semi-supervised approach to help distinguish between true and false null hypotheses. 
Moreover, it is applicable to both competition- and p-value-based multiple testing scenarios, and controls either the FDR or the FDX, as pre-specified in advance.
We evaluate RESET in a range of simulations and real data applications to verify that it is a competitive alternative to existing methods in terms of statistical power and speed.

\section{Background}
\label{sec:background}

We first review the competition-based multiple testing framework in Section~\ref{sec:background_comp}, including its extension to side information in Section~\ref{sec:comp_side}. 
We then describe the p-value-based multiple testing framework in Section~\ref{sec:background_pvalue} and its side-information extensions in Section~\ref{sec:pval_side}. Finally, in Section~\ref{sec:related_work_novelty} we discuss related work in conformal inference.

\subsection{Competition-based multiple testing}
\label{sec:background_comp}
In competition-based multiple testing, each hypothesis $H_i$ is associated with two scores: a test or \textit{target} score, $Z_i$, where larger values indicate greater evidence against the null, and a \textit{decoy} score, $\tilde{Z}_i$, a random draw from the null distribution.
Each pair $(Z_i,\tilde{Z}_i)$ is evaluated in a head-to-head competition yielding a label $L_i \in \{\pm 1 \}$ that indicates which score is higher and a winning score $W_i$:
\[ L_i := \begin{cases} 
      +1 & Z_i > \tilde{Z}_i \\
      -1 & Z_i < \tilde{Z}_i
   \end{cases}, \quad W_i := f(Z_i, \tilde{Z}_i),
\]
where $f$ is symmetric in its arguments and increases with the values of $Z_i$ and $\tilde{Z}_i$, or with the difference between them, e.g., $f(x, y) = \max \{x, y \}$ or $f(x, y) = \lvert x - y \rvert$. Hypotheses with ties (i.e., $Z_i = \tilde{Z}_i$) are either discarded or the ties are randomly broken. 
We refer to a hypothesis with $L_i = 1$ as a target win or a target for short, and $L_i = -1$ as a decoy win or a decoy for short. 
By construction, the false nulls with large $Z_i$ will likely win their competition, resulting in a target win ($L_i = 1$) with large $W_i$.
The other ingredient is an assumption that is essential to type-1 error control guarantees in competition-based multiple testing:
\begin{quote}
(A0)~Conditioned on all the scores $W$ and the labels of the false null hypotheses, the true null labels $\{L_i : i \in N\}$, where $N$ denotes the indices of the true null hypotheses, are \textit{i.i.d.}\ $\pm 1$ uniform random variables.
\end{quote}
Under~(A0), the number of decoy wins ($L_i = -1$) that score above a cutoff $\tau$ can be used to conservatively estimate the number of true null hypotheses with $L_i = 1$ that are also above $\tau$.
Using this strategy, we can report a list of discoveries in one of two ways. 

The first and by far the more common way is to apply \textit{Selective SeqStep+} (SSS+, Algorithm~\ref{alg:SeqStep})~(\cite{barber:controlling}), which controls the FDR under~(A0) at a prespecified threshold $\alpha$.
Assuming the labels $L$ are reranked according to the scores $W$ in descending order, SSS+ reports the following list of discoveries, $R_\alpha$:
\begin{align}
    \label{discovery_fdr}
    R_\alpha := \{i \leq \tau : L_i = 1 \}, \text{ where }\tau := \max \left\{t \in [m] : \frac{\#\{i \leq t, L_i = -1 \} + 1}{ \#\{i \leq t, L_i = 1 \} \vee 1 } \leq \alpha \right\},
\end{align}
where $[m] := \{1, 2, \dots, m \}$. 
Notably, SSS+ can be adjusted to handle a more general setting when the probabilities of the positive and negative labels of the true null hypotheses are no longer uniform~(\cite{barber:controlling,lei:power}).

Alternatively, we can control the FDX by using, for example, \textit{FDP-stepdown} (FDP-SD, Algorithm~\ref{alg:FDP-SD}) (\cite{luo:competition}). 
FDP-SD works by first reranking the labels $L$ according to the scores $W$ in descending order.
Next, it compares the number of decoy wins in the first $i$ hypotheses, denoted as $D_i := \#\{ L_j = -1 : j \leq i\}$,
with a precomputed bound $\delta_i$, which is determined by $\alpha$ and $\gamma$. 
The comparisons are made in a `stepdown' fashion: we test whether $D_i \leq \delta_i$ only if $D_j \leq \delta_j$ for $j < i$. 
For small $i$, it is impossible for $D_i \leq \delta_i$ and so FDP-SD begins this analysis at $i_0$, the smallest possible index for which $D_i \leq \delta_i$ can occur. 
Under~(A0), reporting the following list of discoveries $R_{\alpha,\gamma}$ controls the FDX:
\begin{align}
    \label{discovery_fdp}
    R_{\alpha,\gamma} := \{i \in [m] : i \leq k_{FDP}, L_{i} = 1 \}, \quad k_{FDP} := \max \{i : \Pi_{j = i_0}^i 1_{D_j \leq \delta_j} = 1 \text{ or } i = 0 \}.
\end{align}
\subsubsection{Extension to side information}
\label{sec:comp_side}


To our knowledge, the first `intrinsic' competition-based multiple testing procedure that uses arbitrary side information in a data-driven manner is \textit{Adaptive Knockoffs} (AdaKO) (\cite{ren:knockoffs}), which we compare directly against; a further, more recent such approach is discussed in Section~\ref{sec:related_work_novelty}.
AdaKO is a specialization of the p-value based procedure \textit{AdaPT} (\cite{lei:adapt}) that we introduce in the next section with further details in Section~\ref{sec:summary_of_methods}. 
By `intrinsic', we mean that while other p-value based approaches may be modified to be used in the competition setting, \cite{ren:knockoffs} demonstrate they are substantially less powerful, and are not viable in this sense. 
Adaptive Knockoffs works by iteratively pruning $\mathcal{\hat S}_t$, the candidate set of hypotheses at each step $t$, by selecting the hypothesis it deems most likely to be a true null. 
To do that, it fits a classification algorithm using a strict subset of information that partially masks the data: the scores $W$, the side information $\mathbf{x}$, and the labels $(L_i)_{i \in [m]\setminus \mathcal{\hat S}_t}$ of the hypotheses that were pruned up to that point, as well as $\# \{i \in \mathcal{\hat S}_t : L_i = 1 \}$ and $\# \{i \in \mathcal{\hat S}_t: L_i = -1 \}$. 
In their paper, \cite{ren:knockoffs} investigate this pruning strategy using different types of filters: a logistic regression (LR), a generalized additive model (GAM) which is the default filter in their \texttt{R} package, a random forest (RF), and a two-group model which is fitted using the expectation-maximization algorithm (EM)~(\cite{dempster:maximum}). 
The pruning process terminates at step $\tau$ when the estimated FDR is less than or equal to $\alpha$:
\begin{align}
    \label{discovery_adako}
    \widehat{FDR}_{\mathcal{\hat S}_{\tau}} := \frac{\# \{i \in \mathcal{\hat S}_\tau: L_i = -1 \} + 1}{\# \{i \in \mathcal{\hat S}_\tau: L_i = 1 \} \vee 1 } \leq \alpha.
\end{align}

\noindent The remaining positively labelled hypotheses in $\mathcal{\hat S}_\tau$ are then reported. 
It was shown that Adaptive Knockoffs controls the FDR under the natural side-information-augmented version of~(A0):
\begin{quote}
(A0$'$)~Conditioned on all the scores $W$, the side information $\mathbf{x}$, and the labels of the false null hypotheses, the true null labels $\{L_i : i \in N\}$ are \textit{i.i.d.}\ $\pm 1$ uniform random variables.
\end{quote}

\subsection{P-value based multiple testing}
\label{sec:background_pvalue}

Multiple testing with FDR control was popularized by the BH procedure (\cite{benjamini:controlling}).
Since then, numerous p-value-based methods have been developed to address multiple testing in two main ways.
First, many methods build on the BH framework to improve statistical power by estimating the fraction of true null hypotheses denoted as $\pi_0$ (\cite{storey:strong,benjamini:adaptiveLinear,hu2010false}), reweighting p-values (\cite{genovese2006false,blanchard2008two,roquain2009optimal,habiger2017adaptive,ramdas2019unified}), or by reordering them (\cite{barber:controlling,lei:power,gsell2016sequential,li2017accumulation}).
A second approach focuses on the stricter criterion of FDX control.
One example of this second direction is the stepdown procedure proposed by \cite{guo:generalized}, which we will refer to as GR-SD in our later analyses.

\subsubsection{Extension to side information}
\label{sec:pval_side}

\begin{table}
\caption{\label{table:summary_methods}
Summary of some popular recent p-value based multiple testing procedures that use side information.}  
\centering
\begin{tabular}{ccc}
\hline
\textbf{Method} & \textbf{Error control} & \textbf{Speed} \\
\hline
AdaPT (2018) & Finite FDR & Slow \\
AdaPT\textsubscript{g} (2021) & Finite FDR & Slow \\
AdaPT-GMM\textsubscript{g} (2021) & Finite FDR & Slow \\
ZAP-asymp (2022) & Asymptotic FDR  & Fast \\
AdaFDR (2019) & Asymptotic FDX  & Fast \\
\hline
\end{tabular}
\end{table}

More recent works harness side information in a data-adaptive way to further boost statistical power. 
Such adaptive methods include \textit{Adaptive P-value Thresholding} (AdaPT) and its derivatives (\cite{lei:adapt,chao2021adapt}), \textit{Z-value Adaptive Procedures} (ZAP) (\cite{leung:zap}) and \textit{AdaFDR} (\cite{zhang2019fast}). 
Table~\ref{table:summary_methods} provides a qualitative summary of each method in terms of its type-1 error control, its error control guarantees, and its computational speed.
ZAP-asymp and AdaPT-GMM\textsubscript{g} also have the advantage of operating directly on the test statistic instead of the corresponding p-values, which can boost power.
In this paper, we will focus our comparisons based on the p-value information.
Notably, none of these mentioned tools offer a fast finite type-1 error control (FDR or FDX), and as we will see, some of these methods have variable power when considering a range of simulated and real datasets. 
We provide a description of these methods in Section~\ref{sec:summary_of_methods}.

Our list is not exhaustive. Other methods include \textit{Independent Hypothesis Weighting} (\cite{ignatiadis2016data,ignatiadis2021covariate}) and the \textit{Structure Adaptive Benjamini--Hochberg Algorithm} (SABHA) (\cite{li:multiple}), which both adaptively weight the p-values, as well as \textit{ZAP-finite}, a finite-sample FDR analogue of ZAP-asymp that, like the AdaPT methods, prunes one hypothesis at a time. 
We exclude these from our comparisons: on the same simulations and real datasets, IHW and SABHA were consistently outperformed by methods in Table~\ref{table:summary_methods}, while ZAP-finite is a less powerful alternative to ZAP-asymp that also carries the computational cost of the AdaPT methods.

\subsection{Related work in conformal inference}
\label{sec:related_work_novelty}

Recently, \cite{marandon2024adaptive} developed \textit{AdaDetect} for novelty detection: given null observations $(Y_1, \dots, Y_n) \sim P_0$ and a test set $(X_1, \dots, X_m)$, the goal is to detect which $X_i \not\sim P_0$. 
Much like the method we propose in Section~\ref{sec:RESET_general}, AdaDetect splits the null observations into two parts, one to learn a score that ranks the pooled observations, and one to construct empirical (or \emph{conformal}) p-values~(\cite{bates2023testing,mary2022semi}), to which the BH procedure (or a $\pi_0$ variant) is applied.
AdaDetect's FDR control, however, rests on an exchangeability argument between the null observations and the null portion of the test set, and this argument does not hold in our setting.
The natural analogues of these two sets are the decoys $\{(\mathbf{x}_i, W_i) : L_i = -1\}$ and the true null targets $\{(\mathbf{x}_i, W_i) : L_i = 1, i \in N\}$, but these are generally not exchangeable: the decoys are an indiscernible mixture of true and false nulls, and the true nulls themselves may be heterogeneous.

During the review of this manuscript, we were referred to a related paper by \cite{zhao2025conformalized} on arXiv, proposing a method called \emph{CLAW} for multiple testing with side information.
Like AdaDetect, CLAW relies on a set of null observations but instead applies the Selective SeqStep+ procedure in place of the BH procedure, which is similar to our approach in this regard.
We provide a detailed comparison with our method in Section~\ref{sec:zhao_comparison} of the supplement.

A related strand of the conformal inference literature uses selective inference techniques to combine model selection with valid post-selection inference, which is in a similar spirit to the goal of RESET's data-splitting protocol. 
\cite{bai2024optimized} are one such example, selecting among conformity scores to improve power while re-using the data for valid FDR control in the problem of conformal selection~(\cite{jin2023selection}); see also the references therein.

\section{Methods}
\label{sec:RESET_general}

RESET is designed to control the FDR or the FDX in either the competition or p-value based setting. 
We next describe RESET's main steps, with further details given in Algorithm \ref{alg:reset}. 

\subsection{RESET}

\begin{enumerate}
	\item
	Intrinsically, RESET's input consists of the labels, $L = (L_i)_{i = 1}^m$, the winning scores, $W = (W_i)_{i = 1}^m$, and side information, $\mathbf{x} = (\mathbf x_i )_{i = 1}^m$. 
    Hence, in the p-value setting, each $p_i$ is first converted to a pair $(L_i, W_i)$ in the following way:
    \begin{align}
        \label{eq:convert}
        L_i = \begin{cases} 
            + 1 & p_i \in [0, a) \\
            - 1 & p_i \in (b_1, b_2] \\
         \end{cases}, \quad W_i = \begin{cases} 
         \lvert \Phi^{-1} ( p_i ) \rvert & p_i \in [0, a) \\
         \lvert \Phi^{-1} ( (b_2 - p_i)\cdot\frac{a}{b_2 - b_1} ) \rvert & p_i \in (b_1, b_2] \\
     \end{cases},
    \end{align}
    where $0 < a \leq b_1 < b_2 \leq 1$ and $a \leq 1/2$ determine the cutoff regions for defining a positive and negative label, and $\Phi$ is the standard normal CDF (the motivation for the introduction of $\Phi$ is discussed below).
    Hypotheses with p-values outside of $[0, a) \cup (b_1, b_2]$ are thrown out. 
    The default is $a = b_1 = 1/2$ and $b_2 = 1$.
    \label{step_1}
  \item
	RESET independently and randomly assigns each winning decoy as a \textit{training} decoy with probability $s$ (we used $s = 1/2$ throughout). 
    The complementary set of decoy wins defines the \textit{estimating} decoy set, and the set of \textit{pseudo} targets
	is the set of all estimating decoys and target wins. 
    The \textit{pseudo} labels, $\tilde{L}_i$, denotes whether the $i$th hypothesis is a pseudo target ($\tilde{L}_i = 1$) or a training decoy ($\tilde{L}_i = -1$).
    \label{step_2}
	\item
	Next, RESET applies a user-selected semi-supervised machine learning model which uses the pseudo labels, winning scores, and side information $(\tilde{L}, W, \mathbf{x})$. 
    Note that this is a semi-supervised task because the positive pseudo labels typically contain a mixture of true and false nulls, while the negative pseudo labels are mostly comprised of null hypotheses.

	The output of this step is a rescoring of the training decoys and pseudo targets, denoted as $\widetilde{W}$. 
    Ideally, $\widetilde{W}$ scores many of the false nulls among the pseudo targets
	higher than they were scored originally.
    We provide a specific framework that we developed for this crucial step in Section~\ref{sec:reset_implementation}.
    \label{step_3}
	\item
	With the training complete, the training decoys are thrown out, and the original labels $L$ of the remaining pseudo targets, $J := \{i \in [m]:\tilde{L}_i = 1 \}$, are revealed.
    \item Using the original labels $(L_i)_{i \in J}$ and the new scores $(\widetilde{W}_i)_{i \in J}$ RESET then determines its list of discoveries by applying either 
	\begin{enumerate}
		\item 
		SSS+ (Algorithm~\ref{alg:SeqStep}) at
		the desired level $\alpha$ for FDR control, or 
        \item 
        FDP-SD (Algorithm~\ref{alg:FDP-SD}) at the desired level $\alpha$ and confidence parameter $\gamma$ for FDX control.
	\end{enumerate}
    The two algorithms, SSS+ and FDP-SD, require an additional parameter $c$. 
    This parameter is used to define the expected ratio of the number of null targets to decoys, $\frac{c}{1 - c}$. Typically, this is set to $c = 1/2$ for an expected target-decoy ratio of 1; however, RESET uses approximately half of the decoys for training purposes (by default). 
    Since those decoys are subsequently thrown out, this parameter needs to be adjusted. 
    Indeed, assuming that the labels $L$ of the true nulls are \textit{i.i.d.}\ uniform  $\pm 1$ RVs, and $ s= 1/2$, we set the parameter to $c = c_e := 2/3$, so that the expected target-decoy ratio is $\frac{c}{1 - c} = 2$. 
    More generally, if we know that for a true null $\mathbb{P}(L_i = 1) \leq c_0$, then we set $c = c_e := \frac{c_0}{1 - s\cdot (1 - c_0)}$. For example, in the p-value setting, $c_0 = \frac{a}{a + b_2 - b_1}$. 
    This choice is made rigorous in Lemma~\ref{lemma:ind}.
    \label{step_4}
\end{enumerate}

We note that the above application of $\Phi^{-1}$ in Equation~(\ref{eq:convert}) is not strictly necessary for type-1 error rate control. 
We apply $\Phi^{-1}$ to the p-value $p_i$ or the \emph{mirrored p-value}, $(b_2 - p_i)\cdot\frac{a}{b_2 - b_1}$, because we found that doing so makes the semi-supervised learning task of Step~(\ref{step_3}) to be generally less variable and more powerful (data not shown). 

Crucially, in Step~(\ref{step_4}) the training decoys are thrown out. 
This is important since during the training phase of Step~(\ref{step_3}), the semi-supervised machine learning model will have presumably learned to discriminate between the training decoys and the pseudo targets. 
Hence, true null targets will tend to score higher than the training decoys making the latter unsuitable for estimating the number of true nulls among the top scoring hypotheses.

\subsection{Relationship to Percolator-RESET}
\label{sec:related_work}

We recently developed Percolator-RESET (\cite{freestone2025train}), combining the RESET approach with Percolator, a popular semi-supervised algorithm for FDR control in peptide detection from mass spectrometry data (\cite{kall2007semi-supervised}). 
The present manuscript adds several contributions over \cite{freestone2025train}: (i) the general RESET algorithm, applicable to both the competition and p-value settings; (ii) FDX control, with finite-sample proofs of RESET's FDR and FDX guarantees that also supply the previously-missing proof for Percolator-RESET; (iii) RESET Ensemble as the new semi-supervised engine of RESET's Step~(c); (iv) extensive simulated and real-data comparisons with competing methods, including against Adaptive Knockoffs in peptide detection; and (v) FDX control with side information for peptide detection.

\subsection{RESET controls the FDR or FDX}
\label{sec:fdr_control}

To establish RESET's control of the FDR or FDX in both the competition and p-value settings, we relax~(A0$'$) so that it can be applied to the case where the true null labels are no longer necessarily uniformly or even identically distributed:
\begin{Assumption}
    \label{assumption:tdc_aug_general}
    Let $N$ be the indices of the true null hypotheses. Conditioned on $W$, the side information $\mathbf{x}$, and the labels of the false null hypotheses, the true null labels $(L_i)_{i \in N}$ are independent $\pm 1$ random variables with $\mathbb{P}(L_i = 1) =: c_i \leq c_0$.
\end{Assumption}
We discuss the applicability of this assumption in Sections \ref{sec:exp_comp} and \ref{sec:exp_pval} where we apply RESET in both the competition and p-value settings.

RESET's FDR/FDX control hinges on the following Lemma, which essentially states that the labels $L$ of the true null pseudo targets in Step~(\ref{step_4}) are conditionally independent $\pm 1$ RVs---even after we sample the training decoys in Step~(\ref{step_2}) and the application of the machine learning model in Step~(\ref{step_3}).
\begin{lemma}
	\label{lemma:ind}
	Let $\mathcal{G} := \sigma \left(W, \mathbf{x}, (L_i)_{i \not\in N}, (\tilde{L}_i)_{i \in [m]} \right)$,
	$N$ be the indices of the true nulls, and
	$J := \{i \in [m] : \tilde{L}_i = 1 \}$.
    Then, under Assumption~\ref{assumption:tdc_aug_general}, conditioned on $\mathcal{G}$, $(L_i)_{i \in J \cap N}$ are independent $\pm 1$ RVs with 
	\[
	\mathbb{P} (L_i = -1 \mid \mathcal{G}) \geq 1-c_e  ,
	\]
	where $c_e := \frac{c_0}{1 - s \cdot ( 1 - c_0 )}$.
\end{lemma}
\begin{proof}
	Note that we can express $\tilde{L}_i$ as $\tilde{L}_i=f(L_i,B_i)$ where $f$ is a deterministic function and $B_i$ are Bernoulli$(s)$ RVs
	that are independent given everything else.
	It follows from $\tilde{L}_i=f(L_i,B_i)$ and Assumption~\ref{assumption:tdc_aug_general} that, conditioned on
	$\mathcal{G}_0 := \sigma \left(W, \mathbf{x}, (L_i)_{i \not\in N} \right)$, 
    the pairs in $\{(L_i, \tilde L_i) : i \in N \}$ are independent of one another.

	Moreover, from $\tilde{L}_i=f(L_i,B_i)$ and $(L_i)_{i \not\in N}\in\mathcal{G}_0$ we find that these pairs are still independent conditioned on
	$\sigma\left(\mathcal{G}_0, (\tilde{L}_i)_{i \not \in N} \right)$.
	This independence in turn implies that $(L_i)_{i \in N}$ are independent if we further condition on all the pseudo labels, i.e., on
	$\mathcal{G} = \sigma \left(\mathcal{G}_0, (\tilde{L}_i)_{i \in [m]} \right)$.
    Since $J$ is fixed conditioned on $\mathcal{G}$, the labels $(L_i)_{i \in J \cap N}$ are independent conditioned on $\mathcal{G}$,
	establishing the first part.
	

	It follows from Assumption~\ref{assumption:tdc_aug_general} and the above discussion, that with
	$\mathcal{G}_{-i} := \sigma \left(\mathcal{G}_0, (\tilde{L}_j)_{j \neq i}\right)$,
	$\mathbb{P} (L_i = -1 \mid \mathcal{G}_{-i}) \ge 1 - c_0$.
	Now fix $i \in N$ and condition on $\mathcal{G}_{-i}$ throughout. Then,
    \begin{align*}
        \mathbb{P} (L_i = -1, \tilde{L}_i = 1 \mid \mathcal{G}_{-i}) &= \mathbb{P}(L_i = -1 \mid \mathcal{G}_{-i})  \cdot \mathbb{P} (\tilde{L}_i = 1 \mid L_i = -1, \mathcal{G}_{-i}) \\
        &\geq (1 - c_0) \cdot (1 - s),
    \end{align*}
    Hence,
    \begin{align*}
    \mathbb{P} (L_i = -1 \mid \tilde{L}_i = 1, \mathcal{G}_{-i}) &= \frac{\mathbb{P} (L_i = -1, \tilde{L}_i = 1 \mid \mathcal{G}_{-i})}{\mathbb{P} (\tilde{L}_i = 1 \mid \mathcal{G}_{-i}) }\\
    &\geq \frac{(1 - c_0) \cdot (1 - s)}{(1 - s)\cdot \mathbb{P}(L_i = -1 \mid \mathcal{G}_{-i})  + 1 \cdot \mathbb{P}(L_i = 1 \mid \mathcal{G}_{-i}) } \\
    &= \frac{(1 - c_0) \cdot (1 - s)}{1 - s \cdot  \mathbb{P}(L_i = -1 \mid \mathcal{G}_{-i})} \\
    &\geq \frac{(1 - c_0) \cdot (1 - s)}{1 - s \cdot  (1 - c_0)} \\
	&= 1-c_e.
    \end{align*}
	Additionally, clearly
	\[
	\mathbb{P} (L_i = -1 \mid \tilde{L}_i = -1, \mathcal{G}_{-i}) = 1 \ge 1-c_e.
	\]
	Therefore, the following holds a.s.\
	\[
	\mathbb{P} (L_i = -1 \mid \mathcal{G}) = \mathbb{P} (L_i = -1 \mid \tilde{L}_i, \mathcal{G}_{-i}) \ge 1-c_e.
	\]

%
\end{proof}

It is worth noting that Lemma~\ref{lemma:ind} is not true for an arbitrary splitting of the decoys into the training and estimating sets, e.g., splitting the decoys into two halves of equal size. To see this, suppose that all hypotheses are true nulls, i.e., $N = [m]$. Then by observing the number of training decoys, we know that the number of true null decoys in $J$ must be the same, and the labels $(L_i)_{i \in J \cap N}$ are no longer independent. We are now ready to state our two main theorems.
\begin{theorem}
    \label{theorem:perc_fdr}
	Under Assumption \ref{assumption:tdc_aug_general} RESET controls the FDR at the user-specified threshold $\alpha$.
\end{theorem}
\begin{theorem}
    \label{theorem:perc_fdp}
	Under Assumption \ref{assumption:tdc_aug_general} RESET controls the FDX at the user-specified threshold $\alpha$ and confidence $1-\gamma$.
\end{theorem}

The proofs of Theorem~\ref{theorem:perc_fdr} and \ref{theorem:perc_fdp} follow from Lemma~\ref{lemma:ind} and rely on a technical discussion when the labels $(L_i)_{i \in J \cap N}$, conditioned on $\mathcal G$, are non-identical, which we give in Section~\ref{sec:FDR_FDP_control}.
To briefly see the connection of Lemma~\ref{lemma:ind} to Theorems~\ref{theorem:perc_fdr} and \ref{theorem:perc_fdp}, note that the labels $(L_i)_{i \in J \cap N}$ are independent $\pm 1$ RVs even after observing the machine learning scores $\widetilde W$ in Step~(\ref{step_3}) and the false null labels $(L_i)_{i \notin N}$.
If the labels are identical, i.e., $\mathbb{P} (L_i = -1 \mid \mathcal{G}) = 1-c_e$, then the assumptions of SSS+ in \cite{barber:controlling} and FDP-SD in \cite{luo:competition} are satisfied, and RESET controls the FDR or FDX, respectively.
Intuitively, in the non-identical case, i.e., $\mathbb{P} (L_i = -1 \mid \mathcal{G}) \geq 1-c_e$, FDR or FDX control should be more conservative using these procedures.

\subsection{Semi-supervised approach}
\label{sec:reset_implementation}

In this section we describe our specific framework for RESET's Step~(\ref{step_3}), which can use any semi-supervised machine learning model.
It consists of two main steps: (1) we apply a range of classification algorithms to estimate a high quality positive set from the data, presumably containing many false nulls, and (2) this positive set is then subsequently used for a second application of the classification algorithms to rescore the hypotheses. 
In Section~\ref{sec:further_details_reset}, we outline some additional heuristics that we employ to improve the procedure's speed and power, as well as its applicability to dependent data.

\begin{enumerate}[label=(\roman*)]
    \item We first define the negative and positive sets, as well as the features that are subsequently used by a collection of classification algorithms:
	the negative set is the set of all training decoys, the positive set is a subset of top scoring pseudo targets (Section~\ref{sec:initial_pos_set}), 
    and the features are the combined score and side information $\left(W, \mathbf{x}\right)$.\label{step_i}
    \item We randomly split all the hypotheses into $K$ folds (we used $K = 3$ throughout).
    For each considered classification algorithm, and for each $k \in [K]$, we train the classification algorithm using the features of the hypotheses in the subsets of the positive and negative sets that are in the $[K] \setminus \{k \}$ training folds, and then we apply the trained model to all the hypotheses in the $k$th test fold. 
    We consider a random forest (RF), a generalized additive model (GAM), and a collection of two-layer neural networks (NN) with decay parameter $\lambda \in \{0, 0.1, 1\}$ and a number of hidden layer nodes $h \in \{2, 5, 10 \}$.\label{step_ii}
    \item To reduce variability, we repeat Step~\ref{step_ii} $r$ times (we used $r = 10$).\label{step_iii}
    \item To evaluate each classification algorithm, we apply Selective SeqStep (SSS) (Algorithm~\ref{alg:SeqStep}, which is similar to SSS+ of Equation~(\ref{discovery_fdr}) but without the `+1') at an FDR threshold $\alpha$ to each of the $rK$ test folds, using the scores obtained from the classification algorithm and the pseudo labels. We set $c = c_t := 1 - s\cdot (1 - c_0)$ (recall that $\mathbb{P}(L_i = 1) \leq c_0$ under Assumption~\ref{assumption:tdc_aug_general}). 
    This is to ensure that the ratio $\frac{c}{1-c}$ correctly accounts for the number of null pseudo targets to training decoys (note, this is not a requirement of FDR/FDX control).\label{step_iv}
    \item We record the total number of `discoveries' produced by SSS over the $rK$ test folds from the previous step and select the classification algorithm that maximizes this value.\label{step_v}
    \item Using the chosen classification algorithm, we assign new scores $\widetilde{W}$ to the hypotheses. 
    $\widetilde{W}_i$ is defined as the average decision value for the $i$th hypothesis in the test fold that it appears in, taken over the $r$ repetitions executed in Step~\ref{step_iii}.\label{step_vi}
    \item We reapply Selective SeqStep (SSS) at $\alpha$ using the pseudo labels and the updated scores, $(\tilde{L}, \widetilde{W})$ with $c = c_t = 1 - s \cdot (1 - c_0)$.\label{step_vii}
    \item The positive set is redefined as those pseudo targets that are `discovered' in Step~\ref{step_vii}. 
    Steps~\ref{step_ii}-\ref{step_vi} are then repeated with the new positive set.\label{step_viii}
    \item The final scores $\widetilde{W}$ are then reported.\label{step_ix}
\end{enumerate}

We refer to the general RESET wrapper in combination with the above semi-supervised approach as \textit{RESET Ensemble}. We refer to \textit{RESET RF/GAM/NN} when the ensemble of machine learning methods in Step~\ref{step_ii} is reduced to a random forest, generalized additive model, and a collection of neural networks as described, respectively.
In every one of these variants, ensemble or single-model, the $K$-fold scheme of Step~\ref{step_ii} scores each hypothesis using a model that was not trained on it, which guards against overfitting. We stress that this cross-validation serves power rather than validity: RESET's finite-sample FDR or FDX control follows from the data-splitting of Step~(\ref{step_2}) and holds for any rescoring $\widetilde{W}$ (Lemma~\ref{lemma:ind}), irrespective of the learner or how well it fits the data.

\subsection{Code availability}

Our code is available at \url{https://github.com/freejstone/stat_RESET_paper_code} with an Apache-2.0 licence.
We intend on developing a package for RESET in \texttt{R}.

\section{Experiments: competition setting}
\label{sec:exp_comp}

In the following sections, we explore two applications of competition-based multiple testing.
The first is in selecting relevant variables that are conditionally associated with a response.
The second is in the detection of peptides from mass spectrometry data.

\subsection{Variable selection}
\label{sec:numerical_comp}
We conducted a wide range of simulations in the variable selection problem, where the goal is to identify relevant variables from $\{X_i : i = 1, \dots, p \}$ that are conditionally associated to the response $y$.
The variable selection problem uses a competition framework called the \textit{knockoff filter}~(\cite{barber:controlling,candes:panning}).
Briefly, each variable $X_i$ is paired with an artificial knockoff variable, denoted as $\tilde{X}_i$.
Each variable and knockoff is scored according to how well it correlates with the response $y$, producing $Z_i$ and $\tilde{Z}_i$, respectively.
Then following Section~\ref{sec:background_comp}, the pairs of scores can be converted into a winning score $W_i$ and a label $L_i$, and a competition-based multiple testing procedure can be applied (see Section~\ref{sec:kf_filter} for details).

The following simulations, detailed in Section~\ref{sec:sim_comp}, are based on those introduced by \cite{ren:knockoffs}.
Each setup, is designed to satisfy Assumption~\ref{assumption:tdc_aug_general}, and in each
we compare RESET Ensemble with Adaptive Knockoffs (AdaKO, with its default GAM filter, see Section~\ref{sec:further_details_adako} for details),
as well as with the generic Knockoff Filter (KO) focusing on FDR control.
\begin{itemize}
    \item In simulation 1 we consider a linear model with $n = 1000$ observations with $p = 900$ variables.
    We use one dimensional side-information $\mathbf{x}_i$ where lower values correspond to a greater chance of being a relevant feature ($\beta_i \neq 0$). We vary the number of relevant variables $k \in \{50, 150, 300\}$ and generate and analyze 100 datasets for each.
    \item In simulation 2 we consider a larger linear model with $n = 2000$ and $p = 1800$ with two-dimensional side information.
    The side information is drawn from a joint normal distribution, with different mean vectors for the relevant and irrelevant variables respectively.
    We vary the number of relevant variables $k \in \{150, 300, 450\}$ and generate and analyze 20 datasets for each.
    \item In simulation 3 we consider a high-dimensional logistic model with $n = 1000$ observations and $p = 1600$ variables.
    Relevant and irrelevant variables are determined according to the spatially dependent two-dimensional side information $\mathbf{x}_i = (\mathbf{x}_{i1}, \mathbf{x}_{i2})$ (see Figure~\ref{fig:beta_location}).
    We generate and analyze 100 datasets.
    \item In simulation 4 we consider a linear model with $n = 10K$ observations and $p = 10K$ variables with $k = 0.15p$ relevant variables and three dimensional side information $\mathbf{x_i} = (\mathbf{x}_{i1}, \mathbf{x}_{i2}, \mathbf{x}_{i3})$. 
    Lower values of the $\mathbf{x}_{ij}$'s correspond to a greater chance of being a relevant feature.
    We generate and analyze 5 datasets.
\end{itemize}

\subsubsection{Results}
 
\begin{figure}[h!]
    \centering
    \begin{tabular}{l}
    \includegraphics[width=6in]{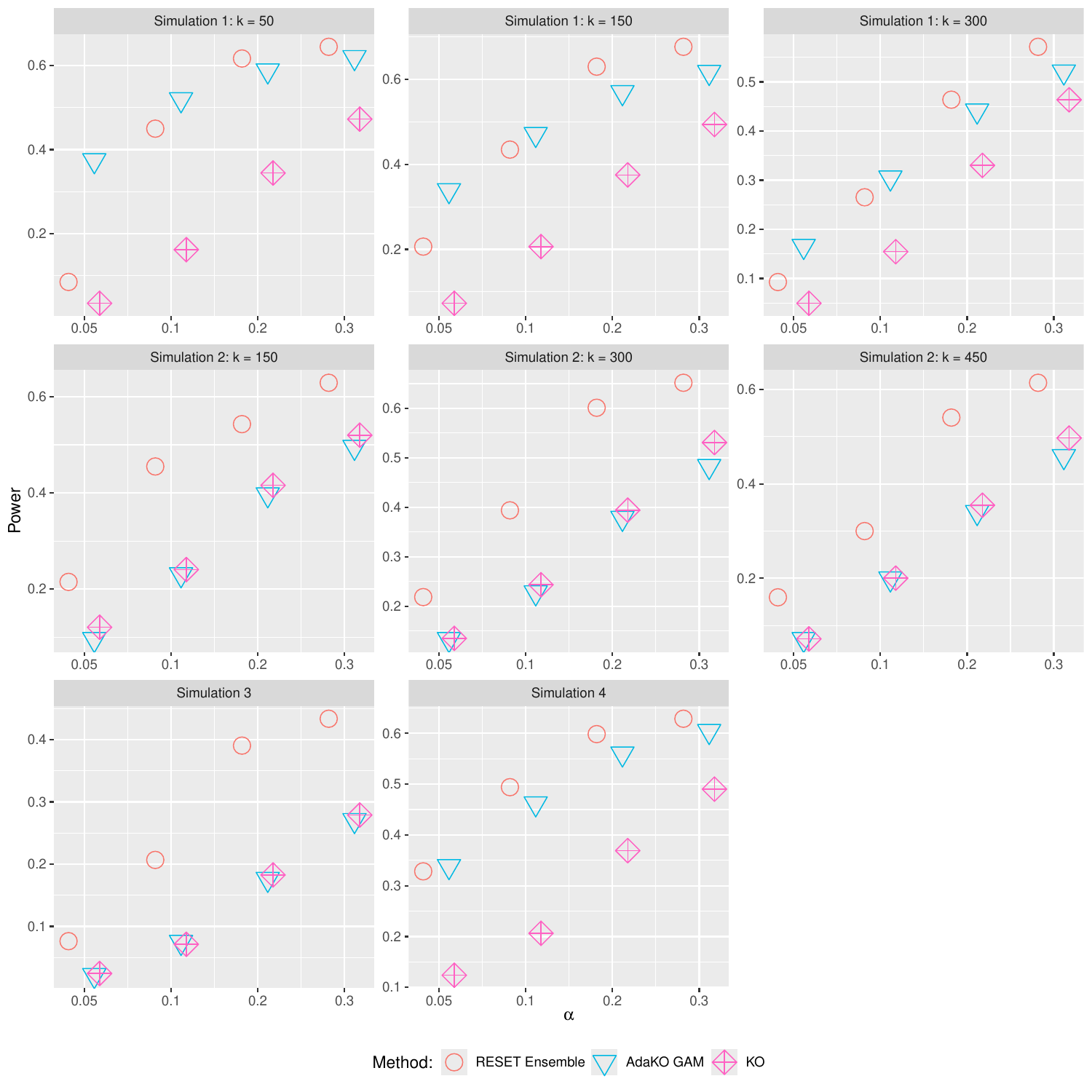}\\
    \end{tabular}
    \caption{Comparing the power of RESET Ensemble, AdaKO GAM and KO.
    Each panel shows the power of each method at FDR thresholds ranging from 5\% to 30\%. The first row corresponds to Simulation 1 with three values of $k \in \{50, 150, 300\}$, the second row corresponds to Simulation 2 with three values of $k \in \{150, 300, 450\}$, and the last row corresponds to Simulation 3 and 4. For readability, the points are jittered in the horizontal direction.}
    \label{fig:reset_adako_gam}
\end{figure}

Figure~\ref{fig:reset_adako_gam} plots each method's power, the proportion of relevant variables correctly discovered, averaged over datasets, against the FDR thresholds $\alpha \in \{0.05, 0.1, 0.2, 0.3\}$. RESET Ensemble uniformly dominates KO, demonstrating that it leverages the side information, while AdaKO GAM mostly improves on KO, with some exceptions.

In Simulation 1, we find that at higher FDR thresholds, RESET Ensemble is the preferred approach in terms of power. 
On the other hand, AdaKO GAM is more powerful than RESET at lower FDR thresholds, particularly in the case of $k = 50$ and $\alpha = 5\%$. 
This is unsurprising: RESET trains on a relatively small positive set of false nulls, and at low thresholds an overfit score that ranks a knockoff win too highly is especially costly---at $\alpha = 5\%$, each knockoff win can cost up to $2/\alpha = 40$ discoveries. 

In Simulations 2 and 3, RESET Ensemble is consistently more powerful than AdaKO GAM, which performs essentially like KO (consistent with Ren and Cand\`es in their analogous simulation): with two or more side-information variables, AdaKO GAM implements no smooth terms and reduces to a linear model that cannot leverage the side information unless the true and false nulls are `easily' separable. 
Finally in Simulation 4, we see that RESET Ensemble is generally the most powerful, with essentially the same power as AdaKO GAM at $\alpha = 5\%$. 
We hypothesize that the greater number of variables in this example, particularly relevant variables, appears to overcome the relatively poor performance of RESET Ensemble in Simulation 1 at low FDR thresholds.

We compared RESET Ensemble with AdaKO GAM because those are the respective default methods.
In Section~\ref{sec:further_comparisons}, we complement this comparison where we summarize the performance of all RESET and AdaKO methods, and also analyze the runtimes of those methods.

The simulations above construct knockoffs assuming the distribution of $X$ is known.
In practice, this distribution is rarely known and knockoffs are only approximately valid, typically constructed using a second-order moment-matching approximation~(\cite{candes:panning}).
In Section~\ref{sec:further_comparisons} of the supplement, we re-run Simulations~1--4 under this more realistic regime, using second-order knockoffs, and find that RESET Ensemble is no more sensitive to the approximation than KO or AdaKO GAM.

\subsection{Application to peptide detection}
\label{sec:real_data}
In peptide detection, a sample is analysed by liquid chromatography mass spectrometry, with each peptide generating a mass spectrum. 
Each hypothesis $H_i$ is the null that a particular \emph{target} peptide is not present in the sample. 
A database search matches each observed spectrum against candidate peptides, scoring the best-matching target peptide ($Z_i$, the evidence against $H_i$) and a paired, artificially constructed \emph{decoy} peptide ($\tilde{Z}_i$).
The competition then yields a winning score $W_i = Z_i \vee \tilde{Z}_i$ and a label $L_i$ equal to $1$ if the target outscored its decoy and $-1$ otherwise. 
Applying SSS+ with $c = 1/2$, known as \emph{target-decoy competition} (TDC) in proteomics, to $(W, L)$, typically at the $1\%$ FDR threshold, returns the list of detected peptides. Full details are given in Section~\ref{sec:peptide_detection}.

In the following results, we show that (a) the Adaptive Knockoff methods are often computationally infeasible in this context,
and (b) feasibility aside, RESET is more powerful than Adaptive Knockoff's default filter, AdaKO GAM, when controlling the FDR.
In Section~\ref{sec:brief_just}, we provide a heuristic justification of Assumption~\ref{assumption:tdc_aug_general}, which is required for RESET's theoretical guarantees. 
In terms of \emph{FDX} control, RESET typically discovers more peptides than the side-information-oblivious FDP-SD (Section~\ref{sec:comp_fdp}).
As outlined in Section~\ref{sec:related_work}, these comparative analyses of RESET Ensemble are new and were not undertaken in \cite{freestone2025train}.

\subsubsection{Runtime comparison}
\label{sec:hek293}

We argue that the application of Adaptive Knockoffs is often too slow to be used effectively in scenarios such as the peptide detection problem that involve a large amount of data. 
We demonstrate this using HEK293~(\cite{chick:mass-tolerant}), a popular dataset used in benchmarking the development of new software~(\cite{kong2017msfragger,yu:identification2,lazear2023sage}) that we downloaded from the Proteomics Identification Database (PRIDE)~(\cite{martens2005pride}). 
Details regarding the HEK293 dataset and the search phase analysis are given in Section~\ref{sec:HEK293_supp}.

\begin{table}
 \caption{\label{tab:HEK293_times}The average runtimes for each method at 1\% FDR on the HEK293 dataset.}
\centering
\begin{tabular}{|c|c|} 
\hline
\textbf{Method} & \textbf{Time} \\
\hline
RESET Ensemble & 17.7 minutes \\
AdaKO GAM & 1.07 days \\
AdaKO EM & 34.17 days \\
AdaKO LR & 3.50 days \\
AdaKO RF & 846.33 days \\
\hline
\end{tabular}
\end{table}

We estimated the time it would take for each method of Adaptive Knockoffs to report discoveries at the 1\% FDR level (see Section~\ref{sec:comp_time}).
Table~\ref{tab:HEK293_times} shows the estimated times averaged over 5 different constructions of the decoy database by randomly shuffling the target peptides.
All the implementations of AdaKO are not practical for analyzing this data, with the worst being AdaKO RF at an estimated 846.33 days, and the default method, AdaKO GAM, taking an estimate of 1.07 days.
In contrast, we were able to run RESET Ensemble in 17.7 minutes on average using 20 cores (Section \ref{sec:comp_specs}).

\subsubsection{Power comparison}
\label{sec:pride}

General applicability aside, we resorted to using small datasets to evaluate the power of Adaptive Knockoffs in the peptide detection context.
Specifically, we focused on the thirteen smallest datasets from PRIDE-20, a collection of twenty datasets that we recently used~\cite{freestone2024analysis}.
For each of the thirteen spectrum files in PRIDE-20, we performed two types of searches: a \textit{narrow} and an \textit{open} search. In a narrow search, each spectrum may match only peptides whose mass is within a small tolerance of the spectrum's measured mass; in an open search this tolerance is several orders of magnitude larger.
Details regarding PRIDE-20 and the search settings are given in Section~\ref{sec:pride_20_supp}.

\begin{figure}[h]
    \centering
    \begin{tabular}{l}
    \includegraphics[width=4in]{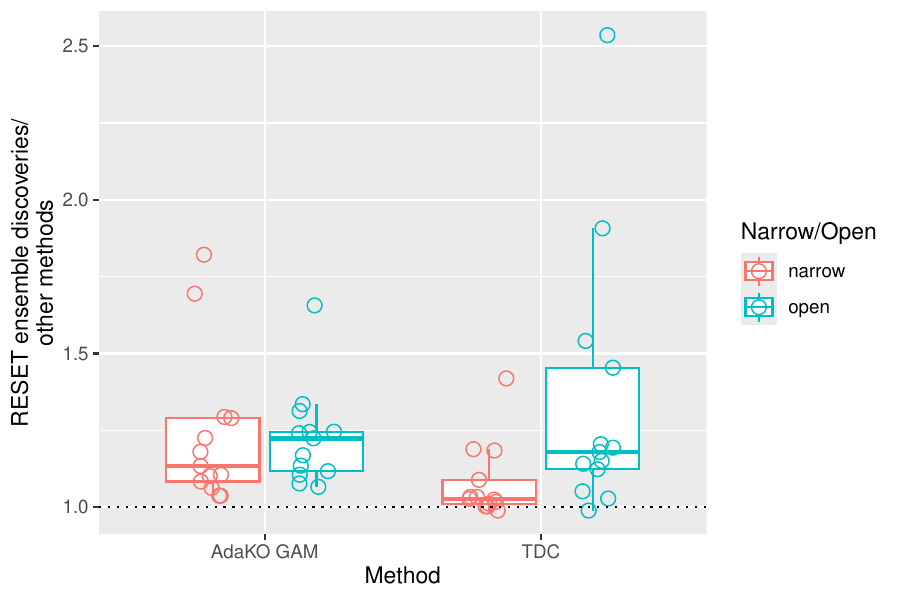}\\
    \end{tabular}
    \caption{Comparison of RESET Ensemble to AdaKO GAM and TDC. 
    For each of the PRIDE-20 datasets, we computed the average number of discoveries using 10 applications of RESET Ensemble, varying RESET's internal seed. 
    The figure shows a boxplot of those RESET Ensemble averages divided by the number of discoveries using AdaKO GAM (left boxplots) and TDC (right boxplots) applied to thirteen PRIDE-20 datasets and two different search modes (there are total of 13 points in each boxplot).
    }
    \label{fig:pride_20}
\end{figure}

Figure~\ref{fig:pride_20} shows the average number of discoveries reported using RESET Ensemble divided by the number of discoveries reported by AdaKO with its default GAM filter (left boxplots) and TDC (right boxplots), using the two different search settings. 
We averaged RESET Ensemble over 10 applications by varying the internal seed which randomly splits the decoys into their training and estimating sets. 
Clearly, RESET Ensemble is overall more powerful than AdaKO GAM and TDC. 
Interestingly, the relative ratios of the number of discoveries between AdaKO GAM and TDC suggest that AdaKO GAM is at times reporting \textit{fewer} discoveries than the generic TDC. 

In Section~\ref{sec:additional_compare}, we compare RESET Ensemble with AdaKO RF since we found AdaKO RF to be the most powerful variant of Adaptive Knockoffs in this setting. 
The results show that RESET Ensemble is comparable to AdaKO RF in terms of power, with the latter still being generally too slow to be practical.

\section{Experiments: p-value setting}
\label{sec:exp_pval}

\subsection{Assumption guaranteeing finite-sample FDR control}

Recall that RESET and Assumption~\ref{assumption:tdc_aug_general} that guarantees its finite-sample error control are formulated in terms
of the labels and winning score $(L,W)$.
To apply RESET in the p-value setting, we first convert each $p_i$ into a pair $(L_i,W_i)$ as in Equation~\eqref{eq:convert}.
Similarly, instead of Assumption~\ref{assumption:tdc_aug_general} we adopt the following assumption formulated by \cite{chao2021adapt} who also
showed it implies the former assumption (the next lemma).
\begin{Assumption}
    \label{assumption:non_decreasing}
    Let $N$ be the indices of the true null hypotheses. Conditioned on the side information $\mathbf{x}$, and the false null p-values $(p_i)_{i \not \in N}$,
    the true null p-values $(p_i)_{i \in N}$ are independent and have a non-decreasing density.
\end{Assumption}
\begin{lemma}[Chao and Fithian]
    \label{lemma:chao}
    Assumption~\ref{assumption:non_decreasing} implies Assumption~\ref{assumption:tdc_aug_general} with $c_0 = \frac{a}{a + b_2 - b_1}$,
	where $[0, a)$ and $(b_1, b_2]$ are the regions corresponding to the positive and negative labels as described in Equation~(\ref{eq:convert}).
\end{lemma}
The above lemma is phrased differently by \cite{chao2021adapt} and is presented as Lemma A.1 in their paper, but it is materially the same. 
Most of the time, we will consider symmetric cutoff regions $[0, 1/2)$ and $(1/2, 1]$. 
For completeness, we provide a proof of this lemma in Section~\ref{sec:proof_of_chao} using our notation.

Because RESET converts each p-value into a label and winning score and then applies the competition-based SSS+ procedure, one might worry that this route is inherently less powerful than working with the p-values directly through BH or its $\pi_0$-adaptive variants (Section~\ref{sec:background_pvalue}).
In fact, all methods in Table~\ref{table:summary_methods} adopt a similar SSS+ type approach, despite being formally developed for the p-value setting.
Regardless, in Section~\ref{sec:pval_equivalence} we demonstrate that even completely ignoring any potential benefit from the side information,
the cost of using SSS+ over other intrinsic p-value based approaches is often small and typically asymptotically negligible.

\subsection{Simulated data}

In this section, we consider the p-value setting, comparing RESET with the procedures we introduced in Section~\ref{sec:pval_side} on a set of simulations introduced by \cite{lei:adapt}. 
Each setup, detailed through Sections~\ref{sec:further_details_adapt}--\ref{sec:further_details_adafdr}, is designed to satisfy Assumption~\ref{assumption:non_decreasing} and hence Assumption~\ref{assumption:tdc_aug_general},
which is required for our theoretical guarantees.

\subsubsection{Simulation 5: p-value based testing with geometric side information}
\label{sec:pval_two_dim}

Based on Simulation 1 of \cite{lei:adapt}, this setup comprises $m = 2500$ hypotheses with two-dimensional side information arranged on a $50 \times 50$ grid, where the false nulls occupy one of three regions of the square: a circle in the middle (a), a circle in the top right (b), or an ellipse (c) (Figure~\ref{fig:mu_location}). The full data-generating process is given in Section~\ref{sec:sim_pval_detail} of the supplement.

\begin{figure}[h]
    \centering
    \begin{tabular}{l}
    \includegraphics[width=6in]{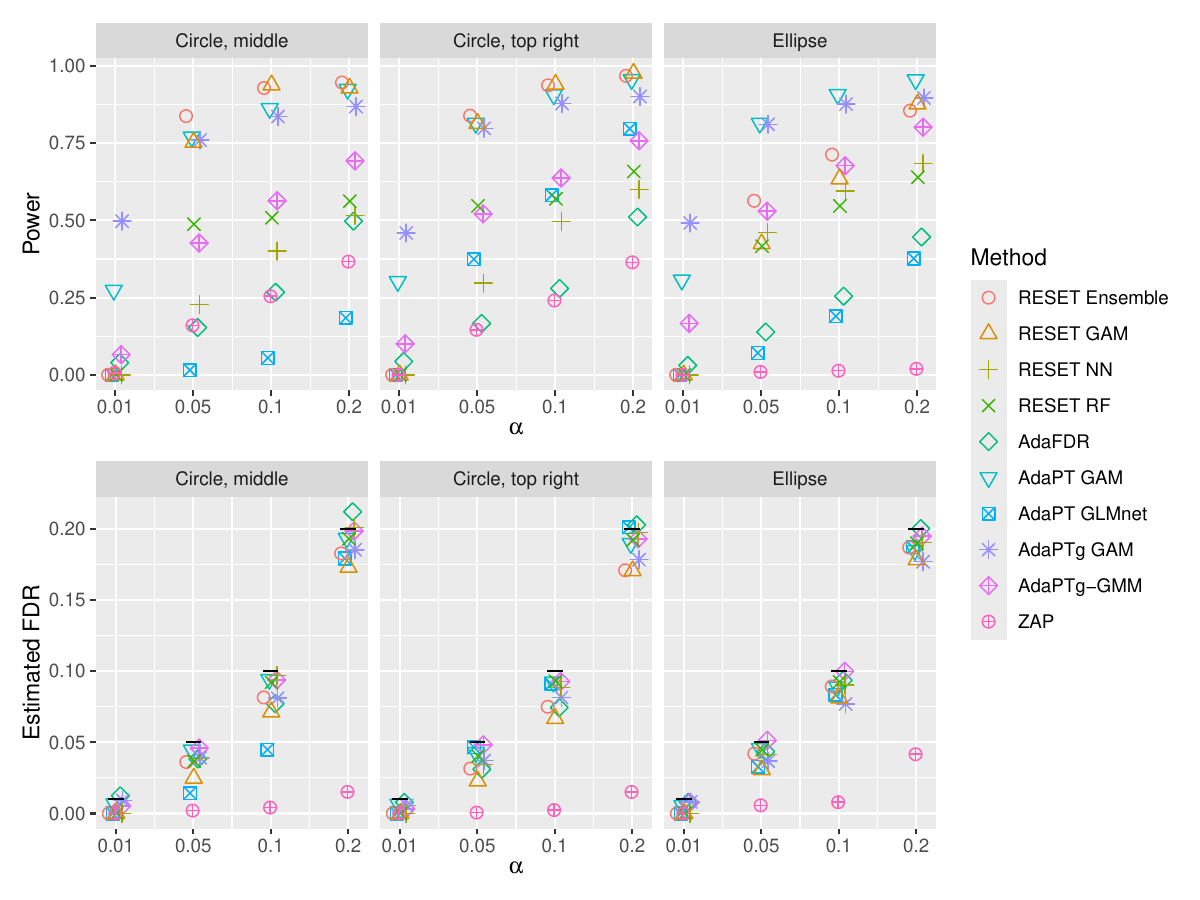}\\
    \end{tabular}
    \caption{RESET vs. other methods using p-values with two-dimensional side information.
    The power (1st row) and the estimated FDR (2nd row) are plotted against FDR thresholds ranging from 1\% to 20\%. Each color and shape combination corresponds to a unique method. The three columns correspond to the hypotheses' relationship to the side information according to Figure~\ref{fig:mu_location}. The black horizontal segments in the second row of panels mark the corresponding FDR thresholds on the y-axis. For readability, each point is jittered in the horizontal direction.}
    \label{fig:two_dim_pvalue}
\end{figure}

Figure~\ref{fig:two_dim_pvalue} shows the results of RESET compared with the p-value based methods reviewed in Section~\ref{sec:pval_side}. 
In the first two settings, (a) and (b), RESET Ensemble is roughly tied with the best method, AdaPT/AdaPT\textsubscript{g} GAM, at the 5\%, 10\% and 20\% FDR levels. 
In (c), AdaPT/AdaPT\textsubscript{g} GAM are clearly more powerful, with RESET Ensemble arguably third at those levels. 
At 1\% FDR all RESET methods have negligible power, as do many of the other methods. 
Interestingly, the finite-FDR procedures (the RESET and AdaPT methods) generally outperform the asymptotic approaches, ZAP and AdaFDR, perhaps indicating a misspecification of the latter's fitted models. 
Figure~\ref{fig:two_dim_pvalue} also shows the estimated FDR of each method. In (a), AdaFDR's estimated FDR at 20\% slightly exceeds the threshold (21.2\% with 50\% power).
This is possibly explained by the fact that AdaFDR only guarantees asymptotic FDX control. 
Consistently, \cite{chao2021adapt} reported inflated estimated FDRs at high thresholds in simulation studies for AdaFDR.

\subsubsection{Simulation 6: p-value based testing with 100-dimensional side information}
\label{sec:pval_100_dim}

Based on Simulation 2 of \cite{lei:adapt}, this setup comprises $m = 2000$ hypotheses with $100$-dimensional side information, of which only the first two coordinates distinguish true from false nulls. The full data-generating process is given in Section~\ref{sec:sim_pval_detail} of the supplement.

\begin{figure}[h]
    \centering
    \begin{tabular}{l}
    \includegraphics[width=6in]{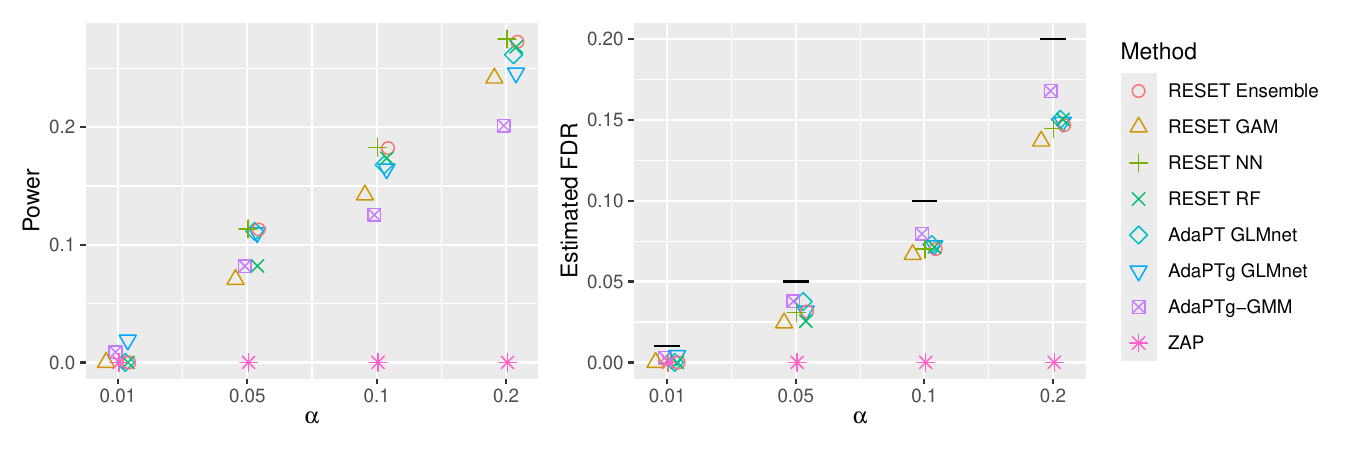}\\
    \end{tabular}
    \caption{RESET vs. other methods using p-values with 100-dimensional side information. The power (left panel) and the estimated FDR (right panel) are plotted as a function of the FDR thresholds: 1\%, 5\%, 10\% and 20\%. Each color and shape combination corresponds to a unique method. The black horizontal segments in the right panel mark the corresponding FDR thresholds on the y-axis. For readability, each point is jittered in the horizontal direction.
    }
    \label{fig:100_dim_pvalue}
\end{figure}

There were a couple of tools we had to omit from the analysis of this data. Surprisingly, we found that AdaFDR was too slow in this simulation. Similarly, we were not able to apply AdaPT GAM or AdaPT\textsubscript{g} GAM as we did in the previous simulation, since it would have required using a smoothing spline on all 100 side-information variables, which is impractical. Hence, we only applied the GLMnet variants of AdaPT. 

Examining Figure~\ref{fig:100_dim_pvalue}, which graphically summarizes this analysis, it appears that RESET Ensemble, RESET NN, and AdaPT GLMnet deliver the most discoveries at the 5\%, 10\%, and 20\% FDR levels. At 1\% FDR, only AdaPTg GLMnet appears to have non-negligible power, albeit a fairly low number of discoveries. Finally, we point out that in this simulation, the model specification of AdaPT/AdaPT\textsubscript{g} perfectly matches the underlying
data generating model: the fitted mixture model used by AdaPT/AdaPT\textsubscript{g} coincides with Equation~(\ref{eq:sim_5}) of Section~\ref{sec:sim_pval_detail}.
Given this, we find it reassuring that RESET Ensemble is still comparable at 5\% and higher thresholds. 

Notably, the performance of AdaPT in these last two simulation setups is highly dependent on the choice of implementation. In Simulation 5, AdaPT GAM significantly outperforms AdaPT GLMnet while in Simulation 6, AdaPT GLMnet is superior and AdaPT GAM is not even practical. In contrast, RESET Ensemble selects from a collection of flexible machine learning models to adapt itself to the problem at hand.

\subsection{Robustness to dependent data}
\label{sec:robustness_to_dependent_data}

In practice, real datasets often exhibit varying degrees of dependency, therefore failing to satisfy Assumptions~\ref{assumption:tdc_aug_general} and \ref{assumption:non_decreasing}.
RESET Ensemble is equipped with a heuristic to account for such data (Section~\ref{sec:handling_dependent_data}).
The following simulations adapted from \cite{zhang2019fast} empirically demonstrate that RESET Ensemble exhibits some robustness to
both weak and strong forms of local dependency.

We consider two simulations adapted from \cite{zhang2019fast}, each with $m = 20{,}000$ hypotheses whose false-null probabilities depend on the side information. \emph{Simulation 7} induces \emph{weak} local dependence among the p-values through a block covariance structure, while \emph{Simulation 8} induces \emph{strong} local dependence through blocks of identical z-scores. The full data-generating processes are given in Section~\ref{sec:sim_pval_detail} of the supplement.

Figure~\ref{fig:pvalue_dependency} shows the estimated FDR calculated from 100 runs of these simulations.
In the weakly dependent case (A), the majority of the methods seem to empirically control the FDR,
with the exception of AdaFDR and RESET GAM which appears to marginally exceed the 5\% threshold.
In the strongly dependent data (B), we see several more methods exceeding their respective thresholds, albeit relatively mildly.
RESET Ensemble seems to be robust and appears to approximately control the FDR in these experiments.

\begin{figure}[h]
    \centering
    \begin{tabular}{l}
    \includegraphics[width=6in]{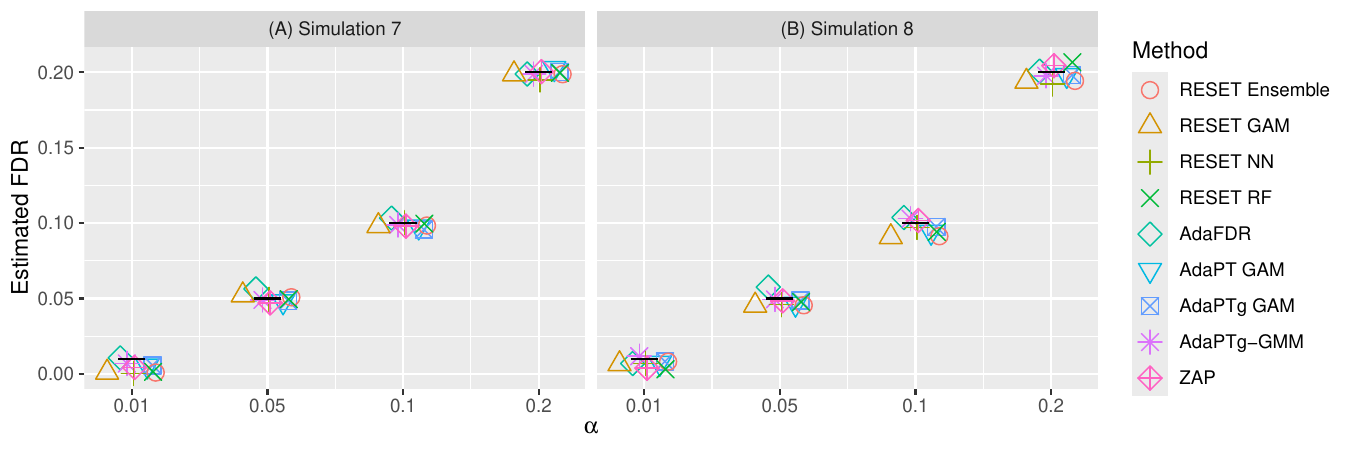}\\
    \end{tabular}
    \caption{Estimated FDR using RESET vs. other methods on dependent p-values. The estimated FDR is plotted as a function of the FDR thresholds: 1\%, 5\%, 10\% and 20\% for weak (A) and strong dependence (B).
    Each color and shape combination corresponds to a unique method. The black horizontal segments mark the corresponding FDR thresholds on the y-axis. For readability, each point is jittered in the horizontal direction.
    }
    \label{fig:pvalue_dependency}
\end{figure}


\subsection{Real data experiments with p-values}
\label{sec:real_data_pval}

We next evaluate RESET and related methods on a collection of publicly available datasets with p-values.
These were investigated by \cite{zhang2019fast}, when introducing AdaFDR, and by \cite{lei:adapt} when introducing AdaPT. 
A description of each dataset can be found in Section~\ref{sec:description}.

Although RESET Ensemble and the other tools show some robustness to deviations from the independence built into Assumptions~\ref{assumption:tdc_aug_general} and~\ref{assumption:non_decreasing}, we removed the two Adipose datasets from this analysis, as RESET's heuristics (Section~\ref{sec:handling_dependent_data}) flagged them for high dependency (cf.~Section~\ref{sec:adipose}).

\begin{figure}[h]
    \centering
    \begin{tabular}{l}
    \includegraphics[width=4in]{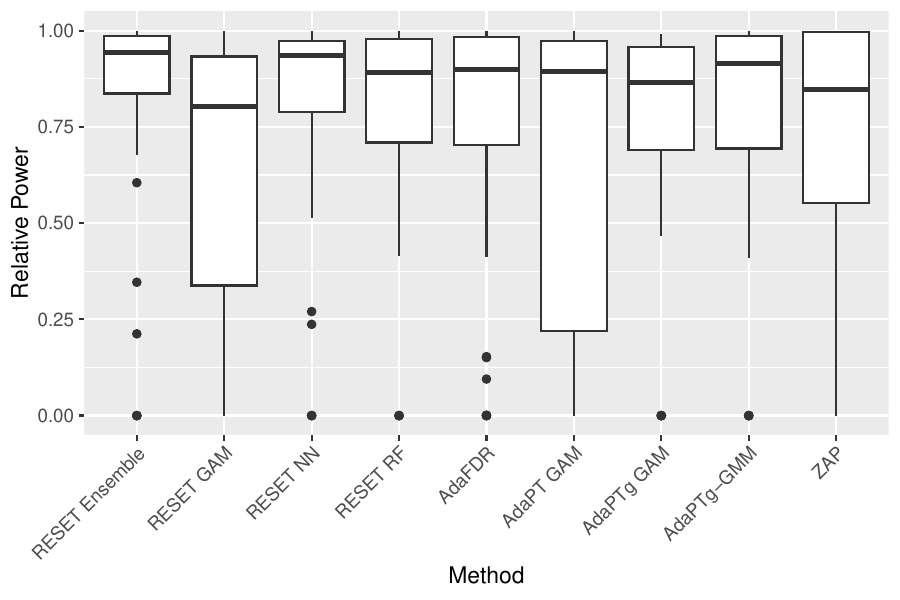}\\
    \end{tabular}
    \caption{Power comparison. Each boxplot is made of 40 data points showing the relative power of a method on all combinations of dataset and FDR threshold.
	There are ten dataset as described in Section~\ref{sec:description} (excluding the two Adipose datasets) and four FDR thresholds: 1\%, 5\%, 10\% and 20\%.
    RESET Ensemble's power and its variants were averaged over 10 applications.
    }
    \label{fig:real_data_pvalue_boxplot}
\end{figure}

For each dataset (excluding the two Adipose datasets) and each FDR threshold in $\{1\%, 5\%, 10\%, 20\%\}$, we computed the maximum number
of discoveries across all methods and determined the power of each method relative to this maximum (Figure~\ref{fig:real_data_pvalue_boxplot}). 
RESET Ensemble (94.4\%), RESET NN (93.6\%) and AdaPT\textsubscript{g}-GMM (91.5\%) have the highest medians, with RESET Ensemble also
arguably having the tightest boxplot.
Figure~\ref{fig:real_data_pvalue} shows the number of discoveries for each method on each of the 40 combinations of dataset and FDR threshold.

There are a couple of implementation comments we highlight here (see Sections~\ref{sec:further_details_adapt}-\ref{sec:further_details_adafdr} for details). 
First, we used the same default options of RESET Ensemble for all datasets, except for the two fMRI datasets which exhibit a spike of p-values close to 1. 
To address this, we followed \cite{chao2021adapt}'s suggestion of defining asymmetric target and decoy regions of $[0, 0.3)$ and $(0.3, 0.9]$ respectively. 
Second, we applied AdaPT\textsubscript{g}-GMM and ZAP using p-values rather than raw test-statistics: AdaPT\textsubscript{g}-GMM accepts p-values directly, while for ZAP we converted each $p_i$ into a plausible z-value $z_i = \pm\Phi^{-1}(p_i/2)$ with \textit{i.i.d.}\ uniform signs, matching the p-value information used by the other tools. 
As noted in Section~\ref{sec:background_pvalue}, our implementation of RESET focuses on p-values, and we leave the test-statistic extension to future work.

\begin{figure}[h]
    \centering
    \begin{tabular}{l}
        \includegraphics[width=5.5in]{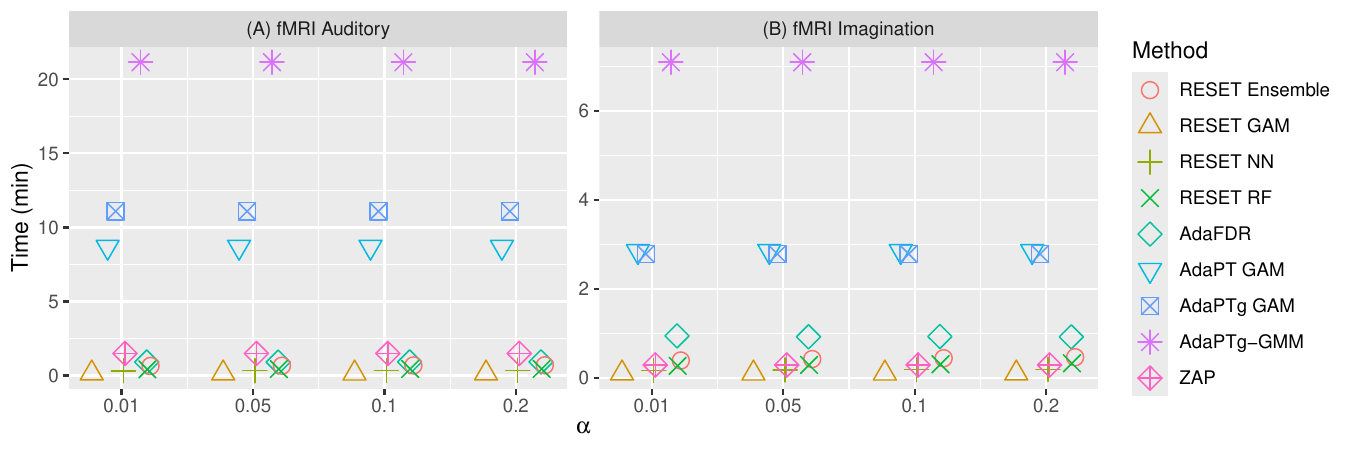}\\    
    \end{tabular}
    \caption{Runtime comparison. In (A) and (B), we compare the runtime of each method on the fMRI datasets. We averaged the times of each RESET method over 10 applications. 
    }
    \label{fig:adipose_fmri}
\end{figure}

To evaluate runtime, we used the two fMRI datasets, which have the most side-information variables and a large number of hypotheses. 
Figure~\ref{fig:adipose_fmri}(A-B) shows that the fastest methods are the RESET variants, ZAP and AdaFDR, with RESET NN, GAM and RF slightly faster than the ensemble variant.

\subsection{FDX control with p-values}
\label{sec:fdp_control_pvals}

\begin{figure}[h]
    \centering
    \begin{tabular}{l}
        \includegraphics[width=6in]{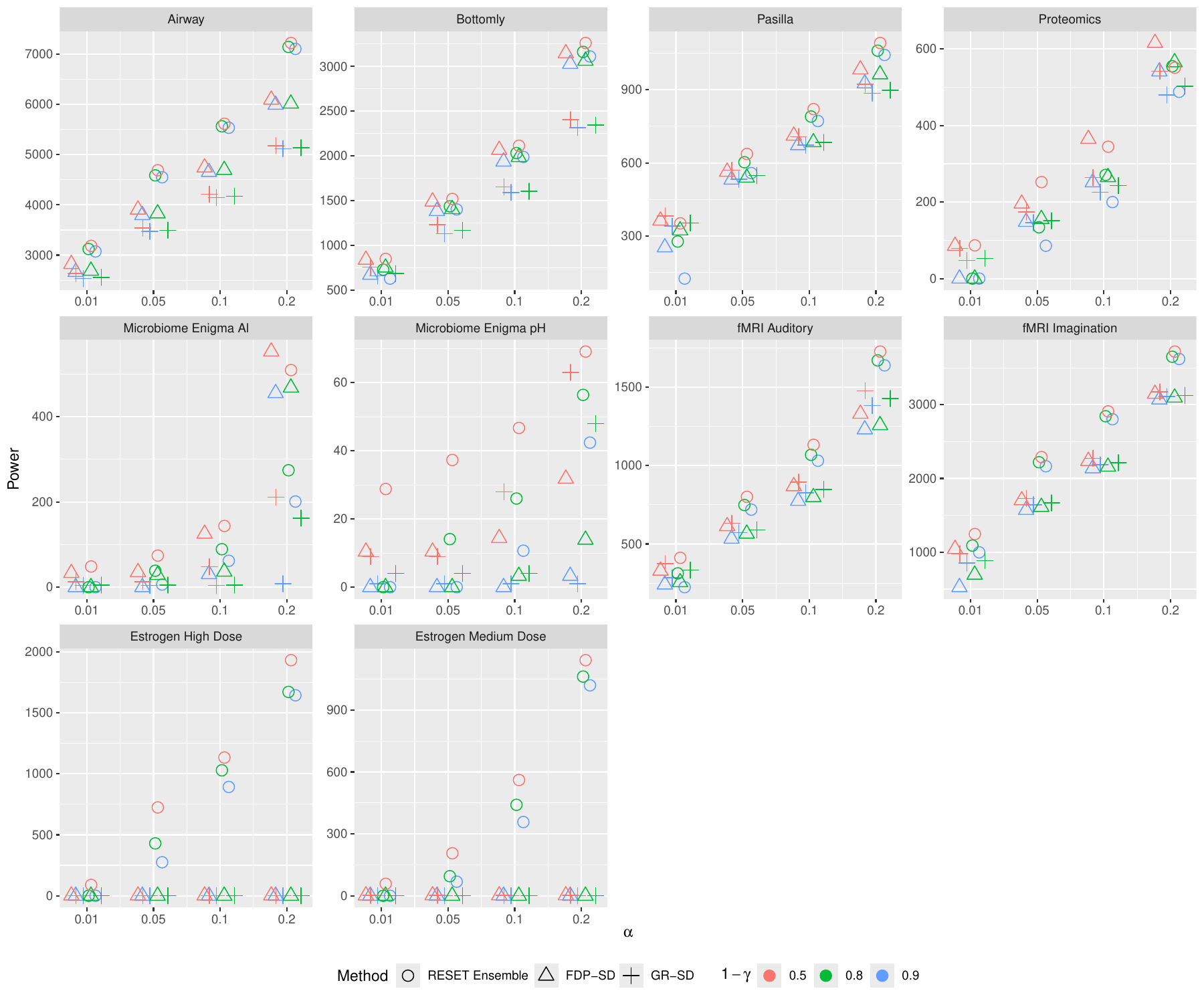}\\
    \end{tabular}
    \caption{Comparison of RESET and other methods with FDX control using a collection of publicly available data.
    Each panel shows the number of discoveries obtained by RESET Ensemble, FDP-SD and GR-SD at a range of FDR thresholds (1\%, 5\%, 10\%, 20\%) with varying confidence $1 - \gamma$ (0.5, 0.8, 0.9).
    RESET Ensemble and FDP-SD were averaged over 10 applications.
    }
    \label{fig:fdp_pval_data}
\end{figure}

In this final section, we demonstrate RESET's FDX control in the p-value setting. 
To the best of our knowledge, there are no existing p-value-based testing procedures that offer finite sample FDX control while utilizing side information.
Therefore, we compare RESET Ensemble against FDP-SD (Algorithm~\ref{alg:FDP-SD}) and GR-SD (Algorithm~\ref{alg:GR-SD}).
For \textit{all} datasets, the labels and scores $(L, W)$ in RESET Ensemble and FDP-SD use asymmetric target and decoy regions, $[0, 0.3)$ and $(0.3, 0.9]$. 
As Section~\ref{sec:further_reset_fdp} explains, a small ratio of estimating decoys to true null targets can cost RESET power under FDX control, which we remedy with a decoy region twice the size of the target region.

Figure~\ref{fig:fdp_pval_data} displays the number of discoveries obtained by each method on the collection of real datasets from the previous section, averaged over 10 runs while varying the FDR thresholds (1\%, 5\%, 10\%, 20\%) and confidence levels (50\%, 80\%, 90\%). 
With few exceptions, RESET Ensemble is generally much more powerful than the alternatives that make no use of the side information.

\section{Discussion}
\label{sec:discussion}

RESET is a general framework for finite-sample FDR and FDX control that can wrap any semi-supervised learning algorithm. Its implementation, RESET Ensemble, is competitive in power with recent methods for both competition- and p-value-based multiple testing with side information, while remaining fast and offering finite-sample FDR or FDX control.

Competing tools such as the AdaPT and Adaptive Knockoffs variants require choosing a variant in advance, since selecting one after seeing its discovery list would generally compromise FDR control; we therefore compared RESET Ensemble either to the default variant (Adaptive Knockoffs' GAM filter) or \textit{separately} to each variant (for AdaPT). Across these comparisons RESET Ensemble achieves comparable or slightly greater power while freeing the user from choosing the `right' method to begin with.

RESET's benefit nonetheless hinges on informative side information. If the covariate carries little signal for distinguishing true from false nulls, RESET (like any side-information-leveraging procedure) may not improve on, and could even lose power relative to, the side-information-oblivious baseline (e.g., SSS+ or BH). This is a power consideration distinct from validity: finite-sample type-1 error control is guaranteed whenever Assumption~\ref{assumption:tdc_aug_general} holds, but uninformative side information cannot create discoveries and may modestly reduce power. 
We therefore recommend prespecifying the candidate side information and assessing its relevance in a preplanned, data-independent manner.
Integrating a fallback procedure to RESET, which switches to a side-information-oblivious procedure when the covariate is uninformative, is left as part of future work.

Each RESET method was also consistently faster than the competing finite-FDR side-information methods. 
ZAP-asymp was marginally faster than RESET Ensemble on one fMRI dataset (Figure~\ref{fig:adipose_fmri}B) but at the cost of only asymptotic FDR control.
The largest gap was in the peptide detection problem, where every Adaptive Knockoffs variant was too slow to be practical.
This speed arises because Adaptive Knockoffs, like the AdaPT variants it is derived from, iterates between training a classifier and deciding which hypothesis to remove from the candidate set $\mathcal{\hat S}_t$ (Section~\ref{sec:comp_side}), whereas RESET Ensemble performs only two iterations and parallelizes heavily.

RESET's decoy split costs little in power: as Section~\ref{sec:decoy_split_cost} shows, discarding half the decoys after training only modestly reduces power. 
RESET can therefore spend those decoys upfront, learning its rescoring on a larger labelled set at once. 
Adaptive Knockoffs and the AdaPT variants, which instead reveal labels sequentially, cannot exploit this labelled data so early.

RESET's compatibility with both FDR and FDX control is also, to our knowledge, unique. 
Combining Adaptive Knockoffs with FDP-SD is problematic, since FDP-SD is a step-down procedure whereas Adaptive Knockoffs is essentially step-up. 
\cite{katsevich:simultaneous} give a general adaptive-ordering method via simultaneous FDP bounds, however it is only a theoretical construction and is conservative as it controls the FDX at every cutoff. 
In the competition setting with side information, RESET is therefore the only reordering method that pairs flexibly with FDP-SD, and we know of no method offering finite-sample FDX control in the p-value setting with side information.

RESET's discovery list is variable because it randomly splits the decoys into training and estimating sets,
albeit this added variance is often marginal relative to the randomness of the data itself (cf.\ Section~\ref{sec:variability}). 
Regardless, the e-value derandomization framework of \cite{ren:derandomised}, which aggregates multiple knockoff runs into a single discovery list with finite-sample FDR control, applies to RESET essentially out of the box, since it relies only on the SSS+ e-values that RESET already uses. 
However, preliminary analysis showed this incurs a substantial power loss so mitigating this loss would require further research.

Lastly, several additional directions remain. First, RESET could be extended to operate directly on the test statistics $z_i$ rather than their p-values, but doing so is not straightforward: there is no single canonical way to feed a per-hypothesis statistic $z_i$ into RESET, and, as \cite{leung:zap} note, converting $z_i$ to a p-value can discard information in the joint structure of $\mathbf{x}_i$ and $z_i$. Accommodating the test statistics directly may therefore require substantial modification of RESET Ensemble.
Second, RESET relies on a handful of prespecified quantities: the probability $s$ of assigning decoys to the estimation set, the target and decoy regions $[0,a)$ and $(b_1,b_2]$, and the mirroring function in Step (\ref{step_1}) that maps larger $p$-values in the decoy region to smaller values in the target region. Since the true null p-values may be non-uniform, other masking functions could be used (\cite{chao2021adapt}); ideally these choices would be data-driven, which we leave to future work.
Finally, we may consider alternatives to our semi-supervised learning engine, RESET Ensemble. One direction is to pair RESET's data-splitting scheme with AdaPT's iterative fitting procedure, which could further increase power (at the expense of additional computation time). Another is to broaden RESET Ensemble's library of classifiers to include the two-group models that are used by many of the competing methods.

\clearpage

\appendix
\renewcommand{\thesection}{A}
\renewcommand{\thealgorithm}{A\arabic{algorithm}}
\renewcommand{\thetable}{A\arabic{table}}
\renewcommand{\thefigure}{A\arabic{figure}}
\renewcommand{\theLemma}{A\arabic{Lemma}}

\section{Algorithms}

\begin{algorithm}[H]
	\caption{ {\bf Selective SeqStep / SeqStep+} (adopted from Selective Sequential Step+ of  \cite{barber:controlling})}
	\label{alg:SeqStep}
	\begin{algorithmic}[1]
		\Require  \begin{tabular}[t]{p{0.8\textwidth}}
		$\bullet$ $(L_i, W_i)_{i = 1}^m$ the list of paired winning labels and scores; \\
		$\bullet$ $c \in (0, 1)$ - the probability of a null target/feature win;\\
		$\bullet$ $\alpha \in (0, 1)$ - the FDR threshold;
		\end{tabular}
		\Ensure A discovery list $R_\alpha$
            \State sort the paired $(L_i, W_i)$ in decreasing order of $W_i$ \Comment{ties are randomly broken}
			\State $A_t \gets \#\{i\le t : L_i = -1 \}$ \Comment{number of decoy/knockoffs wins in top $t$ scores}
			\State $R_t \gets \#\{i\le t : L_i = 1 \}$ \Comment{number of target/original feature wins in top $t$ scores}
		\If{SeqStep}
			\State $\tau \gets \max \{t :  \frac{A_t} {R_t \vee 1} \cdot \frac{c}{1 - c} \leq \alpha \}$	
		\ElsIf{SeqStep+}
			\State $\tau \gets \max \{t :  \frac{A_t + 1} {R_t \vee 1} \cdot \frac{c}{1 - c} \leq \alpha \}$	
		\EndIf
		\State \Return $R_\alpha \gets  \{ i : \text{the (pre-sorted) $W_i$ is in the top $\tau$ ranks and $L_i = 1$}\}$
		\end{algorithmic}
\end{algorithm}

\begin{algorithm}[H]
	\caption{ {\bf FDP-SD} (adopted from \cite{luo:competition})}
	\label{alg:FDP-SD}
	\begin{algorithmic}[1]
		\Require  \begin{tabular}[t]{p{0.8\textwidth}}
		$\bullet$ $(L_i, W_i)_{i = 1}^m$ the list of paired winning labels and scores; \\
		$\bullet$ $c \in (0, 1)$ - the probability of a null target/feature win;\\
		$\bullet$ $\alpha \in (0, 1)$ - the FDP threshold;\\
        $\bullet$ $\gamma \in (0, 1)$ - for a $1 - \gamma$ confidence level;
		\end{tabular}
		\Ensure A discovery list $R_{\alpha, \gamma}$
            \State sort the paired $(L_i, W_i)$ in decreasing order of $W_i$ \Comment{ties are randomly broken}
            \State $\lambda \gets c$ 
            \State $R \gets (1 - \lambda)/(c + 1 - \lambda)$
            \State $i_0 \gets \max \{1, \lceil (\lceil \log_{1 - R}(\gamma) \rceil )/\alpha \rceil \}$
            \For{$i = i_0, \dots, m$}
            \State $A_i \gets \#\{j\le i : L_j = -1 \}$ \Comment{number of decoy wins in top $i$ scores}
            \State $\delta_i \gets \max \{ d \in \{0, 1, \dots, i \} : F_{B( \lfloor (i - d)\alpha + 1 + d, R  \rfloor)}(d) \leq \gamma \}$ \Comment{$F_{B(n, p)}$ denotes the CDF of a Binomial$(n, p)$ RV}
            \EndFor
            \State $i \gets i_0$ and $\delta_{i_{0} - 1} \gets -1$ and  $\bar{\delta}_{i_{0} - 1} \gets 0$
			\While{$i \leq m$}
                \State $k_0 \gets \lfloor (i - \delta_i) \cdot \alpha \rfloor + 1$
                \State $k_1 \gets \lfloor (i - \delta_i + 1) \cdot \alpha \rfloor + 1$
                \State $p_0 \gets F_{B(k_0 + \delta_i, R)}(\delta_i)$
                \State $p_1 \gets F_{B(k_1 + \delta_i + 1, R)}(\delta_i + 1)$
                \State $w_i \gets (p_1 - \gamma)/(p_1 - p_0)$
                \If{$\bar{\delta}_{i - 1} = \delta_i + 1$} 
                    \State $\bar{\delta}_i \gets \bar{\delta}_{i-1}$
                \Else 
                    \If{$\delta_i > \delta_{i - 1}$}
                        \State $w^\prime \gets w_i$
                    \Else
                        \State $w^\prime \gets w_i/w_{i - 1}$
                    \EndIf
                \EndIf
                \State Randomly set $\bar{\delta}_i \gets \delta_i$ or $\bar{\delta}_i \gets \delta_i + 1$ with probabilities $w^\prime$ and $1 - w^\prime$ respectively
                \If{$A_i \leq \bar{\delta}_i$}
                    \State $i \gets i + 1$
                \Else 
                    \State \textbf{break}
                \EndIf
            \EndWhile
            \If{$A_{i_0} \leq \bar{\delta}_{i_0}$}
                \State $k_{FDP} \gets i - 1$
            \Else 
                \State $k_{FDP} \gets 0$
            \EndIf
            \State $R_{\alpha, \gamma} \gets  \{ i : \text{the (pre-sorted) $W_i$ is in the top $k_{FDP}$ ranks and $L_i = 1$}\}$
		\end{algorithmic}
\end{algorithm}

\begin{algorithm}[H]
	\caption{ {\bf GR-SD} (adopted from \cite{guo:generalized})}
	\label{alg:GR-SD}
	\begin{algorithmic}[1]
		\Require  \begin{tabular}[t]{p{0.8\textwidth}}
		$\bullet$ $(p_i)_{i = 1}^m$ the list of p-values; \\
		$\bullet$ $\alpha \in (0, 1)$ - the FDP threshold;\\
        $\bullet$ $\gamma \in (0, 1)$ - for a $1 - \gamma$ confidence level;
		\end{tabular}
		\Ensure A discovery list $R_{\alpha, \gamma}$
            \State sort the p-values $(p_i)$ in ascending order \Comment{ties are randomly broken}
            \For{$i = 1, \dots, m$}
            \State $k_i \gets \lfloor \alpha i \rfloor + 1$ 
            \State $\delta_i \gets F^{-1}[k_i, m - i + 1](\gamma)$ \Comment{$F[\alpha, \beta](\cdot)$ denotes the CDF of a Beta$(\alpha, \beta)$ RV}
            \EndFor
            \State $k_{GR} \gets \max \{i : \prod_{j = 1}^{i} 1_{p_i \leq \delta_i} = 1 \text{ or } i = 0\}$ 
            \State $R_{\alpha, \gamma} \gets  \{ i \in [m] : i \leq k_{GR}\}$
		\end{algorithmic}
\end{algorithm}

\begin{algorithm}[h]
	\caption{ {\bf RESET}}
	\label{alg:reset}
	\begin{algorithmic}[1]
		\Require  \begin{tabular}[t]{p{0.8\textwidth}}
		$\bullet$ $\{(L_i, W_i, \mathbf{x}_i)\,:\,i = 1,\dots,m\}$ - each hypothesis' (label, winning score, side information); \\
            $\bullet$ $\alpha$ - FDR threshold for the discovery list;\\
            $\bullet$ $s$ - the probability of assigning a decoy to the training set (default: $s=1/2$);\\
			$\bullet$ $f$ - a semi-supervised machine learning model;\\
            $\bullet$ $c_0$ - an upper bound probability for a true null label $\mathbb{P}(L_i = 1) \leq c_0$ ($i$ is a true null);\\ 
            $\bullet$ $\texttt{isFDR}$ - a boolean which determines whether FDR or FDX control is desired;\\
            $\bullet$ $\gamma$ - a confidence parameter if FDX control is desired;\\ 
		\end{tabular}
		\Ensure A discovery list $R$
			\State $I \gets$ a subset of the decoy win indices, $\{i : L_i = -1 \}$, determined by independently including each one with probability $s$
            \State $\tilde{L}_i \gets -1$ for $i \in I$ \Comment{training decoys}
            \State $\tilde{L}_i \gets 1$ for $i \in J:=I^c$ \Comment{pseudo-targets}
            \State $(\widetilde{W}_i)_{i = 1}^m \gets f \left( ( W_i, \mathbf{x}_i, \tilde{L}_i )_{i = 1}^m\right)$ \Comment{rescoring using the semi-supervised $f$}
            \If{\texttt{isFDR}}
                \State $R \gets \text{SSS+}((\widetilde{W}_i, L_i)_{i \in J}, c = c_e := \frac{c_0}{1 - s \cdot (1 - c_0)}, \alpha)$ \Comment{Selective SeqStep+}
            \Else 
                \State $R \gets \text{FDP-SD}((\widetilde{W}_i, L_i)_{i \in J}, c = c_e = \frac{c_0}{1 - s \cdot (1 - c_0)}, \alpha, \gamma)$ \Comment{FDP-stepdown}
            \EndIf
			\State \Return $R$
		\end{algorithmic}
\end{algorithm}

\clearpage

\appendix
\renewcommand{\thesection}{B}
\renewcommand{\thealgorithm}{B\arabic{algorithm}}
\renewcommand{\thetable}{B\arabic{table}}
\renewcommand{\thefigure}{B\arabic{figure}}
\renewcommand{\theLemma}{B\arabic{Lemma}}
\renewcommand{\theProposition}{B\arabic{Proposition}}
\renewcommand{\theCorollary}{B\arabic{Corollary}}

\section{Proofs}

\subsection{RESET controls the FDR or FDX}
\label{sec:FDR_FDP_control}

By Lemma \ref{lemma:ind} in the main text, conditional on $\mathcal G$ the true null labels among the pseudo-targets are independent $\pm 1$ random variables, though not necessarily identically distributed.
From this and either Lemma 1 of Barber and Candès \cite{barber:controlling} or Lemma 2 of Lei and Fithian \cite{lei:adapt}, it can be shown that RESET controls the FDR.
On the other hand, FDX control does not immediately follow from existing lemmas; one must separately check that the FDP-SD proof still holds when the true null labels are non-identical.

We therefore introduce a new two-step proof that simultaneously addresses RESET's FDR and FDX control in the non-identical setting.
Our approach may be of independent interest, since the extension to the non-identical setting holds for a generic multiple testing procedure.
Specifically, we show that any procedure which controls FDR or FDX with true null probabilities $\mathbb P(L_i=1)=c_i$ also controls them whenever $\mathbb P(L_i=1)\le c_i$.
Second, we verify RESET's control in the special case of Lemma~\ref{lemma:ind} where the true null labels are identically distributed conditioned on $\mathcal G$, i.e., $\mathbb P(L_i=1\mid\mathcal G)=c_e$, and then invoke the first step to cover the general case, $\mathbb P(L_i=1\mid\mathcal G)\le c_e$.

\begin{Definition}
\label{def:input_template} 
Let $\left(n_i\right)_{i = 1}^{m_0}$ and $\left(a_i\right)_{i = 1}^{m_1}$ be subsequences that partition $[m]$.
Let $\left(c_i\right)_{i = 1}^m$ be a sequence of probabilities in $[0, 1]$.
An input template $\mathcal{I}$ is any configuration of the data such that:
\begin{enumerate}
    \item $\left( H_{n_i} \right)_{i = 1}^{m_0}$ are true null hypotheses and $\left( H_{a_i} \right)_{i = 1}^{m_1}$ are false null hypotheses;
    \item all the scores $\left(W_i\right)_{i=1}^m$ as well as the labels $\left(L_{a_i}\right)_{i=1}^{m_1}$ of the false null hypotheses are fixed;
    \item conditioned on the false null labels $\left( L_{a_{i}}\right) _{i=1}^{m_1}$ and on all the scores $\left( W_{i}\right) _{i=1}^{m}$ the labels of the true null hypotheses are independent with $P(L_{n_{i}}=1)=c_{n_{i}}$.
\end{enumerate}
\end{Definition}

\begin{Lemma}
\label{lemma:extend} Suppose $R$ is a multiple hypothesis selection
procedure (filter) that:
\end{Lemma}
\begin{enumerate}
\item Accepts as input label-score pairs $\left(L_{i},W_{i}\right)_{i = 1}^m$;
\item Reports only a subset of the target wins $\left(L_{i}=1\right)$;
\item For fixed $\left( c_i \right)_{i = 1}^m$, $R$ controls the FDR/FDX for any input template $\mathcal{I}$.
\end{enumerate}
Then $R$ controls the FDR/FDX for a fixed $\left( c^*_i \right)_{i = 1}^m$ in the above sense (3), where $c^*_i=c_i$ for $i\ne j$ and $c^*_j=0$.
\begin{proof}
    Let $\mathcal{I}^*$ be any input template for the sequence $\left( c^*_i \right)_{i = 1}^m$.
    Suppose $H_j$ is a false null hypothesis.
    Then FDR/FDX control trivially follows from the original assumptions since $\mathcal{I}^*$ only depends on $\left( c^*_i \right)_{i = 1}^m$ only through $(c^*_{n_i})_{i = 1}^{m_0} = (c_{n_i})_{i = 1}^{m_0}$.
    Otherwise, there exists $n_i=j$ for some $i\in[m_0]$.
    Define $\mathcal{I}'$ to be the input template that is identical to $\mathcal{I}^*$ except that $H_j$ is a false null hypothesis and $L_j\coloneqq-1$.
    Again, $R$ controls the FDR/FDX on $\mathcal{I}'$ by the original assumptions since  $\mathcal{I}'$  depends on $\left( c^*_i \right)_{i = 1}^m$ only through $(c_{n_i}^*)_{i = 1, n_i \neq j}^{m_0} = (c_{n_i})_{i = 1, n_i \neq j}^{m_0}$.
    Notably, since for $\mathcal{I}^*$ we have $P\left(L_{j}=1\right)=0$ and since we set $L_{j}\coloneqq-1$ for $\mathcal{I}'$, from the perspective of $R$ the two input templates are identical.
    Moreover, because $R$ cannot report decoy wins, the FDP in its list of discoveries is the same for $\mathcal{I}^*$ and $\mathcal{I}'$, hence $R$ also controls the FDR/FDX on $\mathcal{I}^*$ with $\left( c^*_i \right)_{i = 1}^m$.
\end{proof}

\begin{Lemma}
\label{lemma:change_one}Under the assumptions of Lemma \ref{lemma:extend},
$R$ controls the FDR/FDX for a fixed $(c_i^*)_{i=1}^m$ where $c_i^*=c_i$ for $i\ne j$ and $c_j^*\in[0,c_j]$.
\end{Lemma}

\begin{proof}
If $c_{j}=0$, then the Lemma is trivial.
Otherwise, consider any input template $\mathcal{I}^*$ with $c^*_{j} \in [0, c_j]$. 
Define $\mathcal{I}_{1}$ to be the input template that is identical to $\mathcal{I}^*$ except that $c^*_{j}$ is set to $c_j$.
Similarly, define $\mathcal{I}_{2}$ to be the input template that is identical to $\mathcal{I}^*$ except that $c^*_{j}$ is set to $0$.
We can generate data from $\mathcal{I}^*$ by stochastically sampling from the two related templates $\mathcal{I}_{1}$ and $\mathcal{I}_2$.
First, independently of everything else we draw a value $B$ from a Bernoulli distribution with $P(B=1)=c^*_{j}/c_{j}$. 
Next, if $B=1$ we sample from $\mathcal{I}_{1}$, otherwise we sample from $\mathcal{I}_{2}$.

Note that $R$ controls the FDR/FDX on samples generated according
to the $\mathcal{I}_{1}$ template by the original assumptions, and
it similarly controls the FDR/FDX on $\mathcal{I}_{2}$ by Lemma \ref{lemma:extend}.
It therefore follows from the law of total expectation/probability that $R$ controls the FDR/FDX on the template
$\mathcal{I}^*$.
\end{proof}

\begin{Corollary}
\label{corollary:change_all}Under the assumptions of Lemma~\ref{lemma:extend},
$R$ controls the FDR/FDX for a fixed $(c_i^*)_{i=1}^m$ where $c^*_{j}\in[0,c_{j}]$ for all $j\in[m]$.
\end{Corollary}
\begin{proof}
The result follows by induction on $m$ using Lemma~\ref{lemma:change_one}.
\end{proof}

We are now ready to prove that RESET controls the FDR or FDX where we first address the special case where the true null labels are identically distributed conditioned on $\mathcal G$.
We use Corollary~\ref{corollary:change_all} to extend our proof to the general case.

\begin{proof}[Proof of Theorem~\ref{theorem:perc_fdr}]
   For each $i \in J$ where $J = \{i \in [m]: \tilde L_i = 1 \}$, we assign a one-bit p-value:
    \[ \tilde{p}_i = \begin{cases} 
        c_e & L_i = +1 \\
        1 & L_i = -1.
   \end{cases}
    \]
	Recall that the scores $\widetilde{W}$ are themselves a function of $\mathcal G$ and clearly each one bit p-value $\tilde p_i$ is a function of the labels $L_i$. 
    It follows from Lemma~\ref{lemma:ind} of the main text that conditioned on the scores $\widetilde{W}$ and the one-bit p-values of the false nulls, the one-bit p-values of the true nulls $(\tilde p_i )_{i \in J \cap N}$
	are independent with $\mathbb{P}(\tilde{p}_i\le c_e)=\mathbb{P}(L_i = 1) \leq c_e$, i.e., they stochastically dominate the uniform distribution.
	Suppose the labels are identically distributed so that $\mathbb{P}(\tilde{p}_i\le c_e) = \mathbb{P}(L_i = 1) = c_e$ for $i \in J \cap N$.
    Then the conditions of Theorem 3 in Barber and Candès~\cite{barber:controlling} hold and applying, as RESET does, Selective SeqStep+ to the labels $\left( L_i \right)_{i \in J}$ and scores $\left( W_i \right)_{i \in J}$ with $c = c_e$ controls the FDR at level $\alpha$.
    Now suppose the labels are non-identical so that $\mathbb{P}(L_i = 1) \leq c_e$ for $i \in J \cap N$.
    Selective SeqStep+ is a procedure that satisfies the conditions of Lemma~\ref{lemma:extend} and so it follows from Corollary~\ref{corollary:change_all} that RESET also controls the FDR at level $\alpha$.
\end{proof}
\begin{Remark}
    It is worth noting that the original formulation of Selective SeqStep+ in Barber and Candès~\cite{barber:controlling} uses one-bit p-values instead of labels.
    However, these are equivalent since we identify the event $\{\tilde p_i = c_e\}$ with $\{L_i = 1\}$ and $\{\tilde p_i = 1 \}$ with $\{L_i = -1 \}$.
\end{Remark}

\begin{proof}[Proof of Theorem~\ref{theorem:perc_fdp}]
    It follows from Lemma~\ref{lemma:ind} of the main text that conditioned on the scores $\widetilde{W}$ and the labels of the false nulls, the labels of the true nulls $(L_i )_{i \in J \cap N}$	are independent with $\mathbb{P}(L_i = 1) \leq c_e$. 
    Suppose the labels are identically distributed so that $\mathbb{P}(L_i = 1) = c_e$ for $i \in J \cap N$.
    Then this satisfies the assumption of FDP-SD with $c = \lambda = c_e$ (see Section 4 of \cite{luo:competition}) and applying, as RESET does, FDP-SD to the labels  $\left( L_i \right)_{i \in J}$ and scores $\left( W_i \right)_{i \in J}$ controls
	the FDX at $\alpha$ with confidence $1-\gamma$. 
    Now suppose the labels are non-identical so that $\mathbb{P}(L_i = 1) \leq c_e$ for $i \in J \cap N$.
    FDP-SD is a procedure that satisfies the conditions of Lemma~\ref{lemma:extend} and so it follows from Corollary~\ref{corollary:change_all} that RESET also controls the FDX at level $\alpha$ with confidence $1 - \gamma$.
\end{proof}

\subsection{Proof of Lemma~\ref{lemma:chao} in the main text}
\label{sec:proof_of_chao}

We provide a proof of Lemma~\ref{lemma:chao} which is essentially the same as in \cite{chao2021adapt} but has been adapted using our notation.

\begin{proof}[Proof of Lemma~\ref{lemma:chao}]
    By Assumption~\ref{assumption:non_decreasing}, 
    since the pairs $(L_i, W_i)$ are determined by $p_i$ (and vice versa), 
    the null pairs $(L_i, W_i)_{i \in N}$ are independent conditioned on the side information $\mathbf{x}$ and the false null pairs $(L_i, W_i)_{i \not \in N}$.
    Since each $L_i$ only possibly depends on $W_i$, the labels $(L_i)_{i \in N}$ are independent if we further condition on all of $W$, the side information $\mathbf{x}$, and the labels $(L_i)_{i \not \in N}$.
    This proves the first part of Assumption~\ref{assumption:tdc_aug_general}.

    Suppose $\mathbf{x}$ and $(p_i)_{i \not \in N}$ are fixed in the remaining probability statements.
    Then for $i \in N$, since $L_i$ only possibly depends on $W_i$,
    \begin{align*}
        \mathbb{P}(L_i = 1 \mid W_i) &= \frac{\mathbb{P}(p_i \in  [0, a) \mid W_i)}{\mathbb{P}(p_i \in  [0, a) \mid W_i) + \mathbb{P}(p_i \in  (b_1, b_2] \mid W_i)} \\
        & \leq \frac{\mathbb{P}(p_i \in  [0, a) \mid W_i)}{\mathbb{P}(p_i \in  [0, a) \mid W_i) + \mathbb{P}(p_i \in  [0, a) \mid W_i) \cdot \frac{b_2-b_1}{a}} \\
        &= \frac{a}{a + b_2 - b_1} \\
		&= c_0,
    \end{align*}
    where the first line follows since the denominator sums to 1, and the second line follows from the non-decreasing property as explained next.
	
	First, note that the non-decreasing density property from Assumption~\ref{assumption:non_decreasing} implies that
	for a true null p-value $p_i$ ($i\in N$) and a measurable $B \subseteq [0, a)$
    \begin{align}
        \label{eq:mirror_conservative}
        \mathbb{P}(p_i \in B )\leq \frac{a}{b_2 - b_1} \cdot \mathbb{P}(p_i \in q(B) ) ,
    \end{align}
    \noindent where $q$ denotes the inverse transformation of the mirroring that maps p-values in $(b_1, b_2]$ onto $[0, a)$,
	that is $q(p) := b_2 - \frac{b_2 - b_1}{a}\cdot p$.

	Let $A\subseteq\mathbb{R}$ be measurable and let $B \subseteq [0, a)$ be such that $\{W_i \in A \} = \{ p_i \in B\} \cup \{p_i \in q(B)\}$. Then,
    \begin{align*}
        \int_{W_i \in A} 1_{p_i \in [0, a)} d\mathbb{P} = \int 1_{\{p_i \in [0, a) \cap B\}} d\mathbb{P} &= \mathbb{P}(p_i \in  B)\\
        &\leq \frac{a}{b_2 - b_1} \mathbb{P}(p_i \in q(B))\\
        &= \frac{a}{b_2 - b_1} \int 1_{\{p_i \in (b_1, b_2] \cap q(B)\}} d\mathbb{P} \\
        & = \frac{a}{b_2-b_1}\int_{W_i \in A} 1_{p_i \in (b_1, b_2]} d\mathbb{P},
    \end{align*}
    \noindent Since $A$ was arbitrary, it follows that:
    \begin{align*}
        \mathbb{P}(p_i \in  [0, a) \mid W_i) \leq \frac{a}{b_2 - b_1} \mathbb{P}(p_i \in  (b_1, b_2] \mid W_i).
    \end{align*}
\end{proof}
\begin{Remark}
    As part of AdaPT's assumptions, Lei and Fithian~\cite{lei:adapt} directly assume Equation~(\ref{eq:mirror_conservative}), which they call \textit{mirror-conservatism}.
    Lemma~\ref{lemma:chao} is therefore true under a broader class of true null p-values that satisfy mirror-conservatism, not just those that are non-decreasing as
	prescribed by AdaPTg's assumptions~\cite{chao2021adapt} and our Assumption~\ref{assumption:non_decreasing}.
\end{Remark}

\appendix
\renewcommand{\thesection}{C}
\renewcommand{\thealgorithm}{C\arabic{algorithm}}
\renewcommand{\thetable}{C\arabic{table}}
\renewcommand{\thefigure}{C\arabic{figure}}
\renewcommand{\theLemma}{C\arabic{Lemma}}

\clearpage
\section{Supplementary Methods}

\subsection{Summary of methods}
\label{sec:summary_of_methods}

\begin{enumerate}
    \item \underline{AdaFDR}: AdaFDR splits the pairs $(p_i, \mathbf{x}_i)_{i = 1}^m$ into two folds. 
    For each fold, the method fits a mixture model to optimally rescore the hypotheses, and the model is then applied to the other fold where a list of discoveries is obtained. 
    The two list of discoveries are joined together and then reported.
    AdaFDR's mixture model is a composition of Gaussian components to model local `bumps' of genuine signals (false nulls) within the data plus an additional component that captures the monotonic relationship between the side information and the signals.
    \item \underline{AdaPT}: AdaPT works by iteratively pruning a candidate set of hypotheses $\mathcal{\hat S}_t$, similar to Adaptive Knockoffs, which was derived from it. 
    At each iteration, AdaPT fits a two-group mixture model on a strict subset of information: the p-values outside $\mathcal{\hat S}_t$, the \textit{masked} p-values in $\mathcal{\hat S}_t$ given by $\left( p_i \wedge (1 - p_i) \right)_{i \in \mathcal{\hat S}_t}$ where $1 - p_i$ is referred to as the \textit{mirrored} p-value, the side information $\mathbf{x}$, as well as $\#\{i \in \mathcal{\hat S}_t : p_i > 1/2 \}$ and $\#\{i \in \mathcal{\hat S}_t : p_i < 1/2 \}$. 
    It then removes the hypothesis from $\mathcal{\hat S}_t$ that is deemed most likely to be a true null based on the fitted model. 
    To fit the mixture model, the EM algorithm is used. 
    The M-step of the EM algorithm results in a regression problem which allows the user to flexibly use the regression method of their choice. 
    The authors considered a generalized additive model (GAM), a generalized linear model (GLM), and a generalized linear model with $L_1$ regularization (GLMnet). 
    If we identify each p-value $p_i > 1/2$ with a label $L_i = -1$ and each p-value $p_i < 1/2$ with a label $L_i = 1$, then AdaPT terminates this pruning procedure at step $\tau$ when the estimated FDR, as in Equation~(\ref{discovery_adako}) of the main text, is $\leq \alpha$. 
    The rationale is that for true null hypotheses, the p-values are usually uniformly distributed, so $\#\{i \in \mathcal{\hat S}_t : L_i = -1 \}$ estimates the number of true null hypotheses with positive labels, $\#\{i \in \mathcal{\hat S}_t : L_i = 1 \}$. 
    It then reports the remaining positively labelled hypotheses in $\mathcal{\hat S}_\tau$. 
    Note, AdaPT and its variants do not use labels explicitly --- we introduce those here for notational convenience.
    \item \underline{AdaPT\textsubscript{g}}: AdaPT\textsubscript{g} generalizes AdaPT by constructing asymmetric regions to define the positive and negative labels. 
    For example, the p-values $p_i \in \left(0.3, 0.9 \right)$ can be identified with a label $L_i = -1$ and the p-values $p_i < 0.3$ can be identified with a label $L_i = 1$. 
    One advantage of defining these asymmetric regions, is to avoid overestimating the number of true null hypotheses in cases where their p-values are not uniformly distributed and there is a concentration of true null p-values near 1. 
    Subsequently, the estimated FDR needs to be adjusted to account for the fact that we expect twice as many true null p-values with negative labels than positive labels. 
    \item \underline{AdaPT-GMM\textsubscript{g}}: AdaPT-GMM\textsubscript{g} replaces the two group mixture model in AdaPT\textsubscript{g} for a Gaussian mixture model (GMM). 
    Similar to AdaPT/AdaPT\textsubscript{g}, it also allows the user to choose from a collection of regression methods to fit the model at the M-step of the EM algorithm.
    \item \underline{ZAP-asymp}: Lastly, ZAP-asymp directly fits a beta mixture model to the data but is only able to offer asymptotic FDR control. 
    Importantly, ZAP-asymp can operate directly on the test statistic instead of just the corresponding p-values to rerank the hypotheses. 
    This is particularly advantageous for two-sided tests because the test statistic is mapped to its p-value in a non-bijective manner, losing information regarding the sign of the original test statistic along the way. 
    Note that AdaPT-GMM\textsubscript{g} can also operate on test statistics, but in this paper, we focus on the p-value information.
\end{enumerate}

\subsection{Further implementation details of RESET}
\label{sec:further_details_reset}

\subsubsection{Handling hypotheses with a score of zero}
RESET throws out the hypotheses with undefined labels, i.e., those hypotheses with $W_i = 0$.
We try to recover some of this information in the following way.
If the proportion of zero-scoring hypotheses is at least 1\%, then for each hypothesis with $W_i \neq 0$, we identify the $k$-nearest neighbours in terms of the side information (zero-scoring hypotheses included). Then we define an extra side information variable as the number of zero-scoring hypotheses among these identified neighbours (we used $k = 20$).

\subsubsection{Handling noisy side information}

To reduce the effect of `noisy' side information that is unable to distinguish between true and false nulls, we implement an initial side information selection procedure. 
Specifically, we use a generalized additive model to fit a smoothing spline using $W \cdot \tilde{L}$ as the response against each user-provided side information variable one at a time. 
We use \texttt{tryCatch} in case the smoothing spline fails, in which case we fit the response directly to the side information variable, that is, we use a linear fit. 
Again, we try to make use of some of the zero-scoring hypotheses setting $W \cdot \tilde{L} = 0$ in these cases. 
We keep only the side information variables with sufficiently small p-values in the fitted model ($p < 0.01$).

\subsubsection{Positive set for training}
\label{sec:initial_pos_set}

Not all the pseudo-targets need to be included in the initial positive set in Step~\ref{step_i}. 
After all, we expect many of the pseudo-targets to be true nulls. 
Hence, to improve the performance of RESET, we consider the following strategy. 
For each feature in $(W, \mathbf{x})$, reorder the hypotheses according to that feature, and apply SSS to the pseudo labels $\tilde{L}$ at the FDR level $\alpha$. 
Then select the feature that maximizes the number of pseudo discoveries. 
Then using the chosen feature, we define the positive set as those pseudo-targets discovered at $\alpha_0$, again using SSS with the pseudo labels $\tilde{L}$ (we used $\alpha_0 = 50\%$ throughout). 
Usually the score is selected, but in some cases the side information is highly informative and is selected instead.
Users interested in a small FDR level, like $\alpha = 1\%$, could probably benefit from reducing $\alpha_0$ to further improve the performance, and in particular the speed, of RESET (though as we mentioned, we kept $\alpha_0 = 50\%$ across the board). 

To ensure that the positive set is always large enough, we increase $\alpha_0$ by increments of 0.01 until the positive set reaches \texttt{min\_positive} = 50 or until all the pseudo-targets are in the positive set.
This is applied to both the initial positive set and the second round of training.

\subsubsection{Handling dependent data}
\label{sec:handling_dependent_data}

Real data often exhibits dependency to some degree, failing to satisfy Assumption~\ref{assumption:tdc_aug_general},
which guarantees RESET's finite sample FDR/FDX control.
To overcome this challenge, we implement the following two modifications when applied to p-value based data in RESET Ensemble and its variants.

The first modification is in Step~\ref{step_ii}.
For each fold $k \in [K]$, we replace each score $W_i$ from the negative set in the training folds, $[K] \setminus \{k \}$, with scores drawn \textit{i.i.d} from $\lvert \Phi^{-1} ( (b_2 - p_i)\cdot\frac{a}{b_2 - b_1} ) \rvert$ where $p_i \sim U[b_1, b_2]$.
In other words, we replace the scores of the negative set for those obtained from uniform p-values before each training phase.

Note that we only modify the p-values of the negative set in the training folds: all the p-values in the test folds remain unchanged.
This reduces any leakage of information from the training folds into the test folds and in the final stage when the estimating decoys are used to report a list of discoveries.

The following second modification is motivated by the observation that the Random Forest is particularly sensitive to local dependencies.
Specifically, if there are several hypotheses that exhibit essentially the same p-value and side information, then the same observation is essentially split across all folds.
The flexibility of Random Forest enables it to overfit to these local regions of hypotheses and significantly overshoot the FDR threshold.

In the second modification we apply $k$-nearest neighbours to the side information of the set of training decoys.
Next, we sum over the squared differences between each training decoy's score $W_i$ and its $k$-neighbours, denoted as $ss_{\text{obs}}$.
We perform a bootstrap-like procedure in which we shuffle the scores of the training decoys $B = 1000$ times and subsequently compute the sum of squares as previously described for each shuffle, producing $\left(ss_{i}\right)_{i = 1}^B$.
We compute a one-sided p-value by calculating the fraction of bootstrapped sum-of-squares smaller than $ss_{\text{obs}}$.
If the p-value is $\leq 0.01$, the user is notified that the data may exhibit local dependency, and we remove the Random Forest from the ensemble of classification algorithms.
If RESET RF is implemented and the p-value is $\leq 0.01$, then RESET RF switches to RESET NN with default parameters.

Our detection method above compares how close the scores are, through $ss_\text{obs}$, versus the collection of $\left(ss_{i}\right)_{i = 1}^B$ in which the scores are no longer dependent on the side information.
Thus a p-value that is $\leq 0.01$ is likely to catch data exhibiting local dependency.
To make the k-nearest neighbours more efficient, and to be certain that the procedure is only applied to true null hypotheses, we consider only the training decoys that have a small enough mirrored p-value, i.e., $(b_2 - p_i)\cdot\frac{a}{b_2 - b_1} \leq 0.1$ and we restrict number of neighbours to $k = 100 \wedge \lceil 0.01 \cdot m\rceil$, where $m$ is the number of considered hypotheses used in RESET.

The above modifications are always in place, for all simulations and real data experiments.
We demonstrate the robustness of our modifications in Section~\ref{sec:robustness_to_dependent_data} in which we show that RESET Ensemble appears to empirically control the FDR using dependent p-values, along with our analysis of real data in Section~\ref{sec:additional_compare}.
In cases when the data satisfies Assumption~\ref{assumption:tdc_aug_general}, our above modifications do not violate finite sample FDR/FDX control.

\subsubsection{Other technical details}
We point out some additional minor details regarding our implementation of RESET. 
We use the \texttt{randomForest} package, \texttt{nnet} and \texttt{mgcv} to implement the random forest, two-layer neural network and generalized additive model in \texttt{R}, respectively. 
We used the default parameters for the random forest except with $1000$ trees, the default parameters for the two-layer neural network and the default parameters for the generalized additive model (except we use \texttt{drop.intercept = TRUE}). 
If the number of side information variables used by RESET is $\leq 3$, a smoothing spline is fitted with \texttt{mgcv::s} otherwise a natural cubic spline is fitted using 5 degrees of freedom with \texttt{splines::ns} on each side information variable separately (for computational efficiency). 
If GAM fails to fit, we used \texttt{tryCatch} to remove the smooth terms and therefore becoming just a linear model.
Some real datasets contain p-values that are `zero'. Hence when we calculate $W_i = \lvert \Phi^{-1} (p_i) \rvert$ for such p-values, we obtain \texttt{Inf} in \texttt{R}. We replace instances of \texttt{Inf} for the maximum real-valued score in the dataset. 
Unless otherwise stated, we used the same default settings of RESET for \textit{all} simulated and real data experiments, which includes
all of the heuristics described in this section (Section~\ref{sec:further_details_reset}).

\subsection{Competition-based multiple testing: the knockoff filter}
\label{sec:kf_filter}

The competition framework was popularized in the statistics community in the context of variable selection, where the goal is to identify relevant
variables from $\{X_i : i = 1, \dots, p\}$ that are conditionally associated with a response $y$. 
The procedure that achieves this while controlling the FDR is called the \textit{knockoff filter}, which begins by pairing each variable $X_i$ with an artificial variable $\tilde{X}_i$, called a \textit{knockoff}. 
In this paper, we consider the \textit{model-X} version of the knockoff filter~\cite{candes:panning} which makes the following assumptions. 
First, the joint distribution of the variables, $X$, is assumed to be known. 
Second, each observation $(X_{i,1}, \dots, X_{i,p}, y_i)$ is sampled \textit{i.i.d.}\ from the joint distribution $F_{XY}$. 
Because $X$ is random and its distribution known, the knockoffs $\tilde{X}$ are also constructed randomly subject to the following conditions:
\begin{enumerate}
	\item The joint distribution of $\hat{X}:= [X, \tilde{X}]$ is the same if we swap a subset of the variables and their corresponding knockoffs, i.e., $\hat{X} \circ \Pi \overset{d}{=} \hat{X}$, where $\Pi$ is a permutation that swaps the subset of variable indices for their corresponding knockoff indices.
	\item We construct $\tilde{X}$ independently of $y$ by only looking at $X$, i.e., $\tilde{X} \independent y \mid X$.
\end{enumerate}

Once the knockoffs have been constructed, each variable is given a user-defined score $W_i([X, \tilde{X} ], y) \in \mathbb{R}$ and a label $L_i([X, \tilde{X} ], y) \in \{ \pm 1 \}$. 
As in the competition framework described in Section~\ref{sec:background_comp}, the idea is that the label $L_i$ indicates whether the variable $X_i$ or the knockoff $\tilde{X}_i$ has greater evidence for being relevant to the model, and the score $W_i$ indicates by how much. 
Commonly, the score and label are combined by setting $sign(W_i) = L_i$; however, given our presentation of RESET, it is instructive to separate them. 

In this paper, we consider the \textit{Lasso Coefficient Difference} (LCD) score, which is the difference in the magnitude of the coefficients of the Lasso regression~\cite{tibshirani:regression}. 
Specifically, cross-validation is used to determine the $\lambda$ penalty when regressing on the augmented design matrix $[X, \tilde{X}]$. 
We denote the absolute values of the Lasso($\lambda$)-fitted coefficients as $Z_i$ for the variables and $\tilde{Z}_i$ for the knockoffs. 
The scores and labels are then defined as
\begin{align}
    \label{eq:lcd}
    W_i :=  \big\vert  Z_i  -   \tilde{Z_i}  \big\vert, \quad L_i := sign( Z_i  -   \tilde{Z_i} ),
\end{align}
where hypotheses with zero scores are thrown out. 

\subsection{Simulation details for the competition setting}
\label{sec:sim_comp}

\subsubsection{Simulation 1: Linear model with one-dimensional side information}
\label{sec:one_dim}

Our first simulation is based on Simulation 1 of \cite{ren:knockoffs}. 
Specifically, $n = 1000$ independent observations from the joint data distribution of $(X, Y)$ are generated by drawing $X$
from a hidden markov model (HMM) with $p = 900$ variables and $Y|X \sim \mathcal{N}(X\beta, I)$ (linear model). 
The $\beta$'s are selected in the following way. Let $k$ denote the number of nonzero $\beta$'s. 
Then $k$ indices from $\{ 1, 2, \dots, 2k \}$ are sampled without replacement, each with probability proportional to $1/i^2$. 
The $k$ chosen indices are the nonzero $\beta$'s and are assigned a value of $\pm 3.5/\sqrt{n}$, where the signs are chosen \textit{i.i.d.}\ uniformly. 
We consider $k \in \{50, 150, 300\}$ while the original simulation only considers $k = 150$. 
We generate HMM knockoffs using~\cite{sesia2019gene}.
Each hypothesis is supported by the side information $\mathbf{x}_i = i$, which should help in detecting the nonzero $\beta$'s, because smaller values of $i$ are more likely to correspond to a nonzero $\beta_i$. 

We generated 100 independent datasets each consisting of drawing $n$ observations as described above.
The original simulation \cite{ren:knockoffs} sampled the $\beta$'s once and fixed them across the 100 datasets. 
Here we redrew the $\beta$'s for each dataset so that we may average our results over a range of models, as we noticed that all the methods exhibited significant variability between the draws.

\subsubsection{Simulation 2: Linear model with two-dimensional side information}
\label{sec:two_dim_lin}

In a second simulation, we considered a larger linear model where $n = 2000$ observations are taken from the joint distribution $(X, Y)$ as described in Section~\ref{sec:one_dim}, with $p = 1800$. 
Accordingly, we also adjusted the number of false nulls to $k \in \{150, 300, 450\}$. 
Note that unlike the previous simulation, it is not relevant how the $k$ variable indices are selected from $\{1, 2, \dots, p \}$. 
Each hypothesis is equipped with the following two-dimensional side information. 
For the $i$th hypothesis, we draw a pair of observations $\mathbf{x}_i$ from a bivariate normal $\mathcal{N}(\boldsymbol{\mu}, I)$ with $\boldsymbol{\mu} = (0, 0)$ for a true null hypothesis and $\boldsymbol{\mu} = \pm(2, 2)$ for a false null hypothesis, where the signs are chosen \textit{i.i.d.}\ uniformly. 
We generated 20 independent datasets of this larger simulation instead of 100, where each involved drawing those $n$ observations, $\beta$ coefficients, and side information variables. 

\subsubsection{Simulation 3: Logistic model with two-dimensional side information}
\label{sec:two_dim}

\begin{figure}[h]
    \centering
    \begin{tabular}{l}
    \includegraphics[width=3in]{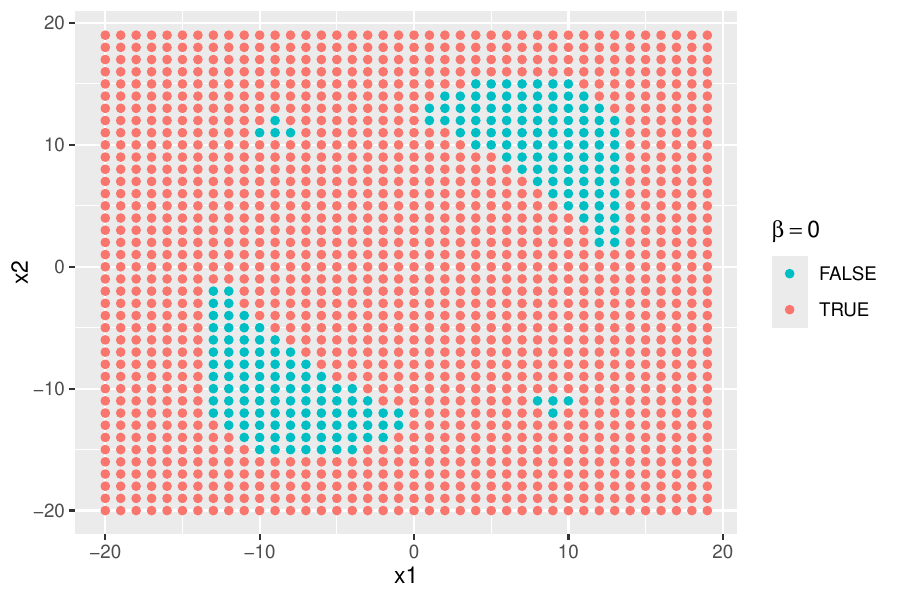}\\
    \end{tabular}
    \caption{Location of nonzero $\beta$'s according to the pair of values $(x, y)$ in the lattice.
    The same image is given in~\cite{ren:knockoffs}.
    }
    \label{fig:beta_location}
\end{figure}

Next, we consider Simulation 2 by \cite{ren:knockoffs}. 
They generate $n = 1000$ observations with $p = 1600$ variables sampled from the joint data distribution of $(X, Y)$, where $X$ is distributed according to a multivariate normal distribution and $Y | X \sim \text{Bernoulli}\left(e^{\beta \cdot X}/\left(1 + e^{\beta \cdot X} \right) \right)$ is a logistic model (see \cite{ren:knockoffs} for further details). 
Each hypothesis' side information is a unique pair of values $\mathbf{x}_i = \left(\mathbf{x}_{i1}, \mathbf{x}_{i2} \right)$ from the lattice $[-20, 19] \times [-20, 19] \subseteq \mathbb{Z}^2$. 
The nonzero $\beta$'s are chosen according to the position of $\left(\mathbf{x}_{i1}, \mathbf{x}_{i2} \right)$ in the lattice. 
An image of the nonzero $\beta$'s is given in Figure~\ref{fig:beta_location}. 
Each nonzero $\beta$ is set to $\pm 25/\sqrt{n}$, where the signs are chosen \textit{i.i.d.}\ uniformly. 
We generate approximate SDP Gaussian knockoffs as described in~\cite{candes:panning}.

We generated 100 independent datasets drawing $n$ observations as above in each but for 2 of those AdaKO EM failed with an error so we used only 98 of the datasets.
The original simulation~\cite{ren:knockoffs} sampled the sign of the $\beta$'s once, fixing them across the 100 datasets. 
Instead, we redraw the sign of the $\beta$'s in each of the 98 datasets since we found that the results of all the methods significantly varied with those draws. 

\subsubsection{Simulation 4: Large $p$ and $n$ and 3-d side information}
\label{sec:three_dim}

\begin{figure*}[h]
    \centering
    \begin{tabular}{l}
    \includegraphics[width=6in]{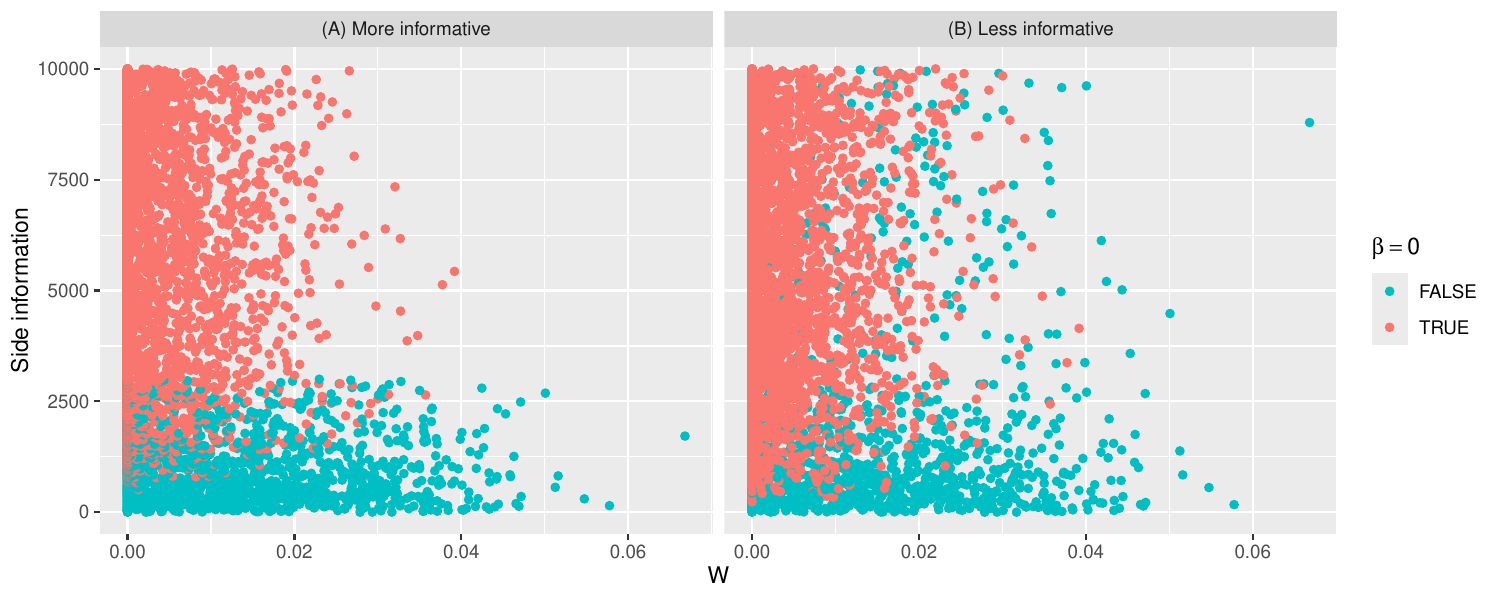}\\
    \end{tabular}
    \caption{\textbf{Scatterplot of each hypothesis according to the LCD score denoted as $W$ (x-axis) and side information variable (y-axis).}
    Each hypothesis is coloured according to whether it is a true (red) or false null hypothesis (blue). In (A), the side information variable is the index of the hypothesis in Simulation 1 (Section~\ref{sec:one_dim}). In (B), we use one of the three side information variables as described in Simulation 4 (Section~\ref{sec:three_dim}).
    }
    \label{fig:side_info}
\end{figure*}

Finally, in the last simulation, $n = 10K$ observations with $p = 10K$ variables and $k = 0.15p$ relevant variables are drawn from the joint data distribution $(X, Y)$ exactly as described in Simulation 1. 
Each variable is equipped with the following three-dimensional side information. 
The first $k$ variables are randomly assigned a value drawn from $i \in \{1, \dots, 2k \}$ taken without replacement and having probability proportional to $1/i^2$. 
The remaining $p - k$ variables are uniquely assigned to one of the remaining values from $i \in \{1, 2, \dots, p\}$ uniformly at random. 
Overall, the final assigment is a permutation of the integers in $\{1, \dots, p\}$, subject to the constraint that the first $k$ variables take values exclusively from $\{1, \dots, 2k\}$.
We expect most of the nonzero $\beta$'s to have small values of $i$, but the effect of the random assignment makes the resulting side information variable less informative than in Simulation 1, as depicted in Figure~\ref{fig:side_info}. 
This process is repeated three times, thus associating three side information variables with each hypothesis, which taken together, should be reasonably informative. 
We generated 5 independent samples, drawing the $n$ observations, $\beta$ coefficients, and side information in each.

\subsection{Further implementation details of Adaptive Knockoffs}
\label{sec:further_details_adako}

There were some small differences between the implementations in the \texttt{adaptiveKnockoff} package in \texttt{R} and the code used in the original manuscript. Accordingly, we used the updated package version. We also fixed a minor bug, which pruned the candidate set $\mathcal{\hat S}_t$ \textit{before} checking the estimated FDR is $\leq \alpha$ --- possibly missing out on a marginally larger discovery list. There are several parameters that may be set before applying the Adaptive Knockoff filters, e.g., the initial proportion of hypotheses revealed, \texttt{reveal\_prop}.
We used the same parameter settings in Simulation 1 here as in Simulation 1 in~\cite{ren:knockoffs} (which uses one-dimensional side information), and the same parameter settings in Simulations 2-4 here as Simulation 2 in~\cite{ren:knockoffs} (which uses two-dimensional side information or more).

In the application to peptide detection, the primary score $W$ can be negative. 
Since Adaptive Knockoffs encodes the labels $L$ by using the sign of the scores $W$, we shifted our peptide scores so that the minimum score was zero in those cases. 
For AdaKO EM, any peptide score $\leq 10^{-3}$ was subsequently assigned a zero score. 
Finally we used, \texttt{reveal\_prop = 10\%} which reveals the hypotheses that are less than or equal to the 10\% of positive scoring hypotheses (which is the same setting used by the simulations). 
All other parameters were default.

\subsection{Competition-based multiple testing: peptide detection}
\label{sec:peptide_detection}

Competition-based multiple testing was first established in the proteomics community where the goal is to infer which proteins, i.e., long chains of amino acids, are present in a biological sample~\cite{elias2007target}. 
Direct protein detection is challenging, and so a bottom-up approach is taken where the proteins are digested into small chains of amino acids, called \textit{peptides}, and the objective is to identify these peptides instead. 
Then, a process called \textit{liquid chromatography tandem mass spectrometry} (LC-MS/MS) is used in which a first round of mass spectrometry isolates each sample peptide and reports the abundance or intensity of each sample peptide.
In a second round of mass spectrometry, each sample peptide that reports a high enough intensity is selected and further fragmented.
The intensity of each fragment is then reported in a corresponding fragmentation spectrum.
Each such spectrum can be thought of as a fingerprint of the peptide that generated the spectrum, allowing us to hypothesize which peptides are present in the sample. 

This is done by searching each spectrum against a database of candidate peptides called \textit{targets} to find the spectrum's optimal peptide-spectrum match (PSM), along with a score indicating the strength of the match.
Unfortunately, due to the incomplete nature of the database and the noise associated with generating the spectra, the PSM can be incorrect.
Hence, we wish to employ a testing procedure for each null hypothesis $H_i$ that `the $i$th target is not present in the sample'. 
In order to decide which null hypotheses to reject, we pair each target peptide with a \textit{decoy} peptide that is generated by randomly shuffling or reversing the target peptide's sequence. 
Accordingly, a second search is performed for each spectrum against the database of decoys to obtain a second PSM and associated score, and only the highest scoring PSM out of the two is retained.

The search phase described above is dictated by a range of user-specified parameters.
The most common configuration of these parameters correspond to two broad types of searches, namely \textit{narrow} and \textit{open} searches, which we consider in this paper.
Briefly a `narrow' search, which is the default search mode, only searches each spectrum in the spectrum file against peptides in the target-decoy database
that have a theoretical mass within a small tolerance of the mass of the sample peptide that generated the experimental spectrum. 
On the other hand, an `open' search makes this mass-tolerance larger so that a spectrum may match with a peptide with a completely different mass. 
This is desirable at times, since sample peptides can undergo \textit{post-translational modifications} which modify the mass of the peptide
so it no longer matches its theoretical mass. However, when searching in an `open' mode, the modified spectrum may still be correctly matched.

The same database peptide, target or decoy, may be matched multiple times with different scores, or have no match at all.
Hence, we associate with $H_i$ the score $Z_i$, defined as the maximum scoring PSM associated with the $i$th target peptide, and by its design the higher the score, the more likely $H_i$ is false, i.e., the peptide is in the sample. 
Similarly, we define the null drawn $\tilde{Z}_i$ as the maximum scoring PSM associated to the $i$th target's corresponding decoy peptide. 
Target or decoy peptides that have no matching spectra are assigned a score of $-\infty$.
Finally, we obtain a winning score $W_i = Z_i \vee \tilde{Z}_i$ and a label $L_i$ indicating whether the winning score was from the $i$th target or its decoy (where ties are randomly broken).
Peptides with a winning score of $-\infty$, indicating that neither the target nor its paired decoy have a matching spectra, are thrown out. 
Finally, a competition-based multiple testing procedure like SSS+ or FDP-SD may be applied to these scores and labels~\cite{nesvizhskii2010survey}.

\subsection{Estimating the computation times of Adaptive Knockoffs}
\label{sec:comp_time}

Running Adaptive Knockoffs (AdaKO) on each of the HEK293 and PRIDE-20 spectrum files is impractical.
We therefore resorted to estimate its runtime, which requires dealing with two challenges.
The first is that the runtime of each iteration of AdaKO, where it determines which hypothesis to reveal next, varies with the number of remaining hypotheses. The second is that we cannot precisely predict the number of iterations AdaKO will take.

Addressing the second challenge first, we note that AdaKO starts by revealing the label of any hypothesis $H_i$ for which $W_i\le q^*_{0.1}$, where $q^*_{c}$ is the $c$-quantile of all positive scoring hypotheses.
Let $m'$ be the number of remaining hypotheses after this initial reveal. 
The number of iterations that AdaKO will make before stopping is then $m'-\ell$, where $\ell$ is the number of hypotheses (target or decoy peptides) that remain when it stops.
We can estimate $\ell$ with $\hat\ell$, the number of target and decoy peptides that score above the 1\% FDR cutoff using RESET Ensemble.
Thus we estimate the number of AdaKO iterations by $m'-\hat\ell$.

To estimate the runtime of each iteration we resorted to taking $k$ measurements. 
Indeed, for each $i \in \{1, \dots, k \}$, we revealed the labels of any hypotheses $H_i$ for which $W_i \leq q_{(0.1\cdot i)}^*$ and determined the time it took for AdaKO to reveal the next label.
We chose $k$ to be the largest positive integer such that the number of remaining labels, after revealing the labels $L_i$ for which $W_i \leq q_{(0.1\cdot k)}^*$, is still more than $\hat{\ell}$.
Thus, we obtain the runtime of $k$ single iterations that are evenly spaced, starting from AdaKO's first iteration and ending at approximately AdaKO's last iteration before it terminates with $\hat{\ell}$ remaining hypotheses.
Next, we computed $\bar{t}_k$, the average of those $k$ measurements.
Thus our overall estimate is:
\[
\bar{t}_k\cdot(m^\prime - \hat\ell) .
\] 

In the analysis of the PRIDE-20 data in Section~\ref{sec:additional_compare}, we considered 10 repeated applications of RESET Ensemble, varying RESET Ensemble's internal seed. 
In this case, we define $\hat\ell$ as the average of the number of winning target and decoy peptides above the cutoff over each application.

\subsection{Computer specifications for computation times}
\label{sec:comp_specs}

Computation times were calculated using an M1 Mac Studio with 20-core CPU and 128GB RAM.

\subsection{Further implementation details of the AdaPT methods}
\label{sec:further_details_adapt}

We applied AdaPT using the \texttt{adaptMT} package in \texttt{R}. 
In Simulations 5 and 6, we used the same settings considered by the authors~\cite{lei:adapt}, i.e.,
we used AdaPT GAM in Simulation 5 and AdaPT GLMnet in Simulation 6 with the same parameters.
Moreover, we considered AdaPT GLMnet in Simulation 5 with the same settings from Simulation 6, demonstrating the importance of selecting the correct working model.
In Simulations 7 and 8, we used the same settings as Simulation 5.
In the real datasets looked at by Zhang et al.\ \cite{zhang2019fast}, we used AdaPT GAM to fit a natural cubic spline using \texttt{splines::ns} with 5 degrees of freedom (knots chosen by default) on each side information variable. 
The use of natural cubic splines is a suggested option from the package documentation. 
There was an exception to this for the GTEx datasets, which flagged an error when we tried to apply the spline basis transformation on the `chromatin state of the SNP'. 
In this case, we applied no transformation to the affected side information variable. 
Zhang et al.\ mention that they employed a ``5-degree for each dimension'', but it is not clear to us if they mean the degree of the piece-wise polynomials that make the spline, or the degrees of freedom of the spline. 
In the latter case, it is not clear how the knots are chosen, or how they overcame the error associated with the `chromatin state of the SNP'. 
Regardless, our results using AdaPT appear to be essentially the same. 
Moreover, Zhang et al.\ omit the application of AdaPT to the two fMRI datasets on account of the categorical side information variable used in these datasets. 
However, Chao and Fithian appear to still apply AdaPT\textsubscript{g} without problem, and so analogously we applied AdaPT as well. 
Lastly, in the gene-drug response data, we followed the same implementation by Lei and Fithian~\cite{lei:adapt}, which used AdaPT GLM and a collection of candidate side information transformations using \texttt{splines::ns} with degrees of freedom ranging from 6 to 10.

\subsection{Further implementation details of the AdaPT\textsubscript{g} methods}

We downloaded AdaPT\textsubscript{g} from the repository \url{https://github.com/patrickrchao/adaptMT} which has all the same implementations as AdaPT while offering the ability to define asymmetric regions for mirroring the p-values. 
Accordingly, we followed the same implementation details as the previous section used by AdaPT and set the so-called `masking parameters' as default.

\subsection{Further implementation details of AdaPT-GMM\textsubscript{g}}

We downloaded AdaPT-GMM\textsubscript{g} from the repository \url{https://github.com/patrickrchao/AdaPTGMM}. 
In Simulations 5, 7 and 8, we used the GAM filter (\texttt{model\_type = `mgcv'}) to fit a smoothing spline using \texttt{mgcv::s} on the side information for the M-Step of the EM algorithm. 
In Simulation 6, we used a generalized linear model filter with regularization (\texttt{model\_type = `glmnet'}) applied directly to the side information variables. 
For the real datasets used by Zhang et al.\ \cite{zhang2019fast}, we used the exact same implementation as Chao et al. from the file \url{https://github.com/patrickrchao/AdaPTGMM_Experiments/blob/main/AdaFDR_experiments/run_all_exp.R}. 
In the gene-drug response data, since only one-dimensional side information is used, we copied the parameters selected from the previous real datasets that also considered just one-dimensional side information.

\subsection{Further implementation details of ZAP}

We used the default parameters of \texttt{zap\_asymp} from the ZAP package \url{https://github.com/dmhleung/zap}. 
In all simulations and real data experiments, we considered a natural cubic spline basis expansion on each side information variable with six degrees of freedom using \texttt{splines::ns}. 
This choice was based on their own analysis on two of the real datasets that we consider here, Airway and Bottomly. 
The exception was in Simulation 6, where we directly used all 100 side information variables rather than applying a spline basis expansion on each of them. 
ZAP requires explicit test-statistics, specifically z-scores, as input.
In Simulations 5, 7 and 8, we used the test statistics as input.
Since the p-values are obtained from a one-sided test, there is a one-to-one correspondence between the p-value information and test statistic information.
Thus our comparisons in Simulations 5, 7, and 8, are still effectively based on the p-values as we addressed in Section~\ref{sec:pval_side}.
In Simulation 6, explicit z-score test statistics were not generated by the simulation. 
Hence, we used the inverse normal CDF $\Phi^{-1}$ to convert the p-values to a z-score test statistic.
In the real data experiments, instead of directly using the $\Phi^{-1}(p_i)$, we converted the p-values to z-values using the transformation $\pm \Phi^{-1}(p_i/2)$ on each p-value $p_i$ where the signs are chosen \textit{i.i.d.}\ uniformly.
Otherwise, applying the transformation $\Phi^{-1}(p_i)$ to true null p-values with a non-decreasing density, or a spike near 1, will produce consistently large z-scores and will be incorrectly discovered by ZAP as an artefact of the data transformation.

\subsection{Further implementation details of AdaFDR}
\label{sec:further_details_adafdr}

We applied AdaFDR using the \texttt{method.adafdr\_test} function from the AdaFDR package \url{https://github.com/martinjzhang/adafdr}. 
We note that installation of the package required setting up a separate conda environment so that we could install an older version of Python (v3.6.13). 
This was because recent versions of Python do not support PyTorch v1.4.0, a requirement of the AdaFDR package \texttt{setup.py} file.

We used the exact same parameters in the vignettes \url{https://github.com/martinjzhang/AdaFDRpaper/tree/master/vignettes} to implement AdaFDR on the 10 datasets that Zhang et al.\ look at.
Interestingly, the two fMRI datasets only consider the Brodmann label as the side information. 
We tried using all the available side information, but the statistical power was worse (data not shown). 
In the gene-drug response data, we used AdaFDR with \texttt{fast\_mode = False}. 
This is in accordance to the discussion from their paper that the fast version of AdaFDR is recommended when few discoveries are expected or if the number of hypotheses is small (which is not the case for both of the gene-drug response datasets).
For the same reasons, we used AdaFDR with \texttt{fast\_mode = True} on Simulation 5 when the number of hypotheses was small, and \texttt{fast\_mode = False} on Simulations 7 and 8 when the number of hypotheses were large.

\subsection{Simulation details for p-value setting}
\label{sec:sim_pval_detail}

\subsubsection{Simulation 5}

\begin{figure}[h]
    \centering
    \begin{tabular}{l}
    \includegraphics[width=6in]{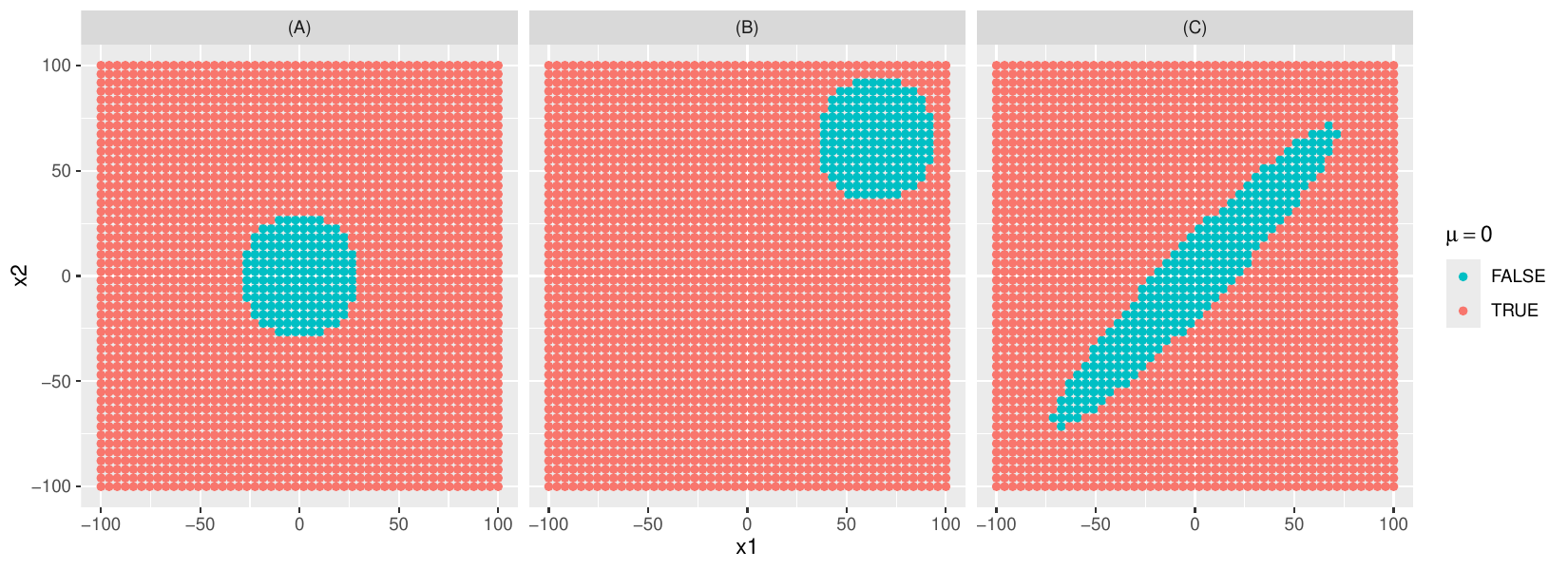}\\
    \end{tabular}
    \caption{Location of false null hypotheses according to the pair of values $\left(\mathbf{x}_{i1}, \mathbf{x}_{i2} \right)$ in the square $\left[-100, 100 \right] \times \left[-100, 100\right]$.
    The same figure is given in~\cite{lei:adapt}.
    }
    \label{fig:mu_location}
\end{figure}

This simulation is based on Simulation 1 of \cite{lei:adapt}. 
For each hypothesis, a p-value is obtained from a one-sided normal test, $p_i = 1 - \Phi(z_i)$ with $z_i \sim \mathcal{N}(\mu, 1)$, where $\mu = 2$ for a false null and $\mu = 0$ for a true null. 
A total of $m = 2500$ hypotheses are used, and each hypothesis' side information is a unique pair $\mathbf{x}_i = (\mathbf{x}_{i1}, \mathbf{x}_{i2})$ drawn from $50 \times 50$ equally spaced values in the square $[-100, 100] \times [-100, 100]$. 
The false nulls occupy one of three regions of the square: a circle in the middle (a), a circle in the top right (b), or an ellipse (c) (Figure~\ref{fig:mu_location}). 
We generated and analysed 100 independent datasets, drawing the 2500 p-values as described in each.

\subsubsection{Simulation 6}

Simulation 6 in Section~\ref{sec:pval_100_dim} models
``noisy'' side information since only the first 2 of the 100 draws are used to distinguish between the true and false null hypotheses. 
Each p-value $p_i$ is drawn \textit{i.i.d.}\ from a two-group beta mixture model, where the true nulls have a uniform distribution and the false nulls have an alternative beta distribution. 
Specifically, the beta mixture model is given by:
\begin{align}
    \label{eq:sim_5}
    f(p_i \mid \mathbf{x}_i) = 1 - \pi(\mathbf{x}_i) + \pi(\mathbf{x}_i) \cdot h(p_i;\mu(\mathbf{x}_i)),\quad h(p_i; \mu(\mathbf{x}_i)) = \left( \frac{1}{\mu(\mathbf{x}_i)} p_i^{\frac{1}{\mu(\mathbf{x}_i)} - 1} \right),
\end{align}
\noindent where $\mathbf{x}_i \in [0,1]^{100}$ denotes the side information of the $i$th hypothesis, $\pi(\mathbf{x}_i)$ denotes proportion of the false nulls in the mixture, $h$ denotes the false null distribution of $\text{Beta}(1/\mu(\mathbf{x}_i), 1)$. 
The $\pi(\mathbf{x}_i)$'s and $\mu(\mathbf{x}_i)$'s are determined by the following relationships with the 100-dimensional side information:
\[ 
\log\left( \frac{\pi}{1 - \pi} \right) = \theta_0 + \mathbf{x}_i^T\theta, \quad \mu_i = \max\{\mathbf{x}_i^T \beta, 1\}, 
\]
\noindent where $\theta = (3, 3, 0, \dots, 0) \in \mathbb{R}^{100}$, $\beta = (2, 2, 0, \dots, 0) \in \mathbb{R}^{100}$ and $\theta_0$ is chosen to satisfy $\frac{1}{m} \sum_{i} \pi(\mathbf{x}_i) = 0.3$. 
Clearly, only the first two values in $\mathbf{x}_i \in [0, 1]^{100}$ contribute to the beta mixture model, as intended. 
We generated 100 independent datasets, where in each we drew the 2000 p-values as described.

\subsubsection{Simulation 7}
    \label{sec:sim_7}

\begin{figure}[h]
    \centering
    \begin{tabular}{l}
    \includegraphics[width=3in]{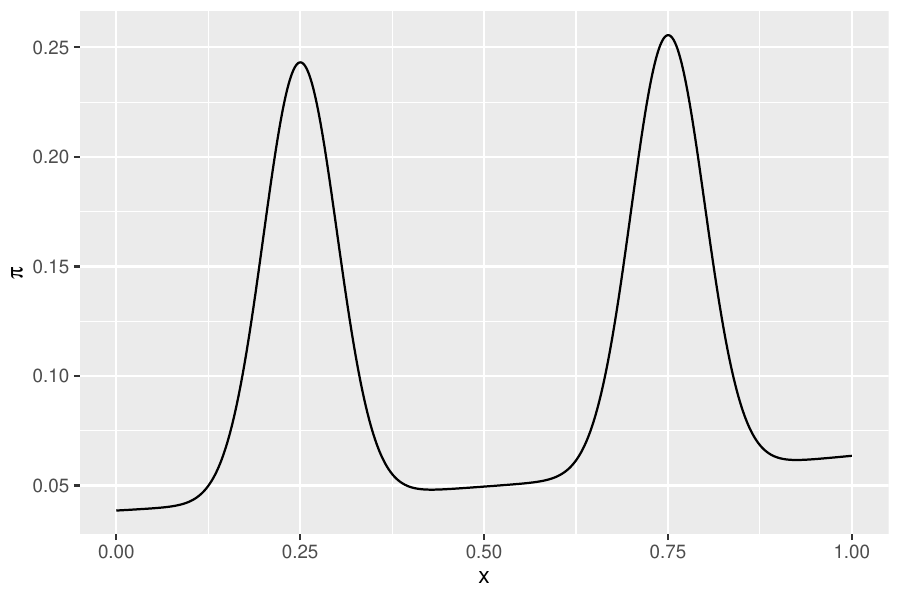}\\
    \end{tabular}
    \caption{The probabilities of a hypothesis being a false null, $\pi$, as a function of the hypothesis' side information, $\mathbf x$.
    }
    \label{fig:plot_pi}
\end{figure}

This simulation is based on Data 4 of \cite{zhang2019fast}. We generated $m = 20{,}000$ hypotheses with \textit{i.i.d.}\ side information $\mathbf{x}_i \sim U[0, 1]$. The $i$th hypothesis is randomly identified as a false null, dependent on its side information $\mathbf{x}_i$, with probability (Figure~\ref{fig:plot_pi})
\[
\pi_i := \pi(\mathbf{x}_i) :=  0.1 \cdot \sum_{j = 0}^2 w_j \cdot f_j (\mathbf{x}_i),
\]
where $(w_0, w_1, w_2) = (0.5, 0.25, 0.25)$, $f_1$ and $f_2$ are normal densities with parameters $(\mu_1, \sigma_1) = (0.25, 0.05)$
and $(\mu_2, \sigma_2) = (0.75, 0.05)$ respectively, truncated between 0 and 1 and normalized such that $\int_{[0,1]} f_1 dx = \int_{[0,1]} f_2 dx = 1$, and $f_0(x) := a \cdot \exp(a x)/(\exp(a) - 1)$ where $a = 0.5$.

For every 10 consecutive hypotheses, starting with $i = 1, \dots, 10$, we drew $z_i$ jointly from $\mathcal{N}(0, \Sigma)$, where $\Sigma$ has upper-triangular entries $\Sigma_{jj} = 1$, $\Sigma_{jk} = 0.25$ for $1 \leq j < k \leq 5$, $\Sigma_{jk} = -0.25$ for $1 \leq j \leq 5 < k \leq 10$, and $\Sigma_{jk} = 0$ otherwise. False-null hypotheses had their z-scores shifted by 2, and a one-sided p-value $p_i = 1 - \Phi(z_i)$ was obtained. The covariance $\Sigma$ induces a local weak dependency among the p-values, affecting both true and false nulls.

\subsubsection{Simulation 8}

This simulation is based on Data 5 of \cite{zhang2019fast}. We used $m = 20{,}000$ hypotheses with the same side information $\mathbf{x}_i$ and false-null probability $\pi_i$ as in Simulation 7. For every 5 consecutive hypotheses, starting with $i = 1, \dots, 5$, we set $z_i = z$ with $z \sim \mathcal{N}(0, 1)$, creating 5 identical z-scores; false-null hypotheses were shifted by 2 and one-sided p-values $p_i = 1 - \Phi(z_i)$ were obtained. This induces a local strong dependency among the p-values, affecting both true and false nulls.

\clearpage

\appendix
\renewcommand{\thesection}{D}
\renewcommand{\thealgorithm}{D\arabic{algorithm}}
\renewcommand{\thetable}{D\arabic{table}}
\renewcommand{\thefigure}{D\arabic{figure}}
\renewcommand{\theLemma}{D\arabic{Lemma}}
\setcounter{figure}{0}

\section{Data preparation}

\subsection{Data preparation and searching with HEK293 data}
\label{sec:HEK293_supp}

We downloaded the human proteome (UP000005640, downloaded on 2023/09/23 from UniProt) and prepared a target peptide database along with 5 randomly shuffled decoy peptide databases using Tide-index within the Crux Toolkit v4.1.6809338~\cite{park2008rapid,kertesz-farkas2023crux} with all options set to default. 
An output containing the target and decoy peptide pairs are conveniently provided using the \texttt{--peptide-list T} option in Tide-index. 
Each of the RAW 24 HEK293 spectrum files~\cite{chick:mass-tolerant} were converted to mzML format using MSConvert 3.0.22314 with the vendor peak-picking filter using the default settings.
For each of the 5 decoy databases, we performed a `narrow' search on each spectrum from the combined 24 mzML spectrum files against the combined target-decoy peptide database using Tide-search~\cite{diament2011faster}, using the options \texttt{--top-match 1 --auto-precursor-window warn --auto-mz-bin-width warn --concat T}. The resulting search files were then converted to so-called pin files using the make-pin function in Crux. Each row in the pin files correspond to the optimal peptide-spectrum match (PSM) for each spectrum, the primary score for quantifying this match, called \textit{XCorr} scores, and a collection of auxiliary information regarding the PSM. 

\begin{table}[h!] 
	\centering
  \begin{tabular}{lp{13cm}}
  \hline
  Side Information & Description\\
  \hline
  deltCn & The difference between the XCorr score of the two top ranked PSMs with respect to the combined database,
  divided by maximum of the top PSM XCorr score and 1\\
  PepLen & The length of the matched peptide, in residues\\
  Charge & The charge of the precursor ion (ranging from +1 to + 5)\\
  lnNumSP & The natural logarithm of the number of database peptides within the specified precursor range\\
  dm & The difference between the calculated and observed mass\\
  absdM & The absolute value of the difference between the calculated and observed mass
  \\
  \hline
  \end{tabular}
  \caption{\textbf{List of described side information used by HEK293 data.} The list of side information used by RESET and Adaptive Knockoffs in the HEK293 data, adapted from \texttt{https://crux.ms/file-formats/features.html}. Note that the Charge state is represented as a one-hot vector, where we left out a Charge state of +1 (due to linearity, since the one-hot vector sums to one).}
  \label{table:side_info} 
\end{table}

We constructed the target and decoy peptide scores, $Z_i$ and $\tilde Z_i$ respectively, according to our description in Section~\ref{sec:peptide_detection} using the XCorr score as the PSM scoring function.
Subsequently we record the winning score $W_i = Z_i \vee \tilde{Z}_i$ along with the label $L_i = sign(Z_i - \tilde{Z}_i)$.
In the case of ties ($Z_i = \tilde{Z}_i$), we randomly set $L_i$ to be $+1$ or $-1$ with equal probability.
To each pair $(W_i, L_i)$ we attribute the side information defined as the auxiliary information $\mathbf{x}_i$ of the underlying PSM that had the XCorr score of $W_i$.
In the case there is more than one PSM with the same XCorr score of $W_i$, we randomly select one of the target PSMs if $L_i = +1$ or one of the decoy PSMs if $L_i = -1$.
The side information that was subsequently used by RESET and Adaptive Knockoffs is given in Table~\ref{table:side_info}.

\subsection{Data preparation and searching with PRIDE-20 data}
\label{sec:pride_20_supp}

We followed the preparation and searching outlined in~\cite{freestone2024analysis}. The only difference is that we used an updated version of the Crux toolkit. 
We provide the following brief description for the reader's convenience. 
Twenty spectrum files and protein databases from separate projects, referred to as the PRIDE-20 dataset, were downloaded (see Table 2 in \cite{freestone2024analysis}).
Tide-index was used to prepare the target peptide database along with 10 randomly shuffled decoy peptide databases for each of the 20 spectrum files. 
Tide-search was used to search each spectrum file against the 10 combined target-decoy databases using two modes: `narrow' and `open' search options as outlined in~\cite{freestone2024analysis}.
The search files were then converted to pin files using the make-pin function in Crux and subsequently filtered for the top 1 (the optimal) peptide-spectrum match for each spectrum.  

For each spectrum file and search type, we obtained the triples $(L, W, \mathbf{x})$ in the following way. 
For each spectrum file that had no modifications (we explain this in more detail next), we proceeded essentially in the same way as in Section~\ref{sec:HEK293_supp}. 
That is, for each peptide in the pin file, we define $Z_i$ (if a target peptide) or $\tilde{Z_i}$ (if a decoy peptide) as the maximum \textit{Tailor} score of all the PSMs associated to the target peptide~\cite{sulimov2020tailor}. 
Here we are using the more sensitive \textit{Tailor} score and relegate the \textit{XCorr} score as one of the side information variables. 
Then we compete each target and its corresponding paired decoy in the same way as Section~\ref{sec:HEK293_supp}, to obtain the triple $(L_i, W_i, \mathbf{x}_i)$ where the side information $\mathbf{x}_i$ is obtained from the auxiliary information of the underlying PSM that had the Tailor score of $W_i$. 
A description of the side information used is given in Table~\ref{table:side_info_pride}.

\begin{table}[h]
	\centering
  \begin{tabular}{lp{13cm}}
  \hline
  Side Information & Description\\
  \hline
  deltLCn & The difference between the XCorr score of the top scoring PSM and the fifth/last ranked PSM with respect to the combined database, divided by the maximum of the top PSM’s XCorr score and 1\\
  deltCn & The difference between the XCorr score of the two top ranked PSMs with respect to the combined database,
  divided by maximum of the top PSM XCorr score and 1\\
  Xcorr & The SEQUEST cross-correlation PSM score\\
  PepLen & The length of the matched peptide, in residues\\
  Charge & The charge of the precursor ion (ranging from +1 to + 5)\\
  lnNumSP & The natural logarithm of the number of database peptides within the specified precursor range\\
  dm & The difference between the calculated and observed mass\\
  absdM & The absolute value of the difference between the calculated and observed mass
  \\
  \hline
  \end{tabular}
  \caption{\textbf{List of described side information used by PRIDE-20 data.} The list of side information used by RESET and Adaptive Knockoffs in the PRIDE-20 data, adapted from \texttt{https://crux.ms/file-formats/features.html}. Note that the Charge state is represented as a one-hot vector, where we left out a Charge state of +1 (due to linearity, since the one-hot vector sums to one). XCorr is used as side information since we define $W$ in terms of Tailor scores here instead.}
  \label{table:side_info_pride} 
\end{table}

We next consider spectrum files that were searched with variable modifications (\texttt{---auto-modifications T}). 
Here we need to make a slight adjustment to account for some dependency within the data. 
Specifically, using variable modifications creates several `copies' of the peptides in the target-decoy database that are distinguished only by slight alterations to the mass of some of the amino acids. 
As an example, \texttt{PEPTIDE} may generate the following `copies': \texttt{PE[16]PTIDE}, \texttt{PEPTIDE[16]} and \texttt{PE[16]PTIDE[16]}, where \texttt{[16]} indicates an increase of 16 Daltons to the amino acid on the left. 
Consequently, the scores of all these peptides are usually correlated: if one of these is scored high, than often they all are. 
Hence in order to satisfy Assumption~\ref{assumption:tdc_aug_general}, we follow the protocol outlined in~\cite{freestone2024analysis}. 
That is, we identify all the copies as being the same peptide, `\texttt{PEPTIDE}', with each of the variable modifications ignored. 
Then we proceed with the same steps as the above paragraph, by determining the maximum score associated with \texttt{PEPTIDE} and recording it as $Z_i$ (or $\tilde{Z}_i$ if it is a decoy). 
We define the labels, winning scores and side information in the same way to obtain our triplets $(L, W, \mathbf{x})$.
 
Due to runtime considerations, we used 13 of the spectrum files to assess Adaptive Knockoff's power as described in Section~\ref{sec:pride} of the manuscript (each resulting in $< 33K$ PSMs), and only one of the 10 target-decoy databases. 
To fairly compare with Adaptive Knockoffs, we used the same target-decoy databases for RESET ensemble. 
All 20 spectrum files and all 10 target-decoy databases for each spectrum file were used in comparing RESET Ensemble's FDX control with FDP-SD's.

\subsection{Description of real data with p-values}
\label{sec:description}

We provide the following description of each dataset used in Section~\ref{sec:real_data_pval} in the main text and in Section~\ref{sec:adipose}. The following three RNA-seq datasets, the proteomics dataset, the two microbiome datasets, the two fMRI datasets, and the two eQTL datasets were downloaded from \url{https://github.com/patrickrchao/AdaPTGMM_Experiments/tree/main/AdaFDR_experiments/data_files}.
The last gene-drug response data was obtained from the \texttt{adaptMT} package.

\begin{itemize}
    \item Three RNA-seq datasets (Airway~\cite{himes2014rna}, Bottomly~\cite{bottomly2011evaluating}, Pasilla~\cite{brooks2011conservation}): The goal is to identify genes that are associated with varying gene expression levels in response to a change of conditions. In the Airway data, the differential expression analysis is conducted in response to a drug, dexamethasone; in the Bottomly data, the analysis is conducted between two mouse strains; and in the Pasilla data, the analysis is conducted between normal and so-called `Pasilla knockdown' conditions~\cite{huber2016pasilla}. In all cases the side information consists of the log-normalized gene expression read counts.
    \item A proteomics dataset (Proteomics~\cite{dephoure2012hyperplexing}): The goal is to identify proteins that have different protein abundances when treated with rapamycin versus DMSO in yeast cells. 
    Each protein consists of side information equal to the natural log of the number of peptides belonging to the protein that were `quantified' in all six samples.
	\item Two microbiome datasets (Microbiome Enigma Ph and Microbiome Enigma Al~\cite{smith2015natural,korthauer2019practical}): The goal is to identify microbiome organisms, recorded as OTUs (Operational Taxonomic Units), that are associated with certain conditions (pH and Al). Each hypothesis (OTU) is equipped with a two-dimensional side information --- the ubiquity, which is defined as the percentage of samples detected with the OTU, and the mean nonzero abundance, which is defined as the average abundance of each OTU across the samples in which it was detected.
    \item Two fMRI datasets (Auditory and Imagination)~\cite{tabelow2011statistical}: The goal is to identify voxels that are activated in response to different stimuli. In the first dataset, a single individual received auditory stimulus and in the second dataset, a single individual was instructed to imagine playing tennis. Each hypothesis (voxel) is equipped with four dimensional side information: the spatial position of the voxel (in three dimensions) and a categorical variable with the Brodmann label~\cite{brodmann1909vergleichende}, which is used to delineate different areas of the brain with different functions.
    \item Two eQTL datasets (Subcutaneous and Omentum~\cite{gtex2015genotype,gtex2020gtex}): The goal is to identify single nucleotide polymorphisms (SNPs) that are associated with changes in gene expression levels. Such an association is referred to as an expression quantitative trait locus (eQTL). Each hypothesis (SNP) is equipped with the following four side information variables: (1) the distance between the SNP and the gene transcription's start site (TSS), (2) the log gene expression level, (3) the alternative allele frequency (AAF) of the SNP and (4) the chromatin state of the SNP.
    \item Gene-drug response data (Estrogen~\cite{davis2007geoquery,li2017accumulation}): The goal is to identify genes that are differentially expressed in response to a low-dosage estrogen treatment applied to breast cancer cells. Each hypothesis (gene) is assigned a `rank' that was generated using an analysis of the gene expression levels in response to (1) a high-dosage estrogen treatment and (2) a medium-dosage estrogen treatment. Here, the ranks are in terms of the strength of association between the expression level of each gene and the level of dosage in (1) and (2). Thus, we can analyse the data twice, once for each side information variable based on (1) and (2).
\end{itemize} 

\subsection{Preparation of data with p-values}
\label{sec:zhang_supp}

\begin{figure*}[h]
    \centering
    \begin{tabular}{l}
 \includegraphics[width=6in]{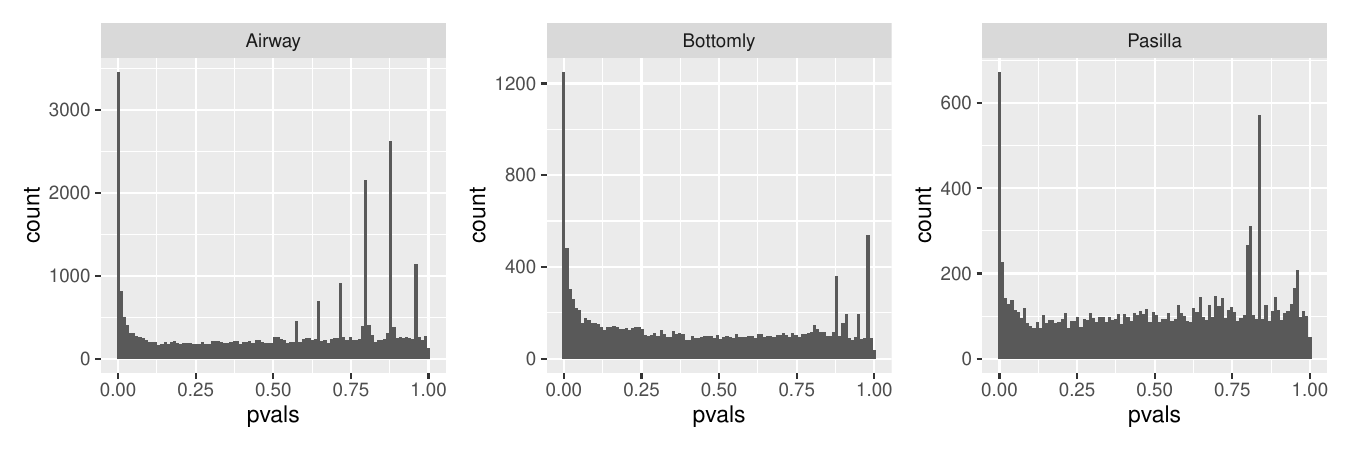} \\
 \includegraphics[width=6in]{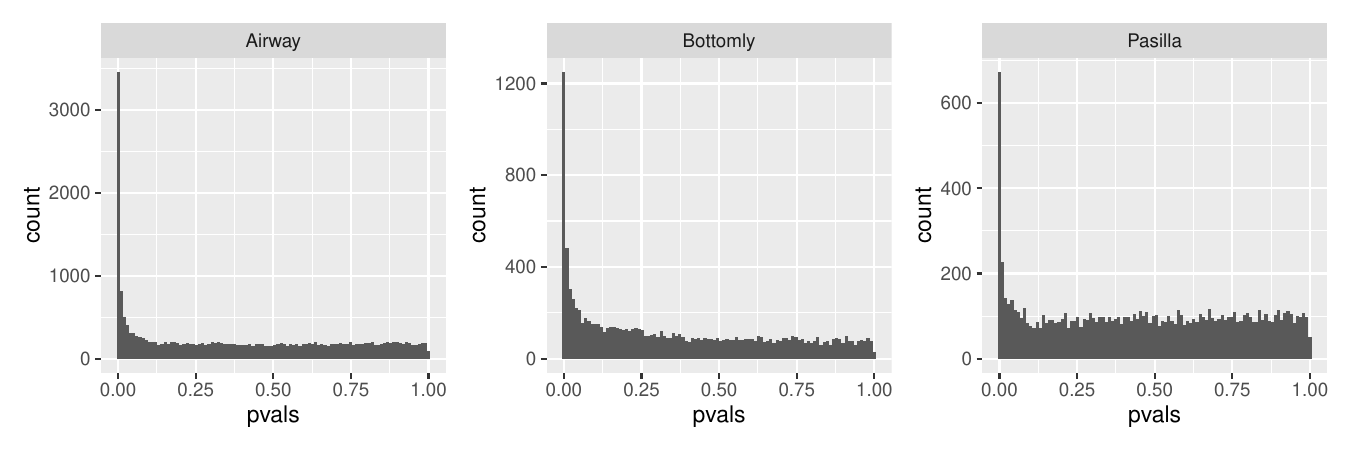} \\
    \end{tabular}
    \caption{\textbf{Histograms of p-values in the RNA-seq data before and after removal of genes with negative log-normalized read count.} Top row: the original histograms of the p-values. Bottom row: the corresponding histograms after the removal of genes with low read counts.}
    \label{fig:rna_hist}
\end{figure*}

All pairs of p-values and side information from the above datasets were used to report a list of discoveries using each p-value based method with the following exceptions.
In the RNA-seq datasets, Airway, Bottomly and Pasilla, there exists `spikes' of high p-values (Figure~\ref{fig:rna_hist}, top row).
These spikes are associated with genes with low log-normalized read count.
Hence, removing all genes with log-normalized read count less than zero --- whose information content is questionable anyhow --- also eliminated those undesirable
spikes (Figure~\ref{fig:rna_hist} bottom row). 
The number of genes removed is given in Table~\ref{tab:removed}.

\begin{table}[h]
    \centering
        \begin{tabular}{lcc}
        \hline
        Dataset & \# of genes removed & \% of genes removed \\
        \hline
        Airway & 10616 & 32\% \\
        Bottomly & 2203 & 16\% \\
        Pasila & 1575 & 13\%\\
        \hline
    \end{tabular}
    \caption{\textbf{The number and percentage of genes removed from the RNA-seq datasets.} }
  \label{tab:removed}
\end{table}

\appendix
\renewcommand{\thesection}{E}
\renewcommand{\thealgorithm}{E\arabic{algorithm}}
\renewcommand{\thetable}{E\arabic{table}}
\renewcommand{\thefigure}{E\arabic{figure}}
\renewcommand{\theLemma}{E\arabic{Lemma}}
\setcounter{figure}{0}
\setcounter{table}{0}

\clearpage
\section{Supplementary results}

\subsection{Further comparisons between RESET and AdaKO in simulations}
\label{sec:further_comparisons}

\begin{figure}[h]
    \centering
    \begin{tabular}{l}
    \includegraphics[width=6in]{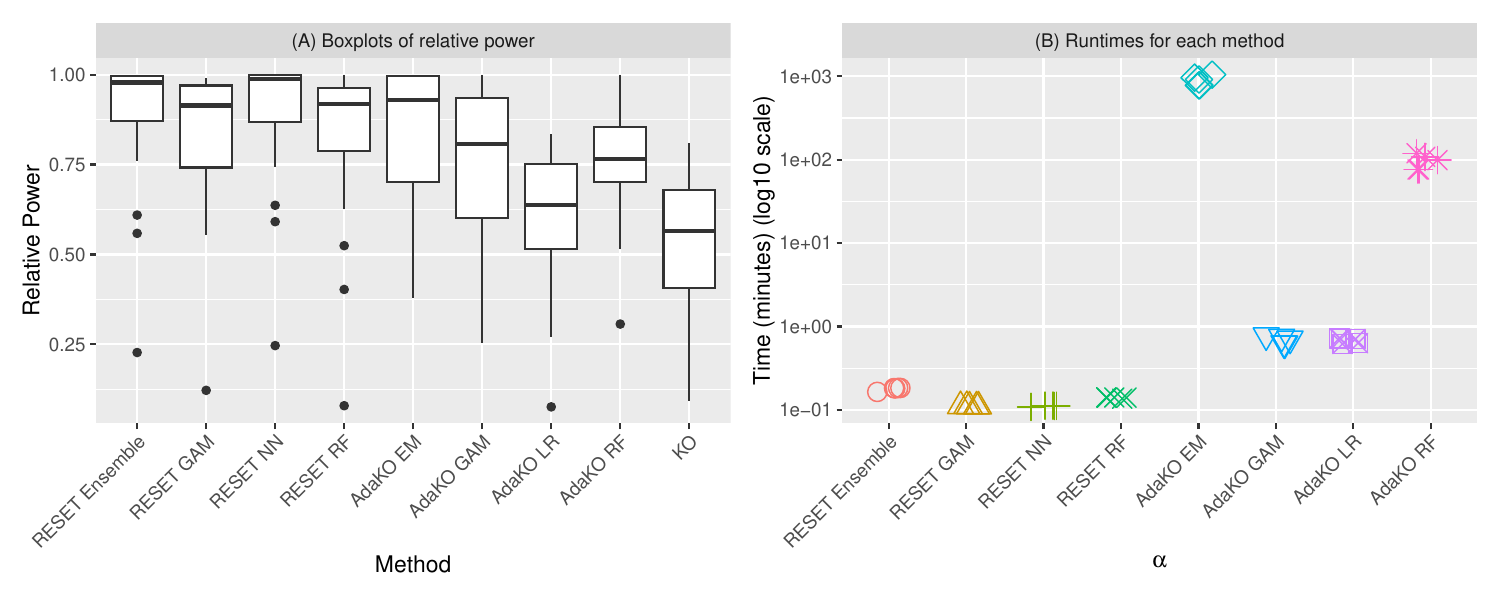}\\
    \end{tabular}
    \caption{\textbf{Relative power and runtimes for each method.}
    (A) Each boxplot describes the relative power for each method, computed across the combination of simulations and FDR thresholds in Figure~\ref{fig:reset_adako_gam}. (B) The log10 runtimes of each method applied at the 5\% FDR level for 5 runs of Simulation 4. For readability, the points are jittered in the horizontal direction.}
    \label{fig:reset_comp_all_methods}
\end{figure}

\begin{figure}[h]
    \centering
    \begin{tabular}{l}
    \includegraphics[width=6in]{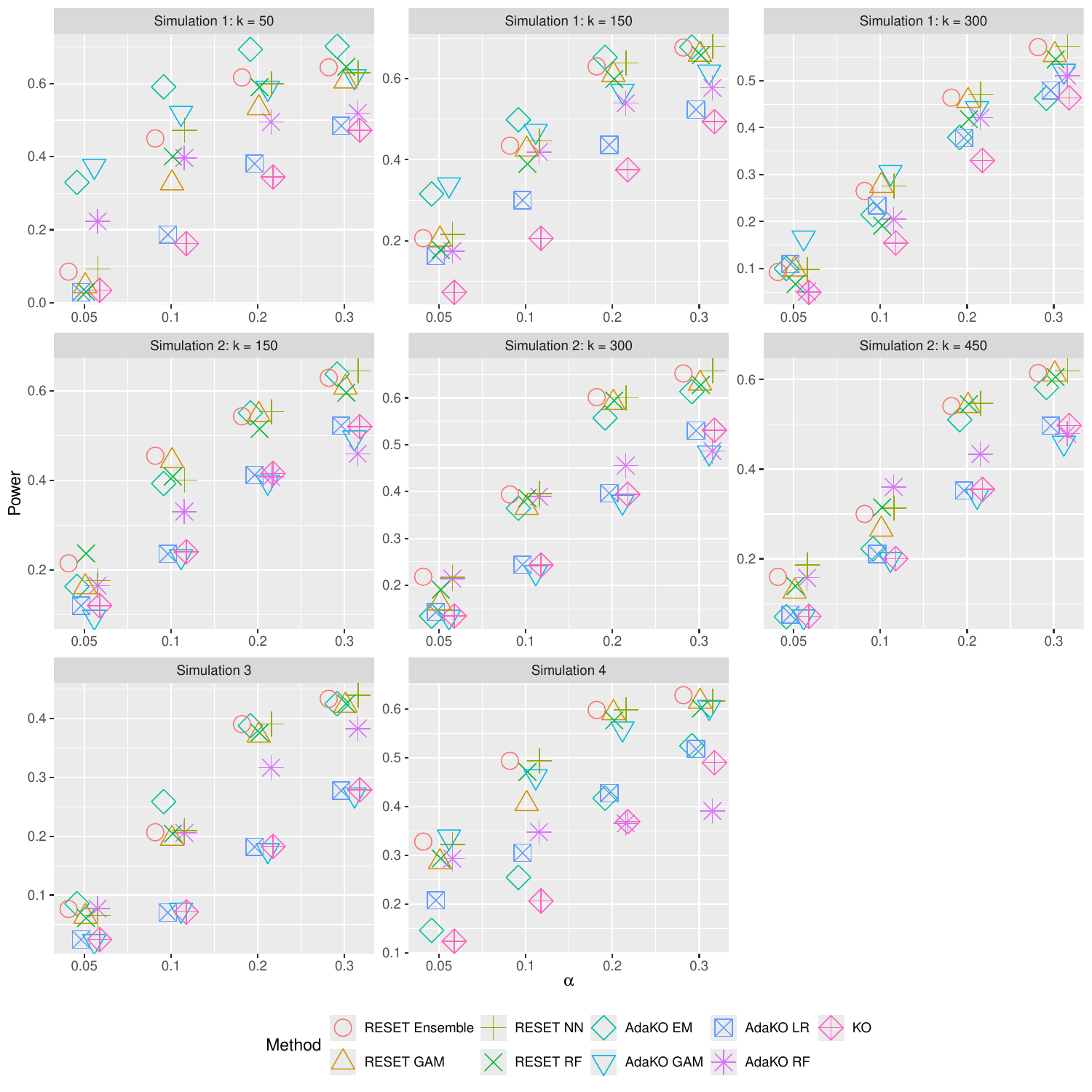}\\
    \end{tabular}
    \caption{\textbf{Estimated power of each method in numerical simulations.}
    Each panel shows the power for each method at FDR thresholds ranging from 5\% to 30\%. The first row corresponds to Simulation 1 with three values of $k \in \{50, 150, 300\}$, the second row corresponds to Simulation 2 with three values of $k \in \{150, 300, 450\}$, and the last row corresponds to Simulations 3 and 4. For readability, the points are jittered in the horizontal direction. A description of the AdaKO methods and KO can be found in Section~\ref{sec:background_comp}.}
    \label{fig:all_power}
\end{figure}

\begin{figure}[h]
    \centering 
    \begin{tabular}{l}
    \includegraphics[width=6in]{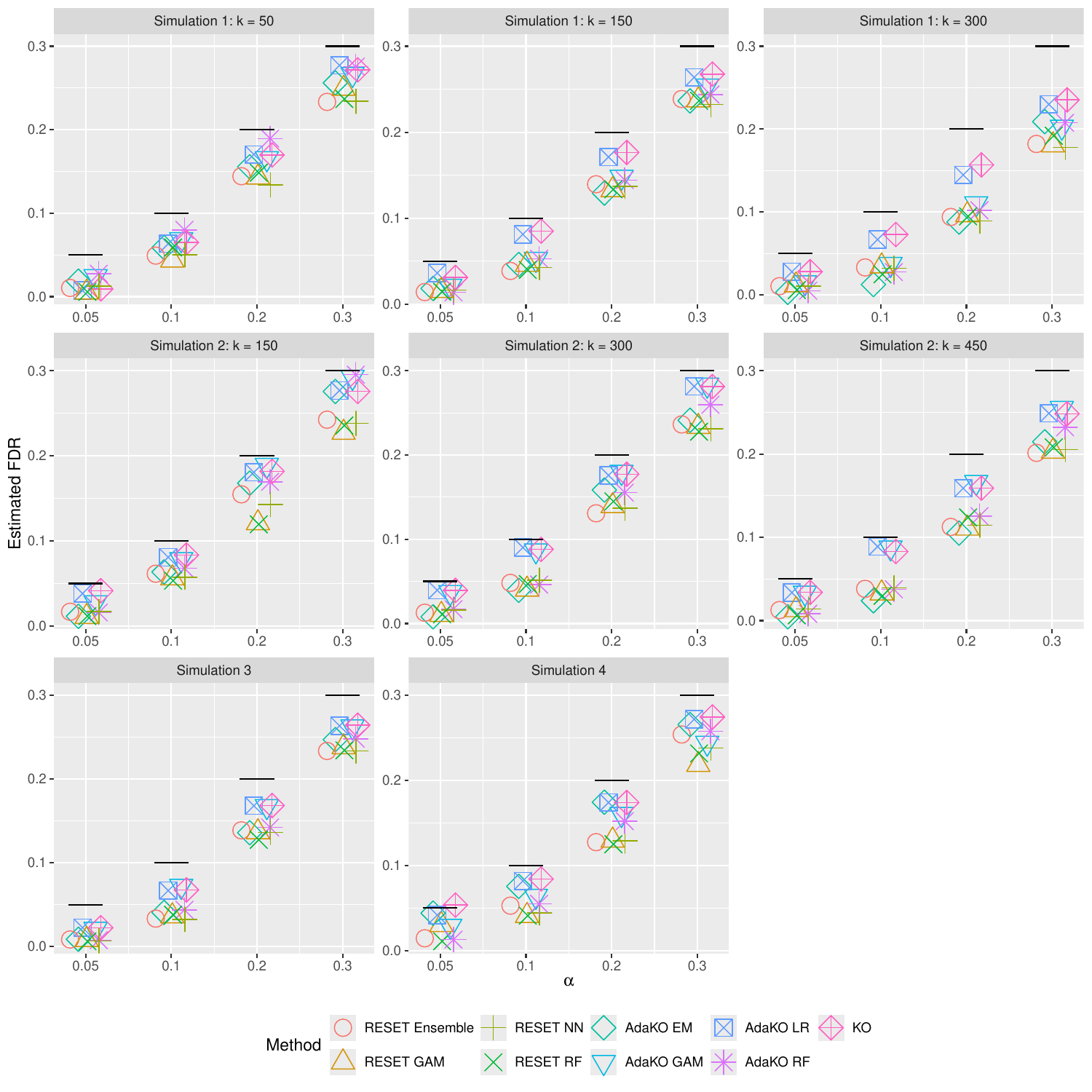}\\
    \end{tabular}
    \caption{\textbf{Estimated FDR of each method in numerical simulations.}
    Each panel shows the estimated FDR for each method at FDR thresholds ranging from 5\% to 30\%. The estimated FDR is calculated as the proportion of irrelevant variables in the discovery list, averaged over the runs. The first row corresponds to Simulation 1 with three values of $k \in \{50, 150, 300\}$, the second row corresponds to Simulation 2 with three values of $k \in \{150, 300, 450\}$, and the last row corresponds to Simulations 3 and 4. For readability, the points are jittered in the horizontal direction.}
    \label{fig:all_fdr}
\end{figure}

For each simulation and FDR threshold in Figure~\ref{fig:reset_adako_gam} of the main text, we determined the maximum power across \textit{all} RESET and AdaKO methods and computed the relative power of each method to this maximum. The resulting relative powers are reported as boxplots in Figure~\ref{fig:reset_comp_all_methods}A, giving an indication of the performance of each method across the simulations as a whole.
Additional plots showing the performance of each method in each separate simulation are given in Figures~\ref{fig:all_power} (power)
and \ref{fig:all_fdr} (estimated FDR).

While in Figure~\ref{fig:all_power} we often find a version of AdaKO that is more powerful, or at least comparable to RESET Ensemble, this varies between the versions. 
On the other hand, RESET Ensemble's performance is consistent across these simulations, and overall typically achieves the highest relative power when the results are aggregated. 
Interestingly, RESET NN seems to be marginally more powerful than RESET Ensemble. 
While this may be the case, we still recommend RESET Ensemble as the default approach to alleviate the burden of selecting any single classification algorithm, as for example, RESET Ensemble appears to perform marginally better in the p-value setting (see Figure 6 of the main text).

One advantage of RESET is its computational speed.
Indeed, our implementation of RESET, as described in Section~\ref{sec:reset_implementation} of the main text, has many steps that can be parallelized: the $r$ evaluations of each classification algorithm on the $K$ folds can all be done in parallel using multiple cores. In contrast, 
Adaptive Knockoffs must iteratively apply a classification algorithm for every nonzero scoring hypothesis in the candidate set $\mathcal{\hat S}_t$ until the estimated FDR is $\leq \alpha$, or all hypotheses are revealed. Figure~\ref{fig:reset_comp_all_methods}B demonstrates this advantage showing the (log10) runtimes of RESET (using 20 CPUs) and of Adaptive Knockoffs at the 5\% FDR level across 5 runs of Simulation 4. We find that each version of RESET is significantly faster than each version of Adaptive Knockoffs.
RESET Ensemble is around $10^3$ to $10^4$ times faster than AdaKO EM, $10^2$ to $10^3$ times faster than AdaKO RF, and about 3-5 times faster than AdaKO LR and AdaKO GAM.

\begin{figure}[h]
    \centering
    \begin{tabular}{l}
    \includegraphics[width=5in]{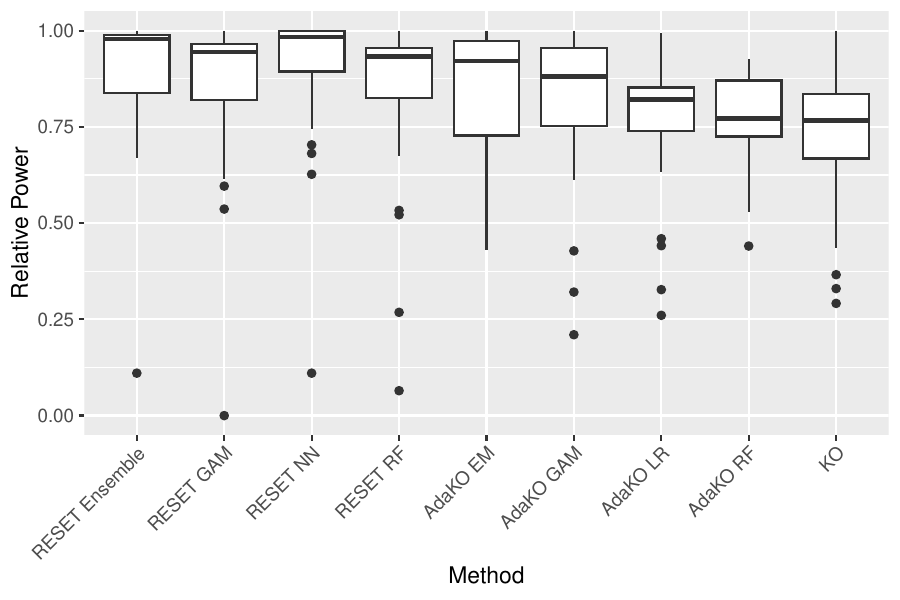}\\
    \end{tabular}
    \caption{\textbf{Relative power of each method under approximately valid knockoffs.}
    Each boxplot describes the relative power of a method, computed across the combination of simulations and FDR thresholds shown in Figure~\ref{fig:all_power_approx}: for each simulation--threshold combination, the per-method power is divided by the maximum power achieved by any RESET or AdaKO method at that combination, and the resulting ratios are aggregated into the boxplot.
    This figure is the approximately-valid-knockoff analogue of Figure~\ref{fig:reset_comp_all_methods}A.}
    \label{fig:reset_comp_all_methods_approx}
\end{figure}

\begin{figure}[h]
    \centering
    \begin{tabular}{l}
    \includegraphics[width=6in]{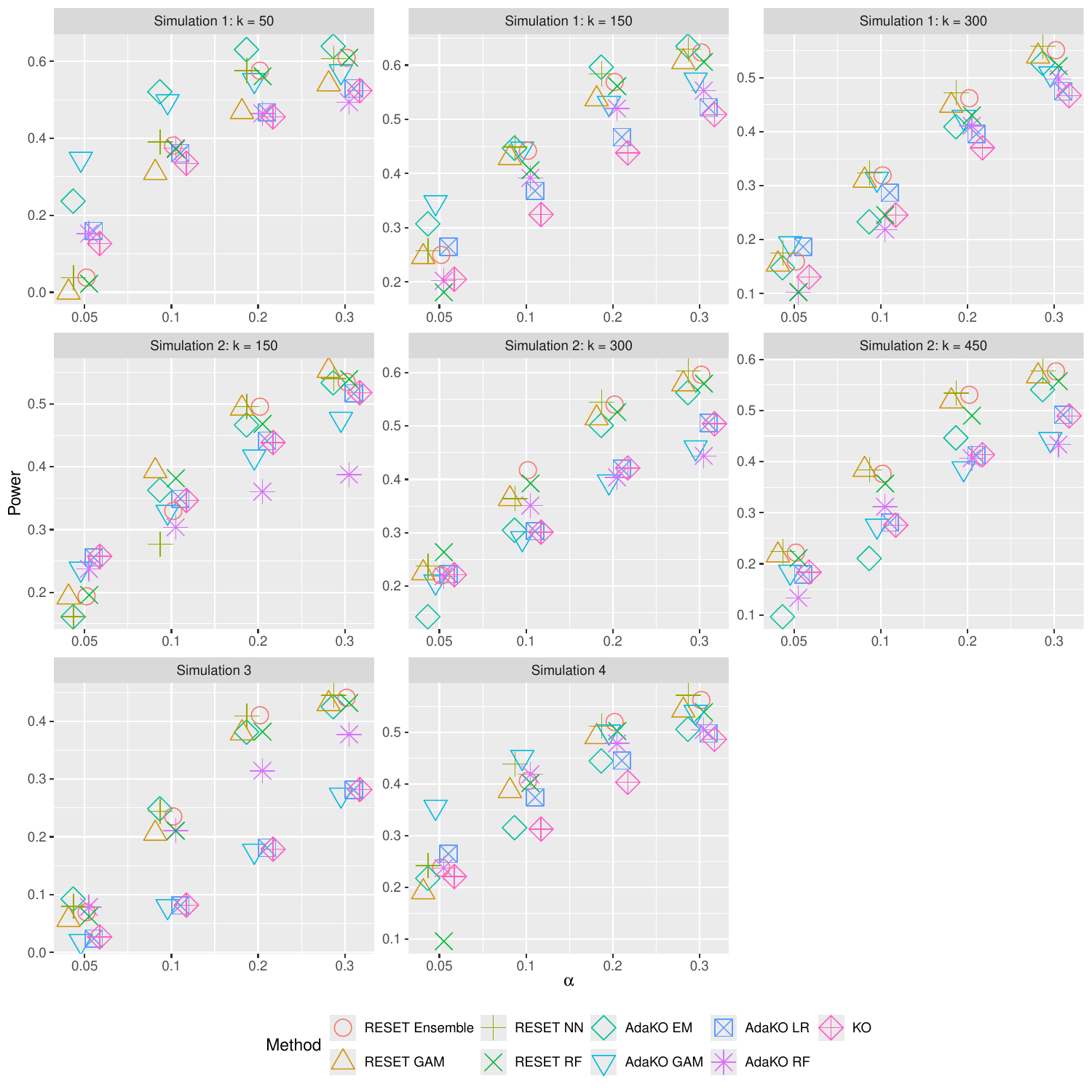}\\
    \end{tabular}
    \caption{\textbf{Estimated power of each method under approximately valid knockoffs.}
    Each panel shows the power for each method at FDR thresholds ranging from 5\% to 30\%, computed in the analogues of Simulations~1--4 in which the design $X$ is Gaussian (or unchanged, for Simulation~3) and knockoffs are constructed using the second-order approximation of~\cite{candes:panning} via \texttt{create.second\_order} with \texttt{shrink~=~TRUE}.
    The first row corresponds to Simulation 1 with three values of $k \in \{50, 150, 300\}$, the second row corresponds to Simulation 2 with three values of $k \in \{150, 300, 450\}$, and the last row corresponds to Simulations 3 and 4.
    For readability, the points are jittered in the horizontal direction.}
    \label{fig:all_power_approx}
\end{figure}

\begin{figure}[h]
    \centering
    \begin{tabular}{l}
    \includegraphics[width=6in]{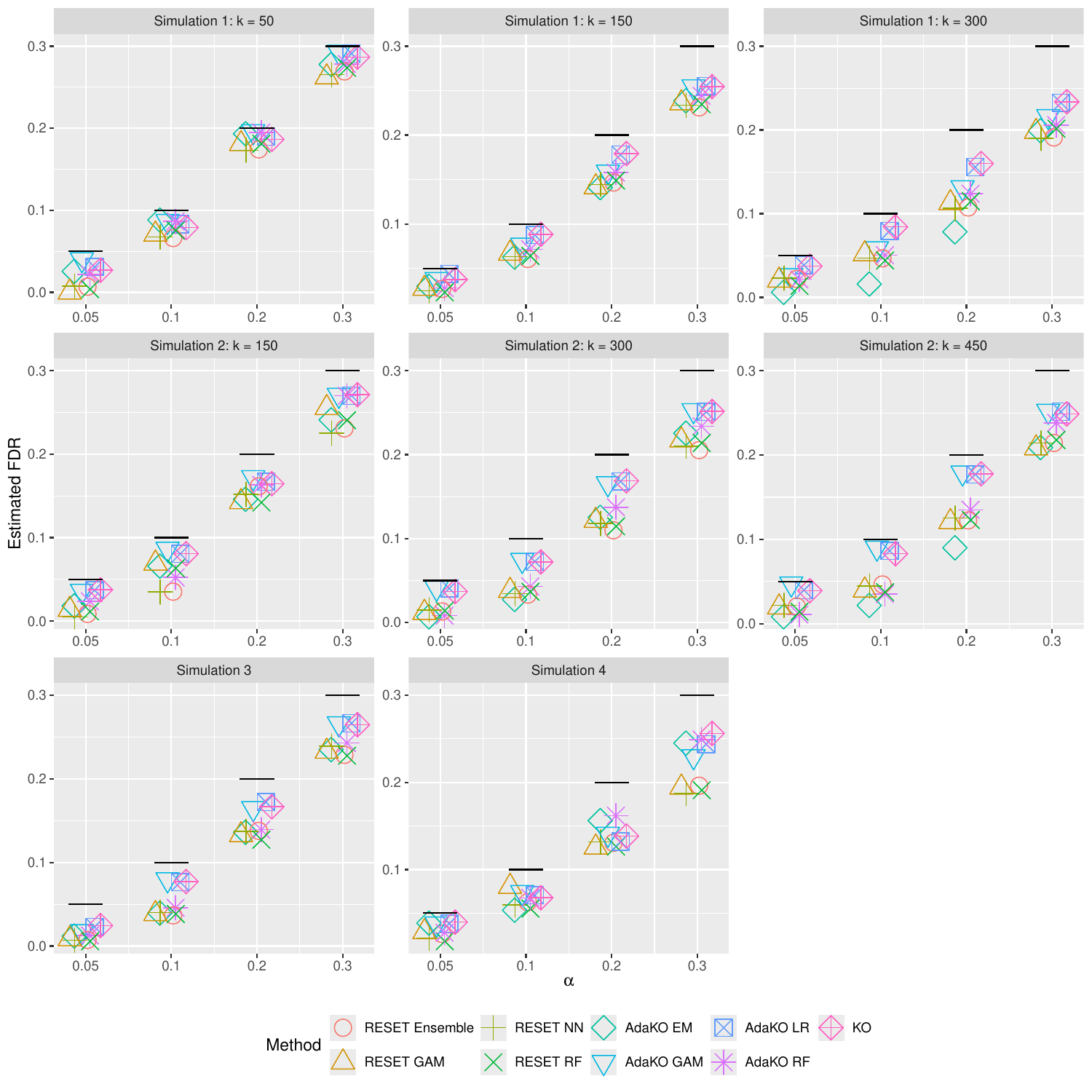}\\
    \end{tabular}
    \caption{\textbf{Estimated FDR of each method under approximately valid knockoffs.}
    Each panel shows the estimated FDR for each method at FDR thresholds ranging from 5\% to 30\%, computed as the proportion of irrelevant variables in the discovery list averaged over the runs, under the same approximately-valid-knockoff setup as in Figure~\ref{fig:all_power_approx}.
    The black horizontal segments mark the corresponding nominal FDR threshold on the y-axis.
    The first row corresponds to Simulation 1 with three values of $k \in \{50, 150, 300\}$, the second row corresponds to Simulation 2 with three values of $k \in \{150, 300, 450\}$, and the last row corresponds to Simulations 3 and 4.
    For readability, the points are jittered in the horizontal direction.}
    \label{fig:all_fdr_approx}
\end{figure}

The simulations in Figure~\ref{fig:all_power} generate knockoffs assuming the true distribution of the design $X$ is known, with HMM knockoffs in Simulations~1,~2 and~4 and exact Gaussian knockoffs in Simulation~3.
In practice, however, the distribution of $X$ is rarely known and knockoffs are at best only \emph{approximately valid}, typically constructed using a second-order approximation that matches only the first two moments of $X$~\cite{candes:panning}.
To assess RESET's sensitivity to this approximation, we repeated Simulations~1--4 from Section~\ref{sec:sim_comp} with two modifications:
\begin{itemize}
    \item In Simulations~1, 2 and~4, we replaced the HMM-distributed $X$ with a Gaussian $X$ having zero-mean and a Toeplitz covariance structure with $\Sigma_{ij} = 0.5^{|i-j|}$.
    Simulation~3 already uses a Gaussian $X$, so its design was left unchanged.
    \item In all four simulations, knockoffs were constructed using the second-order approximation of~\cite{candes:panning} via \texttt{create.second\_order} with shrinkage enabled (\texttt{shrink~=~TRUE}).
    Under this construction, the empirical mean and covariance of $X$ are estimated from the data; in particular, the knockoff generator has no access to the true distribution of $X$.
    In Simulation~4, we set $n = p = 2000$ since the second-order approximation is computationally infeasible for $n = p = 10K$.
\end{itemize}
All other aspects of the simulations (sample sizes, $\beta$ generation, side information, number of repetitions) were unchanged.
The relative-power summary across all methods is reported in Figure~\ref{fig:reset_comp_all_methods_approx} (mirroring Figure~\ref{fig:reset_comp_all_methods}A for the exact-knockoff case), and the per-simulation power and estimated FDR are reported in Figures~\ref{fig:all_power_approx} and~\ref{fig:all_fdr_approx} (mirroring Figures~\ref{fig:all_power} and~\ref{fig:all_fdr}).
Comparing with the corresponding exact-knockoff results, the qualitative picture is essentially unchanged: the relative power ordering of the RESET, AdaKO and KO methods is preserved, with RESET Ensemble and RESET NN achieving the highest median relative power across the simulation-by-threshold combinations, and the estimated FDR of every method remains at or below the nominal threshold across all simulations and FDR levels.

\clearpage
\subsection{FDX control in mass spectrometry data}
\label{sec:comp_fdp}

\begin{figure}[h]
    \centering
    \begin{tabular}{l}
    \includegraphics[width=6in]{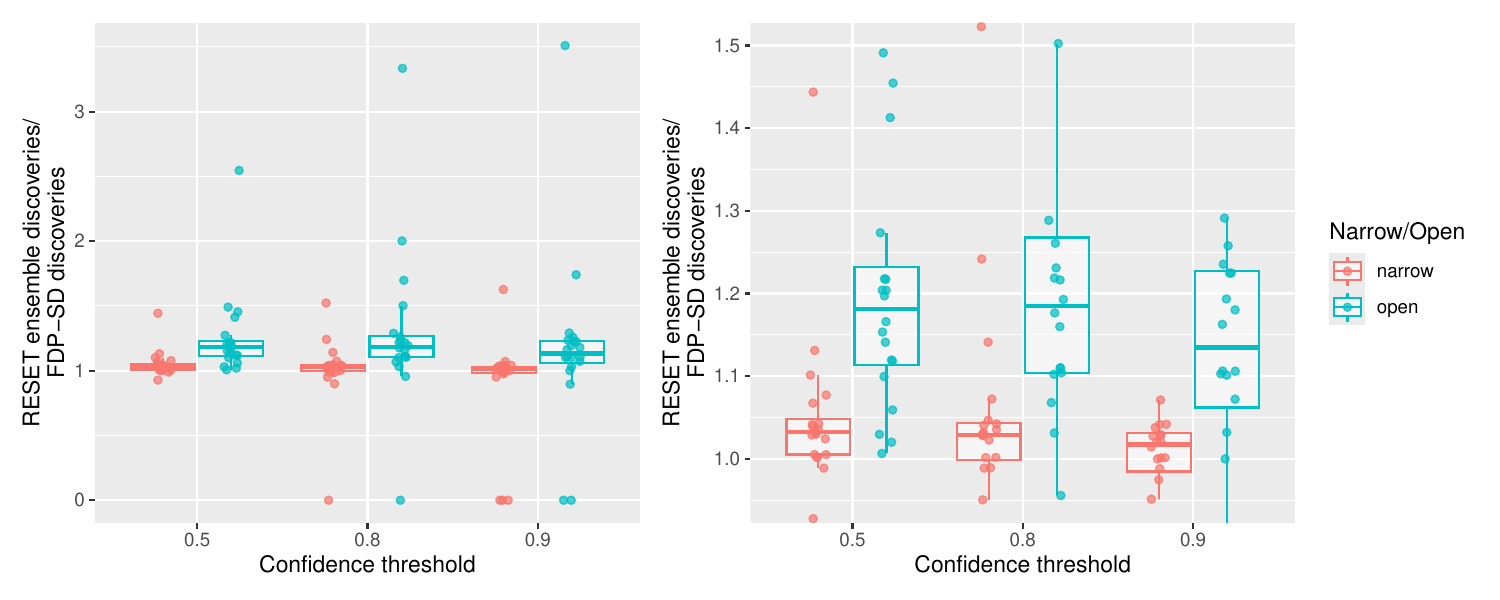}\\
    \end{tabular}
    \caption{\textbf{Comparison of RESET Ensemble and FDP-SD.} The left panel shows boxplots of the average number of discoveries using RESET divided by the average number of discoveries using FDP-SD (the right panel is just a `zoomed-in' version).
    We used all twenty PRIDE-20 datasets at an $\alpha = 1\%$ FDP threshold and varying confidence parameters with both narrow and open search modes.
    The averages are taken over 10 applications, one for each of the 10 target-decoy database pairs we generated for each of the PRIDE-20 dataset.
    }
    \label{fig:pride_20_fdp}
\end{figure}

We evaluated RESET Ensemble's FDX control using the HEK293 and all twenty PRIDE-20 spectrum files (see Sections~\ref{sec:HEK293_supp} and \ref{sec:pride_20_supp} for data preparation and search settings). 
As in Section~\ref{sec:hek293}, the joint HEK293 spectrum files were searched against 5 distinct target-decoy databases, where each target-decoy database was generated by concatenating the target database to a decoy database consisting of shuffled target peptides. 
Similarly, each PRIDE-20 spectrum file was searched against 10 distinct target-decoy databases. 
We used the \texttt{coinflip} version of FDP-SD~\cite{luo:competition}, which provides a uniform increase in the number of discoveries (at the cost of some randomness in the discovery list, see Supplementary Algorithm~\ref{alg:FDP-SD}). Our results using RESET Ensemble and FDP-SD are given in Table~\ref{tab:HEK293_FDP} and Figure~\ref{fig:pride_20_fdp}. 

\begin{table}
    \caption{\label{tab:HEK293_FDP}\textbf{The average number of (HEK293) discoveries while controlling the FDX.} The average number of discoveries reported for RESET Ensemble and FDP-SD on the combined HEK293 spectrum files at different confidence values $1 - \gamma = 0.5, 0.8, 0.9$ and an FDP threshold of 1\%. The averages are taken over 5 applications of each method, once for each of the 5 combined target-decoy databases.}
    \centering
    \begin{tabular}{|c|c|c|c|} 
        \cline{2-4}
        \multicolumn{1}{c|}{} & \multicolumn{3}{c|}{\textbf{Confidence $1-\gamma$}} \\
        \hline
        \textbf{Method} & 0.5 & 0.8 & 0.9 \\
        \hline
        RESET Ensemble & 82930 & 82677 & 82508 \\
        \hline
        FDP-SD & 75977 & 75450 & 75255 \\
        \hline
    \end{tabular}
\end{table}

RESET Ensemble delivers on average approximately 9-10\% more HEK293 discoveries than FDP-SD across all confidence levels (Table~\ref{tab:HEK293_FDP}).
The picture is somewhat murkier when considering the PRIDE-20 datasets: while RESET Ensemble appears to be generally more powerful (Figure~\ref{fig:pride_20_fdp}, zoomed-in right panel),
it fails to make any discoveries at high confidence levels in a few datasets (Figure~\ref{fig:pride_20_fdp}, left panel).
Upon a closer look, the numbers of discoveries made by FDP-SD in these datasets are relatively small, with the largest average number of discoveries among these datasets being 417. 
When the number of discoveries is expected to be small and the confidence parameter is high, RESET may yield less power, even if the side information is reasonably informative ---  a phenomenon we dwell on in Section~\ref{sec:further_reset_fdp}.

RESET Ensemble advantage over FDP-SD is clearer when applying the two procedures to open as opposed to narrow searches.
As discussed in Section~\ref{sec:pride_20_supp}, an open search allows for the detection of peptides whose masses have been altered by \textit{post-translational modifications} (PTMs). 
In this setting, side information variables such as \texttt{dm}, which records the mass difference between the peptide and the spectrum, can be highly informative given that some differences will correspond to masses of common PTMs.
Thus, it is not surprising that RESET, which can take advantage of such side information, improves on FDP-SD even further across all confidence levels in this context.

\clearpage
\subsection{Comparison of AdaKO RF and RESET Ensemble using PRIDE data}
\label{sec:additional_compare}

\begin{figure}[h]
    \centering
    \begin{tabular}{l}
    \includegraphics[width=6in]{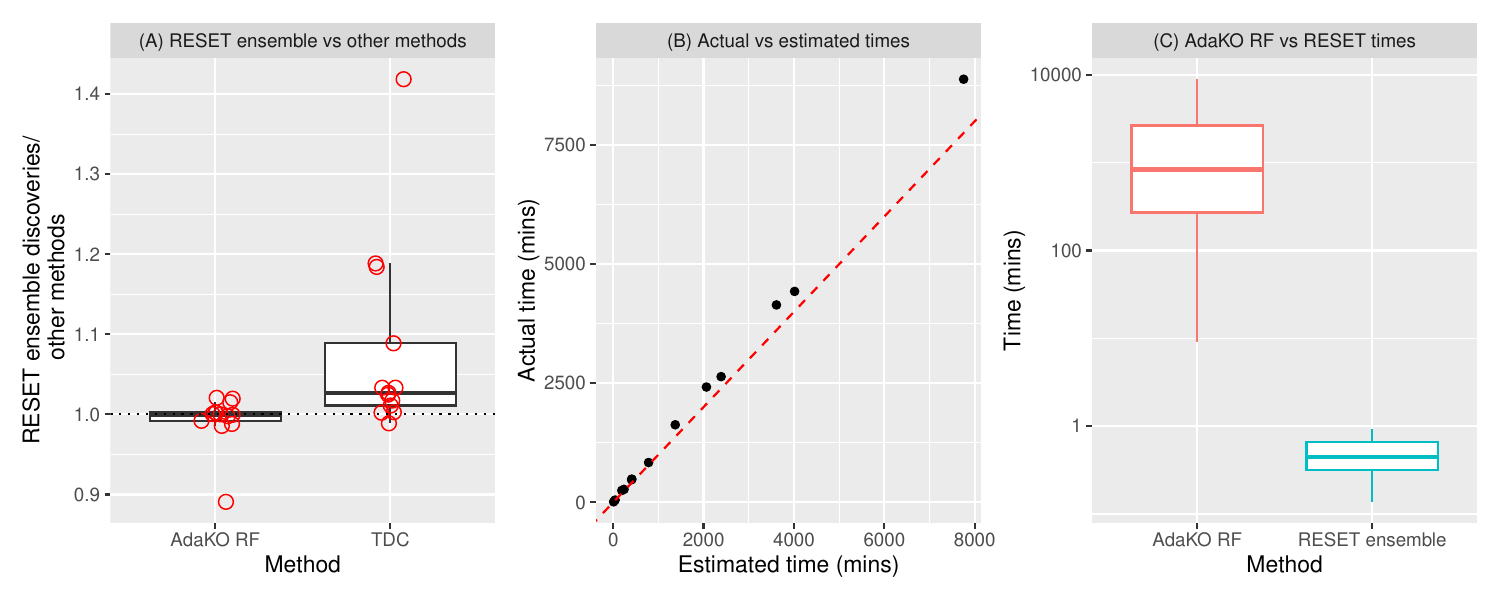}\\
    \end{tabular}
    \caption{\textbf{Comparison of RESET Ensemble and AdaKO RF.} For each of the 13 PRIDE datasets, we computed the average number of discoveries using 10 applications of RESET Ensemble, varying RESET's internal seed. The left panel shows a boxplot of those RESET Ensemble averages divided by the number of discoveries using AdaKO RF (left boxplot) and TDC (right boxplot) where each circle corresponds to one of the 13 smaller PRIDE-20 datasets.
	In the centre panel, we compare the estimated and actual runtimes for AdaKO RF. Lastly, in the right panel, we compare the actual runtimes of RESET Ensemble and AdaKO RF (in log10 scale) on the same 13 PRIDE-20 datasets.
    }
    \label{fig:pride_13_RF}
\end{figure}

We looked at the performance of RESET Ensemble versus AdaKO RF because in this setup it was the most powerful implementation of Adaptive Knockoffs.
Due to the computational constraints of AdaKO RF, we focused on the results using a narrow search.
In Figure~\ref{fig:pride_13_RF}A, the left boxplot displays the average number of discoveries by RESET Ensemble, divided by the number of discoveries by AdaKO RF using thirteen of the PRIDE-20 spectrum files. 
In six of the thirteen PRIDE spectrum files, we found RESET Ensemble to be the superior method, while in the remaining seven spectrum files, AdaKO RF produced the same or more discoveries. 
In one dataset, AdaKO RF discovered 1,217 peptides --- approximately an 11\% increase over the average of 1,084 peptides discovered by RESET Ensemble.

We also assessed the computational times of AdaKO RF in Figure~\ref{fig:pride_13_RF}B, using both the actual and estimated runtimes as outlined in Section~\ref{sec:comp_time}. 
Reassuringly, our estimated times were less than the actual times taken by AdaKO RF to complete. 
Although we find that, when analysing these smaller spectrum sets, AdaKO RF is the most powerful method of Adaptive Knockoffs,
and it is competitive to RESET Ensemble power-wise, it is again clear that AdaKO RF is generally too slow with some runtimes exceeding multiple days even for these small datasets. 
In fact, on the dataset where AdaKO RF outperforms RESET Ensemble by about 11\%, it takes 3.07 days for AdaKO RF to discover its peptides while RESET Ensemble takes about half a minute to complete. 
Figure~\ref{fig:pride_13_RF}C shows that in the worst cases, RESET takes roughly 1 minute while AdaKO RF takes over 6 days.
The issue of runtimes is especially problematic considering the fact that multiple spectrum files are often analysed jointly, as in our HEK293 analysis in Section~\ref{sec:hek293}.

\subsection{The cost of RESET's decoy split}
\label{sec:decoy_split_cost}

RESET's split of the decoys has two competing effects on power: the decoys spent on training are no longer available for estimation, while in return they enlarge the negatively labelled data which is used in learning the improved score function. 
The removal of the training decoys has minimal effect on the overall power of the procedure.
Indeed the remaining estimating decoys are multiplied by a factor ($c_e/(1 - c_e)$) to balance-out the reduced decoy count. While the ``$+ 1$'' offset in the numerator is also multiplied by this factor, the resulting impact is small when the number of discoveries is large.
For example, assuming the labels $L$ of the true nulls are \emph{i.i.d.}\@ $\pm 1$ RVs, and $s = 1/2$, the offset $+ 1$ is multiplied by $c_e/(1 - c_e) = 2$ and so the resulting estimated FDR is increased by only $1/(R(\tau) \vee 1)$, where $R(\tau)$ is the number of discoveries at threshold $\tau$.
Splitting the decoys does, however, increase the variability of this estimate, which in turn may cost some power. 
We assess the net effect empirically below.

To illustrate this effect, we compare three procedures using the same data-generating process of Simulation~1: \emph{Ordinary}, the usual knockoff filter run on all of the statistics at level $\alpha$; \emph{Halved}, the same filter after discarding each decoy with probability $s = 1/2$ and run at level $\alpha/2$; and \emph{RESET Ensemble}.
The rationale of the \emph{Halved} procedure is to match the reduced decoy count of RESET without the rescoring benefits. 

\begin{figure}[h]
    \centering
    \includegraphics[width=5in]{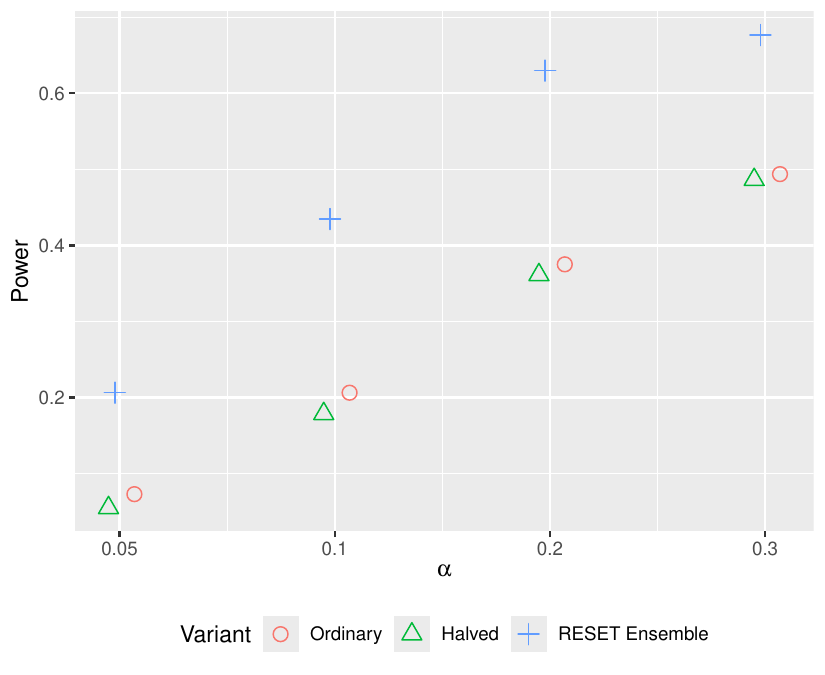}
    \caption{\textbf{The cost of removing decoys is small, and is outweighed by RESET's use of them in rescoring.}
    Power averaged over 100 replications at FDR thresholds $\alpha \in \{5\%, 10\%, 20\%, 30\%\}$ using Simulation~1.
    \emph{Ordinary} is the knockoff filter using all of the statistics at level $\alpha$; \emph{Halved} discards roughly half of the decoy wins and runs the same filter at level $\alpha/2$; \emph{RESET Ensemble} instead rescores the hypotheses using the side information. Because the first two do not incorporate side information, the gap between them isolates the cost of the decoy split. Points are jittered horizontally for readability.}
    \label{fig:kf_decoy_experiment}
\end{figure}

Figure~\ref{fig:kf_decoy_experiment} reports the resulting power. 
While the \emph{Halved} knockoff filter performs uniformly worse than the \emph{Ordinary} knockoff filter, the power loss is relatively mild. 
RESET Ensemble, by contrast, is substantially more powerful than either, roughly doubling the power of the ordinary knockoff filter at $\alpha = 10\%$. 
The modest cost of RESET's decoy split is therefore far outweighed by the gain from using those decoys, together with the side information, in the rescoring step.

Finally, the experiment bears on RESET's advantage over iterative procedures such as Adaptive Knockoffs and AdaPT, which reveal the labels sequentially and so cannot draw on a large pool of confirmed decoys for training upfront, whereas RESET's split supplies exactly that at only the mild estimation cost observed above. 
It is tempting to attribute RESET's greater empirical power to this extra training data. 
We stop short of claiming so outright, since a head-to-head comparison also varies the learner, RESET Ensemble against, for instance, the filters used by Adaptive Knockoffs, so the two effects are confounded. 
Nonetheless, provided both procedures make adequate use of the data, we would expect the additional training decoys to be the decisive advantage.

\clearpage
\subsection{Result-related figures}

\begin{figure}[h]
    \centering
    \begin{tabular}{l}
    \includegraphics[width=6in]{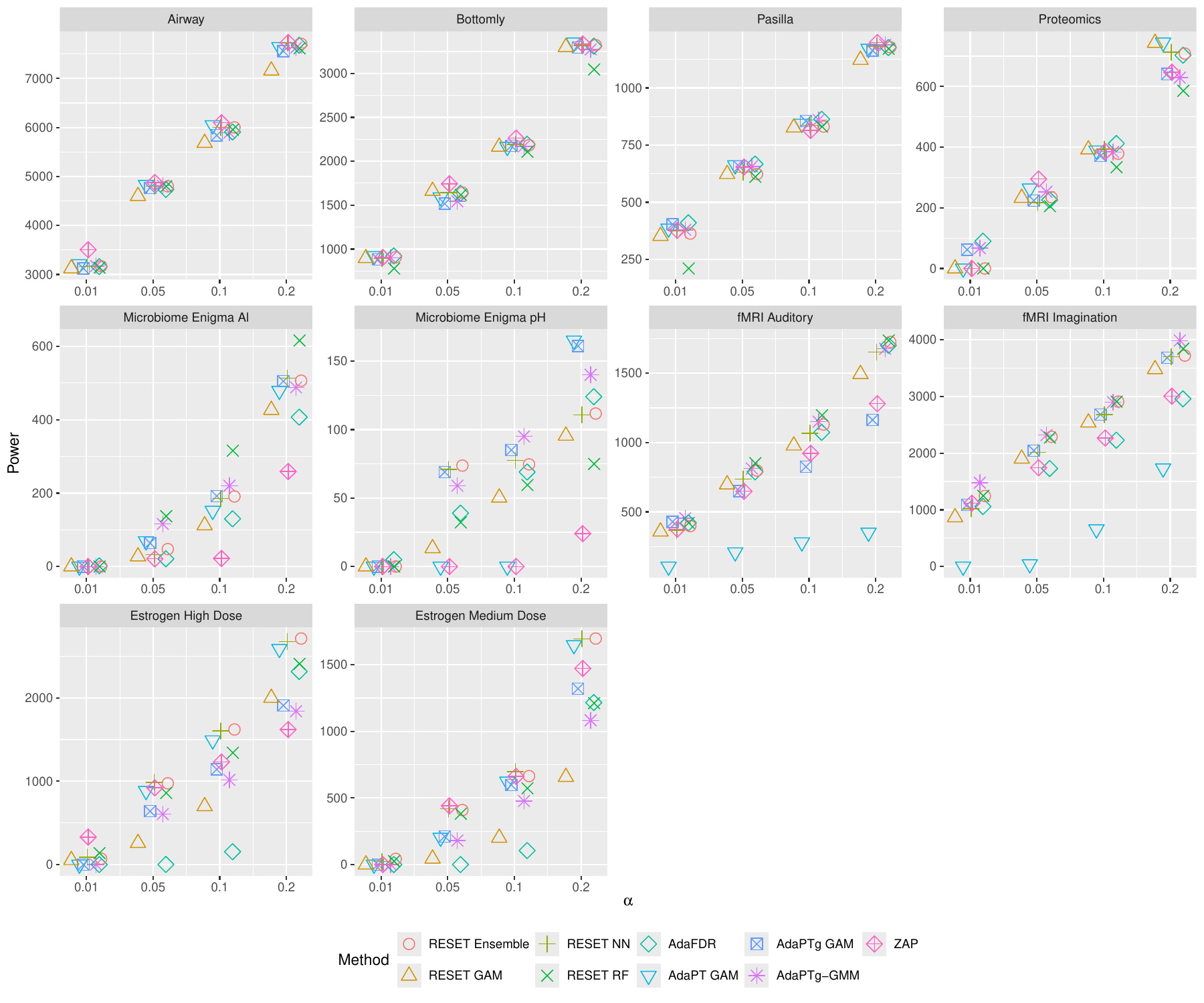}\\
    \end{tabular}
    \caption{\textbf{Comparison of RESET and other methods using a collection of publicly available datasets.} Each method was evaluated at the FDR thresholds of 1\%, 5\%, 10\%, and 20\%. The number of discoveries reported by each RESET method is averaged over 10 applications at FDR threshold and dataset. Each colour and shape combination corresponds to a unique method. For readability, each point is jittered in the horizontal direction. 
    }
    \label{fig:real_data_pvalue}
\end{figure}

\appendix
\renewcommand{\thesection}{F}
\renewcommand{\thealgorithm}{F\arabic{algorithm}}
\renewcommand{\thetable}{F\arabic{table}}
\renewcommand{\thefigure}{F\arabic{figure}}
\renewcommand{\theLemma}{F\arabic{Lemma}}
\renewcommand{\theProposition}{F\arabic{Proposition}}
\renewcommand{\theCorollary}{F\arabic{Corollary}}
\setcounter{figure}{0}
\setcounter{table}{0}
\setcounter{Lemma}{0}
\setcounter{Proposition}{0}
\setcounter{Corollary}{0}

\clearpage
\section{Supplementary analyses}

\subsection{Relationship to CLAW}
\label{sec:zhao_comparison}

During the review of this manuscript, we were made aware of a recent paper by \cite{zhao2025conformalized} that appeared on arXiv, proposing a method (CLAW) for multiple testing with side information.
Like RESET, CLAW constructs a covariate-adaptive score and applies the Selective SeqStep+ procedure of \cite{barber:controlling} for finite-sample FDR control. 
Indeed, in their Section~D.5 the authors situate CLAW within this Barber--Candès (SSS+) family and contrast it with existing procedures in that family.

There are several important differences between RESET and CLAW.
First, RESET works in scenarios when no explicit decoy scores are given (a \emph{null training set} in their paper), but scores and labels may still be obtained regardless, e.g., the p-value setting. 
Second, RESET's Assumption~\ref{assumption:tdc_aug_general} of the main text requires only the one-sided condition $\mathbb{P}(L_i = 1) \leq c_0$ on the true-null labels. 
This assumption makes it compatible in situations when the labels $L_i$ are determined by decoy scores $\tilde Z_i$ that are not necessarily pairwise exchangeable with the target score $Z_i$, e.g., when $\tilde Z_i$ stochastically dominates $Z_i$.
In contrast, CLAW's FDR control is proved on the stronger assumption of pairwise exchangeability which is more akin to when the resulting labels are uniform \emph{i.i.d.}\@ $\pm 1$ RVs (like Assumption (A0$'$)). 
Third, RESET optionally delivers finite-sample FDX control through FDP-SD, whereas CLAW only controls the FDR.
Fourth, the two methods construct their side-information-aware scores differently.
RESET is model-agnostic: its implementation, RESET Ensemble, selects from a library of off-the-shelf machine-learning algorithms with no working model for the data, the requisite symmetry being preserved by its training/estimating decoy split. 
On the other hand, CLAW's implementation builds its score from a conformalized empirical-Bayes model, via density-ratio and null-proportion estimation.

\subsection{Brief justification of the assumptions used with HEK293 and PRIDE-20 datasets}
\label{sec:brief_just}

TDC (equivalently SSS+) has become the \emph{de facto} method of FDR analysis in the proteomics literature since its introduction by \cite{elias2007target}. 
An assumption of TDC for valid FDR control is that the label $L_i$, indicating whether an incorrect target or its decoy scored higher, is uniformly $\pm 1$ independently of all the scores as well as of the labels corresponding to the correct targets (cf.\ the standard competition-based assumption discussed in Section~\ref{sec:background_comp} of the main text).
This assumption relies heavily on the decoy construction which we now outline.

There are two main approaches to decoy construction: randomly shuffling or reversing each target peptide sequence.
This random shuffling or reversing is typically applied to the non-terminal amino acids, i.e., the amino acids between the first and last.
The rationale is that the terminal amino acids of target peptides are typically of a certain type, e.g., lysine or arginine,
and therefore they can be used to distinguish target and decoy peptides if they are not preserved.
By either shuffling or reversing in this way, the result is a set of decoy peptides that have the exact same length, amino acid profile,
and terminal amino acids as their corresponding target peptides, thus creating realistic competition for incorrect target discoveries
in an attempt to satisfy the above assumption.
While both approaches are commonly used, in this paper we use randomly shuffled decoys so that our results can be averaged across different decoy generations.

Importantly, the above assumption has been corroborated empirically through \textit{entrapment experiments}~\cite{granholm2011using}.
In these experiments, a sample of known peptides is used so that the actual FDP can be gauged.
This strategy has allowed others~\cite{he:theoretical,lin2022improving} to show that the FDR is controlled using TDC, suggesting that this assumption is reasonable in practice.

RESET \textit{additionally} requires this statement to be true if we further condition on the side information $\mathbf{x}$. 
This is obviously true for side information variables that are constant between the target and corresponding decoys --- e.g., PepLen, Charge, lnNumSP, dm, absdM. 
The other side information variables are score-related --- e.g., Xcorr, deltCn and deltLCn, and so conditioning on them assumes no more than the `standard assumption' that already conditions on such a score in the first place.
Hence, we believe Assumption~\ref{assumption:tdc_aug_general} with $P(L_i = 1) = c_0 = 1/2$ to be reasonably satisfied.
Moreover, in our recent publication, we empirically demonstrated through entrapment experiments that RESET (combined with Percolator, a semi-supervised algorithm designed for peptide detection) controls the FDR~\cite{freestone2025train}.

\subsection{Comparison of SSS+ to intrinsic p-value methods}
\label{sec:pval_equivalence}

\begin{figure}[h]
    \centering
    \begin{tabular}{l}
    \includegraphics[width=6in]{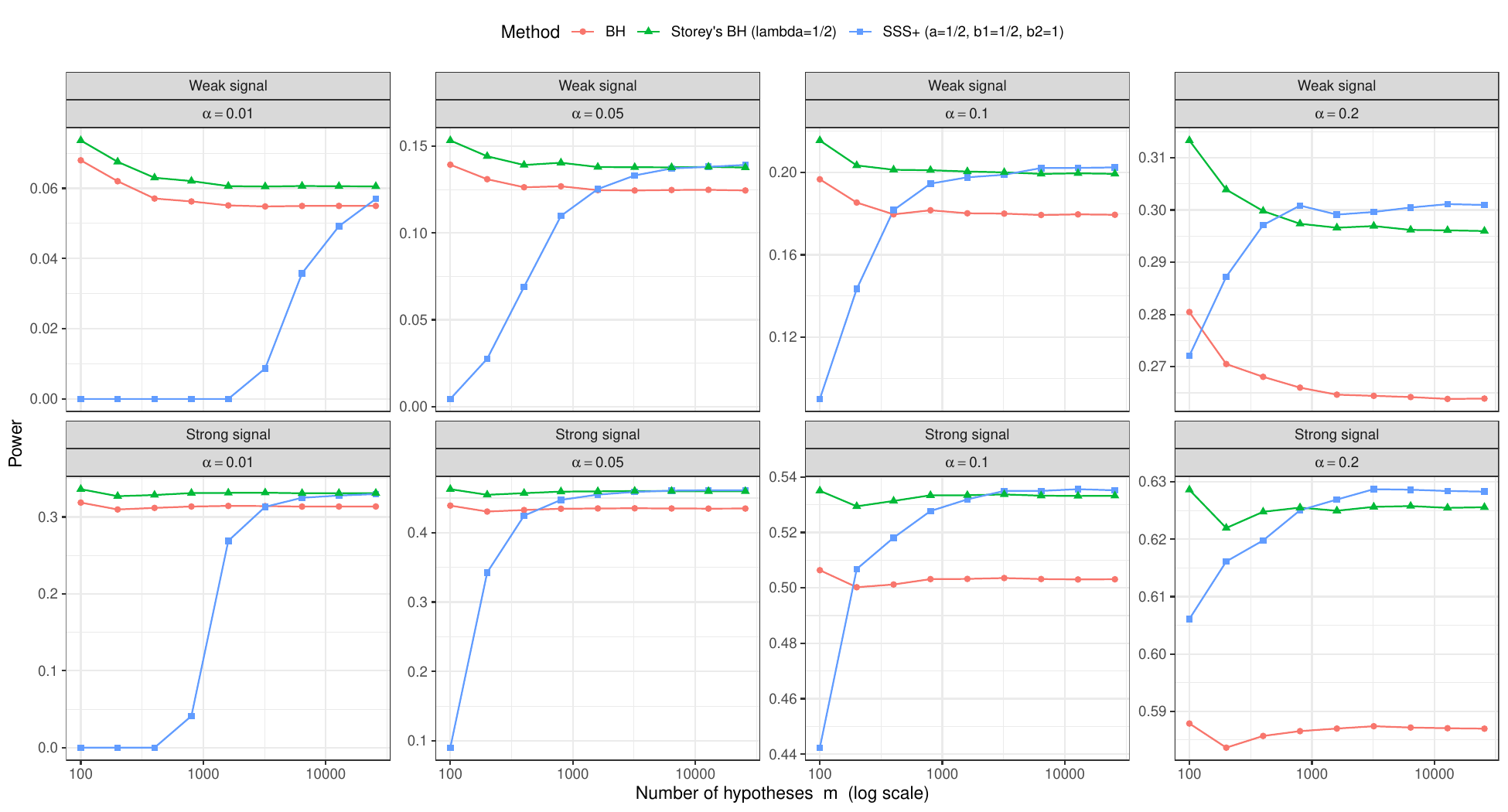}\\
    \end{tabular}
    \caption{\textbf{Power of BH, Storey's BH and SSS+ applied to the p-values}
    Rows: weak ($\mu = 3$) and strong ($\mu = 6$) signal; columns: nominal level
    $\alpha$. All three procedures control the FDR (not shown). SSS+ converges to the
    $\pi_0$-adaptive Storey's BH and exceeds BH, and does so at smaller $m$ under the stronger
    signal. The exception is at smaller $\alpha$ and smaller $m$, where the ``$+1$'' forces SSS+ to accumulate
    at least $\lceil 1/\alpha \rceil$ target wins before it can make any discovery.}
    \label{fig:sss_storey_m}
\end{figure}

We compare SSS+ applied directly to the p-values, Storey's BH with $\lambda = 1/2$, and the BH procedure.
The data follows the same two-group model in Simulation~6 of Section~\ref{sec:pval_100_dim} in the main text.
Given that we are not considering side-information, we fix the true null proportion to $\pi_0 = 0.7$ (matching roughly the average proportion in Simulation~6) and we set $\mu \in \{3, 6\}$ to vary the signal. 
We vary the number of hypotheses $m \in \{100, 200, 400, \dots, 25600\}$ and report the mean power and empirical FDR over $1000$ replicates at $\alpha \in \{0.01, 0.05, 0.1, 0.2\}$ (Figure~\ref{fig:sss_storey_m}).

As $m$ grows, SSS+ converges to Storey's BH, in fact marginally surpassing it at larger values of $m$ and $\alpha$. 
In the case of small $\alpha$ and small $m$, the ``$+1$'' inherent to SSS+ means it makes no discoveries until at least $\lceil 1/\alpha \rceil$ target wins accumulate, so it is briefly powerless.
Hence methods that utilise the SSS+ procedure (e.g., RESET, AdaPT, AdaFDR, ZAP), may not be greatly affected by this conversion depending on the configurations of $m$, $\alpha$, and the signal strength; and their utilisation of side-information clearly outweighs this cost.

\subsection{Analysis of Adipose datasets}
\label{sec:adipose}

\begin{figure}[h]
    \centering
    \begin{tabular}{l}
    \includegraphics[width=6in]{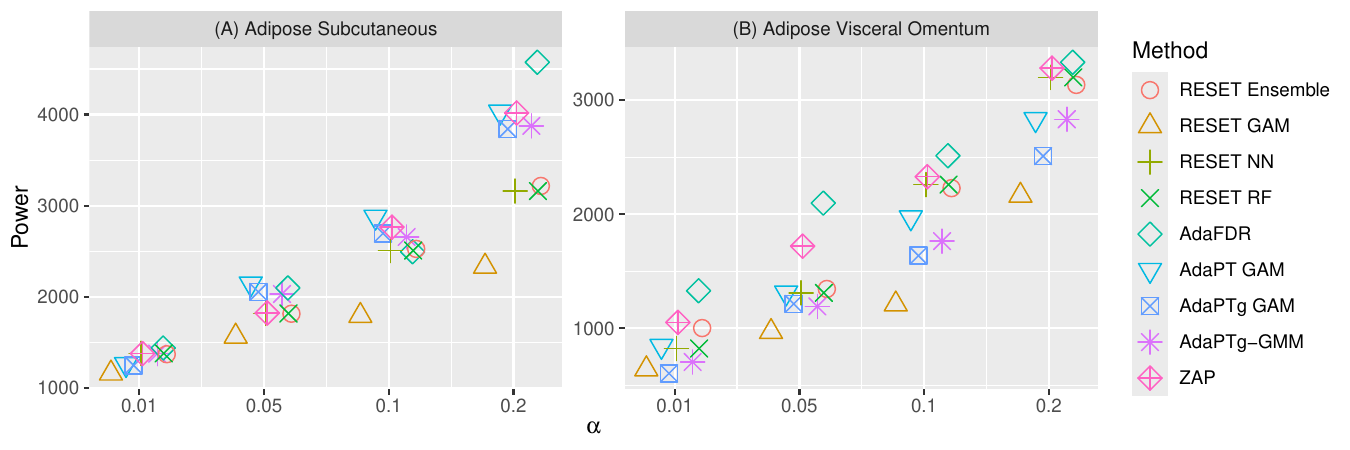}\\
    \end{tabular}
    \caption{\textbf{Comparison of power in Adipose data.} In (A) and (B), we compare the number of discoveries using the two eQTL datasets. The number of discoveries reported by each RESET method is averaged over 10 applications at a given FDR threshold and dataset. In both panels, we evaluated each method at FDR thresholds of: 1\%, 5\%, 10\%, 20\%.
    }
    \label{fig:adipose}
\end{figure}

Notably, we have excluded from Figure~\ref{fig:real_data_pvalue_boxplot} the analysis of the two eQTL studies for the following reason.
Due to linkage disequilibrium these datasets exhibit a high degree of p-value as well as side-information dependency,
even among the true null hypotheses.
Hence these datasets violate Assumption~\ref{assumption:tdc_aug_general} which guarantees our --- as well as the other considered methods' --- type-1 error control.
It is thus unclear whether our comparisons would remain fair, particularly if one method violates FDR control more than the others.
Notably, RESET Ensemble flagged this data as exhibiting suspicious dependence using the feature outlined in Section~\ref{sec:handling_dependent_data}.

\begin{figure}[h]
    \centering
    \begin{tabular}{l}
    \includegraphics[width=4in]{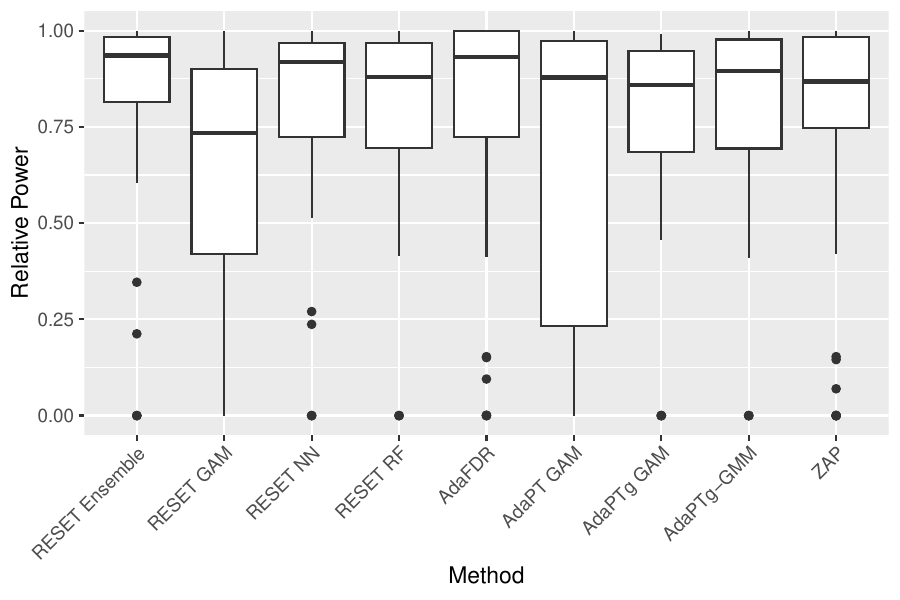}\\
    \end{tabular}
    \caption{\textbf{Relative power of each method.} The boxplots show the relative power of each method, calculated for each dataset described in Section~\ref{sec:description} (including the two Adipose datasets) and across FDR thresholds of 1\%, 5\%, 10\% and 20\%.
    }
    \label{fig:real_data_pvalue_boxplot_all}
\end{figure}

Figure~\ref{fig:adipose}(A-B) shows that overall RESET Ensemble is able to report a similar number of discoveries to other methods,
with the exception of $\alpha = 20\%$ in the Adipose Subcutaneous dataset, where RESET Ensemble reports substantially less discoveries.
Even if we overlook the dependency in the data, and include the two eQTL studies in our power analysis in Figure~\ref{fig:real_data_pvalue_boxplot}, RESET Ensemble still ranks the highest.
Indeed, Figure~\ref{fig:real_data_pvalue_boxplot_all} shows the analogous results with the eQTL studies included.
RESET Ensemble has the highest median at 93.6\%, followed by AdaFDR at 93.2\% and RESET NN at 91.8\%.

\subsection{Further analysis of RESET with FDX control}
\label{sec:further_reset_fdp}

We point out the following limitation on the topic of RESET's FDX control. When the number of discoveries is small, e.g., when the confidence is too high, RESET may yield less power, even if the side information is reasonably informative. To illustrate this phenomenon, we computed the bounds, $\delta_i$ on the number of decoy wins in the top $i$ scores from Algorithm~\ref{alg:FDP-SD}, which are used to determine the reported list of discoveries in FDP-SD (with $c = 1/2$) and RESET (using the default $c = 2/3$). We considered a  confidence level of $1- \gamma = 90\%$ and an FDP threshold of $\alpha = 1\%$. Since we removed approximately half the decoys for training purposes in RESET, the bounds need to be doubled to fairly compare between the two approaches. 

\begin{figure}[h]
    \centering
    \begin{tabular}{ll}
    \includegraphics[width=3in]{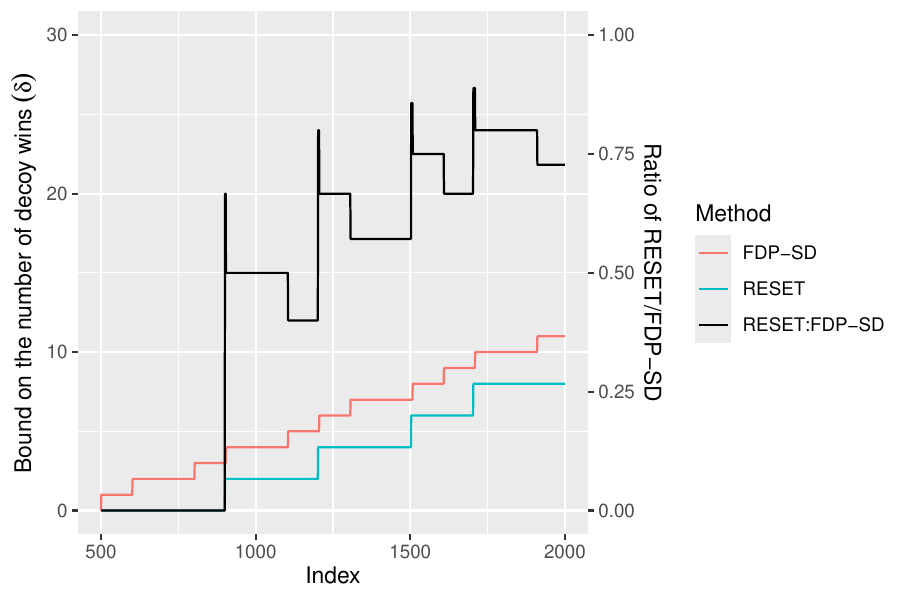} & \includegraphics[width=3in]{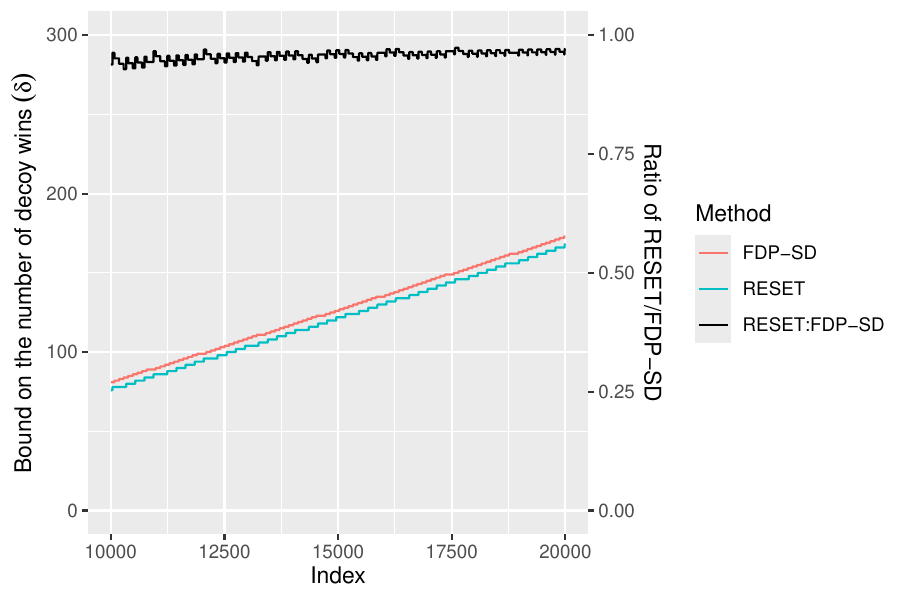} \\
    \end{tabular}
    \caption{\textbf{RESET's and FDP-SD's bounds on the number of decoy wins.}
    We recorded, on the left y-axis, the bounds used to determine the list of FDP-SD discoveries and twice the bounds used to determine RESET's discoveries at a range of indices at $\alpha =1\%$ and $1 - \gamma = 90\%$. On the right y-axis, we plot the ratio of these bounds (in black). The left panel looks at indices between 501 to 2K, while the right panel looks at indices from 10K to 20K. An index corresponds to the number of top scoring hypotheses (regardless of their labels).}
    \label{fig:bounds}
\end{figure}

Figure~\ref{fig:bounds} shows the bounds for FDP-SD, and double the bounds for RESET, along with the ratio of the two at a range of different indices. We can see that for smaller indices, this ratio is much smaller, initially at 0\% at an index of 501 and 50\% at an index of 1K, while at 10K this ratio is much higher at 94\%. If the order of the hypotheses is the same, then RESET's smaller bounds imply it will report fewer discoveries because RESET employs FDP-SD to control the FDX by searching for the largest index $i$ s.t. $A_{i_0} \leq \delta_{i_0}, \dots, A_i \leq \delta_i$, and reports the top $i$ scoring positively labeled hypotheses. In other words, if RESET is to report roughly the same number of discoveries as FDP-SD, then RESET needs to rearrange the targets and estimating decoys successfully enough so as to make up for a 50\% difference in the bounds when considering the top 1K scores, while only 6\% when considering the top 10K. If the side information is not rich enough, 
then it is possible RESET will report fewer discoveries than FDP-SD in these cases. We intend to investigate this as part of future work.

\subsection{Variability of RESET}
\label{sec:variability}

\begin{figure}[h]
    \centering
    \begin{tabular}{l}
    \includegraphics[width=6in]{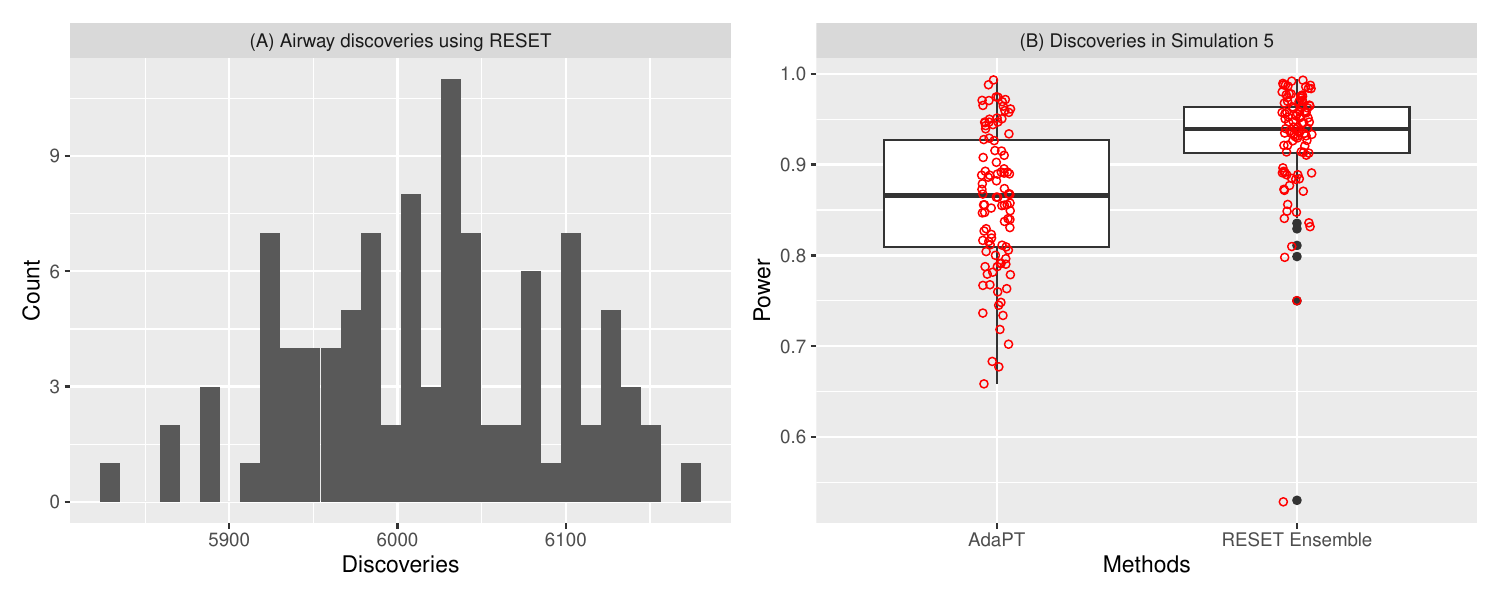} \\
    \end{tabular}
    \caption{\textbf{Variability in the number of discoveries.}
    (A) Histogram of the number of RESET Ensemble discoveries applied 100 times to the Airway data with a 10\% FDR threshold and varying its internal seed. (B) Boxplots of RESET's and AdaPT's power in 100 applications of the `circle in the middle' setup of Simulation 5 using a 10\% FDR threshold. There is one major outlier in RESET's case (which occurred when the signal was considerably low).}
    \label{fig:reset_variance}
\end{figure}

RESET's discovery list is variable due to the random split of the decoys. This variability is demonstrated in the histogram of the number of discoveries in Figure~\ref{fig:reset_variance}A using 100 applications of RESET Ensemble to the Airway dataset at an FDR threshold of 10\%. 
The resulting number of discoveries has a mean of 6018 with a standard deviation of about 75.

While this is undesirable, this analysis does not take into account the randomness of the data, and therefore that the added variance from RESET may be marginal.
Indeed, in Figure~\ref{fig:reset_variance}B, we find that in the case of Simulation 5 from Section~\ref{sec:pval_two_dim}, RESET is overall arguably less variable than AdaPT, even though AdaPT does not employ any internal randomization. 
In practice, users may get around RESET's randomization by setting the random number generator to be a function of the input data and parameters. We plan on adding this feature to a future update.

\clearpage

\bibliographystyle{rss}
\bibliography{copy-of-refs}

\end{document}